\documentclass[pdftex]{article}
\usepackage[pdftex]{graphicx,color}
\usepackage[pdftex,colorlinks]{hyperref}
\usepackage{epstopdf}
\usepackage{amsmath}
\usepackage{amssymb}
\usepackage{amsthm}
\usepackage{array}
\usepackage{xy}
\usepackage{fancyhdr}
\usepackage{marsden_article}
\usepackage[square,numbers,sort&compress]{natbib}

\newcommand\vcent[1]{\ensuremath{\vcenter{\hbox{{#1}}}}}

\makeatletter
\newcommand*\nobreakhyphen{\hbox{-}\nobreak\hskip\z@skip}
\makeatother

\newtheorem*{remarks*}{Remarks}
\newtheorem*{remark*}{Remark}
\newtheorem*{example*}{Example}
\newtheorem*{lemma*}{Lemma}

\hypersetup{%
pdftitle={Geometric, Variational Discretization of Continuum Theories},
pdfauthor={Evan Gawlik et al.},
pdfsubject={},
pdfstartview={FitH},
urlcolor=blue,
citecolor=blue,
linkcolor=red
}%

\begin{document}

\title{Geometric, Variational Discretization of Continuum Theories}

\author{E. S. Gawlik
\and
P. Mullen
\and
D. Pavlov
\and
J. E. Marsden
\and
M. Desbrun
}

\maketitle

\begin{abstract}
This study derives geometric, variational discretizations of continuum theories arising in fluid dynamics, magnetohydrodynamics (MHD), and the dynamics of complex fluids.  A central role in these discretizations is played by the geometric formulation of fluid dynamics, which views solutions to the governing equations for perfect fluid flow as geodesics on the group of volume-preserving diffeomorphisms of the fluid domain.  Inspired by this framework, we construct a finite-dimensional approximation to the diffeomorphism group and its Lie algebra, thereby permitting a variational temporal discretization of geodesics on the spatially discretized diffeomorphism group.  The extension to MHD and complex fluid flow is then made through an appeal to the theory of Euler-Poincar\'{e} systems with advection, which provides a generalization of the variational formulation of ideal fluid flow to fluids with one or more advected parameters.  Upon deriving a family of structured integrators for these systems, we test their performance via a numerical implementation of the update schemes on a cartesian grid.  Among the hallmarks of these new numerical methods are exact preservation of momenta arising from symmetries, automatic satisfaction of solenoidal constraints on vector fields, good long-term energy behavior, robustness with respect to the spatial and temporal resolution of the discretization, and applicability to irregular meshes.
\end{abstract}

\tableofcontents

\section{Introduction}

Fluids, magnetofluids, and continua with microstructure are among the most vivid examples of physical systems with elaborate dynamics and widespread appearance in nature.  For systems of this type -- whose motions are governed by collections of coupled partial differential equations that include the Navier-Stokes equations, Maxwell's equations, and continuum counterparts to Euler's rigid body equations -- numerical algorithms for simulation play an indispensable role as predictors of natural phenomena.  The aim of this paper is to design a family of \emph{structured integrators} for magnetohdyrodynamics (MHD) and complex fluid flow simulations using the framework of variational mechanics and exterior calculus.

Naive discretizations of physical theories can, in general, fail to respect the physical and geometric structure of the system at hand~\cite{Hairer2006}.  A classic example of this phenomenon arises via the application of a standard forward-Euler integrator to a conservative mechanical system, where some of the most basic structures of the continuous flow -- energy conservation, volume preservation, and the preservation of any quadratic or higher-order invariants of motion like angular momentum -- are lost in the discretization.  Even high-order integration schemes, including popular time-adaptive Runge-Kutta schemes for ordinary differential equations, are prone to structure degradation, unless special care is taken to design the integrator with the appropriate goals in mind~\cite{Marsden2001}.

Conventional numerical schemes for MHD and complex fluid flow bear similar defects.  It is well known to the numerical MHD community, for instance, that failure to preserve the divergence-freeness of the magnetic field during simulations can lead to unphysical fluid motions~\cite{Brackbill1980}.  A vast array of MHD literature over the past few decades has been devoted to this issue.  Proposed solutions include divergence-cleaning procedures, the use of staggered meshes, the use of the magnetic vector potential rather than the magnetic field, and even modification of the MHD equations of motion themselves~\cite{Zachary1994,Liu2001,Clarke1986,Powell1994}.

For a variety of physical systems, the problem of structure degradation in numerical integration may be remedied through the design of \emph{variational integrators}~\cite{Lew2004}.  Such integration algorithms made their debut in mechanics, where a discretization of the Lagrangian formulation of classical mechanics allows for the derivation of integrators that are symplectic, exhibit good energy behavior, and inherit a discrete version of Noether's theorem guarantees the exact preservation of momenta arising from symmetries~\cite{Marsden2001,Hydon:2005:MCL}.  An extension of these ideas to the context of certain partial differential equations may be made through an appeal to their variational formulation~\cite{Marsden1998}, and the associated integrators can often be elegantly formulated in the language of exterior calculus on discrete manifolds~\cite{Hiptmair2002,Bochev2006,Arnold:2006:FEEC,Desbrun2008,Arnold:2010:FEEC}.

The fields of magnetohydrodynamics and complex fluid dynamics provide intriguing arenas for the design of structure-preserving integration algorithms.  The former field lies at the confluence of two major domains of the physical sciences -- fluid dynamics and electromagnetism -- and its equations of motion comprise the union of two celebrated systems of equations in modern physics: the Navier-Stokes equations, describing the motion of fluids, and Maxwell's equations, describing the spatial and temporal dependence of the electromagnetic field~\cite{Goedbloed2004}.  The theory of complex fluid flow is equally interdisciplinary, entwining fluid dynamics with continuum versions of Euler's rigid body equations.

\paragraph{Progress in the Structured Integrators Community.}
Conveniently, structured integrators for fluid, electromagnetism, and rigid body simulations have been subjects of active research as we briefly review.

For fluids simulation, Perot et al.~\cite{Perot:2000:CPUSMS,Perot:2002:ACP3D} and Mullen et al.~\cite{Mullen2009} have developed time-reversible integrators that preserve energy exactly for inviscid fluids. Elcott and co-authors~\cite{Elcott2007} have proposed numerically stable integrators for fluids that respect Kelvin's circulation theorem. Cotter et al.~\cite{Cotter:2007:MFF} have provided a multisymplectic formulation of fluid dynamics using a ``back-to-labels'' map, i.e., the inverse of the Lagrangian path map. More recently, Pavlov and co-authors~\cite{Pavlov2009} have derived variational Lie group integrators for fluid dynamics by constructing a finite-dimensional approximation to the volume-preserving diffeomorphism group -- the intrinsic configuration space of the ideal, incompressible fluid. The latter advancement shall serve as the backbone of our approach to the design of integrators for MHD and complex fluid flow.

In the electromagnetism community, Stern and co-authors~\cite{Stern2008} have derived multisymplectic variational integrators for solving Maxwell's equations, proving, in the process, that the popular Yee scheme~\cite{Yee1966} (and its extension to simplicial meshes~\cite{Bossavit:1999:YSTM}) for computational electrodynamics is symplectic. The integrators of Stern et al.~\cite{Stern2008} are formulated in the framework of discrete exterior calculus, giving the added benefit that the integrators may be easily made asynchronous.

Structured rigid body integrators have a longer history, due in large part to the simpler nature of finite-dimensional systems.  Among the many milestones in structured rigid body simulation are Moser \& Veselov's~\cite{Moser1991} integrable discretization of the rigid body, Bobenko \& Suris's~\cite{Bobenko1999} integrable discretization of the heavy top, and Bou-Rabee \& Marsden's~\cite{BouRabee2009} development of Hamilton-Pontryagin-based Lie group integrators.  For a more comprehensive survey, see reference~\cite{BouRabee2009}.

In this report, we combine techniques from the structured fluid, structured electromagnetism, and structured rigid body communities to design a family of variational integrators for ideal MHD, nematic liquid crystal flow, and microstretch continua.  The integrators we derive exhibit all of the classic hallmarks of variational integrators known to the discrete mechanics community: they are symplectic, exhibit good long-term energy behavior, and conserve momenta arising from symmetries exactly.  Moreover, their formulation in the framework of discrete exterior calculus ensures that Stoke's theorem holds at the discrete level, leading to an automatic satisfaction of divergence-free constraints on the velocity and magnetic fields in the resulting numerical schemes.

\paragraph{Layout.}

This paper consists of five main sections. In Section~\ref{section:diff}, we describe the geometric formulation of ideal fluid flow, and we explain the role played by the diffeomorphism group in this formulation;  we summarize some key Lie group theoretic aspects of the diffeomorphism group, and proceed to construct a finite-dimensional approximation of the diffeomorphism group in the manner laid forth by Pavlov and co-authors~\cite{Pavlov2009}. In Section~\ref{section:EP}, we present the theory of \emph{Euler-Poincar\'{e} systems with advected parameters}, which provides the variational framework for all of the subsequent continuum theories presented in this paper.  We then derive a variational temporal discretization of the Euler-Poincar\'{e} equations with advected parameters. In Section~\ref{section:applications}, we state precisely the geometric formulation of ideal fluid flow and proceed to discretize it using the tools developed in Sections~\ref{section:diff}-\ref{section:EP}.  We then discretize three continuum theories: magnetohydrodynamics in 3D, as well as nematic liquid crystal flow and microstretch fluid flow in 2D.  We present these discretizations as methodically and comprehensively as possible in order to highlight the systematic nature of our approach. In Section~\ref{section:cartesian}, we specialize to the case of a cartesian mesh and record the cartesian realizations of the numerical integrators derived in Section~\ref{section:applications}.  Those readers most interested in computational implementation may wish to proceed directly to the end of this section for a concise catalogue of our novel numerical schemes. Finally, in Section~\ref{section:numerical}, we implement our structured integrators on a variety of test cases adapted from the literature.  We focus primarily on our MHD integrator since, relative to complex fluid dynamics, the field of computational MHD is replete with well-established numerical test cases and existing integrators for comparison.  We show numerically that our integrators exhibit good long-term energy behavior, preserve certain conserved quantities exactly, respect topological properties of the magnetic field that are intrinsic to ideal magnetohydrodynamic flows, and are robust with respect to the spatial and temporal resolution of the grid.

Our exposition is largely self-contained, but assumes a working knowledge of Lie groups and Lie algebras.  For the reader's convenience, we give a brief summary of those aspects of Lie theory most relevant to our study in Appendix~\ref{appendix:lie}.

\section{The Diffeomorphism Group and its Discretization} \label{section:diff}

Pioneered by Arnold~\cite{Arnold1966}, the variational formulation of ideal fluid flow stems from the recognition that the governing equations
\begin{align}
\frac{\partial \mathbf{v}}{\partial t} + (\mathbf{v} \cdot \nabla) \mathbf{v} &= - \nabla p \label{fluid1a} \\
\nabla \cdot \mathbf{v} &= 0 \label{fluid2a}
\end{align}
for ideal, incompressible fluid flow \emph{describe geodesic motion on the group $\mathrm{Diff}_{\mathrm{vol}}(M)$ of volume-preserving diffeomorphisms of the fluid domain $M$}.  Equivalently, in the language of mechanicians,~(\ref{fluid1a}-\ref{fluid2a}) are \emph{Euler-Poincar\'{e}} equations on $\mathrm{Diff}_{\mathrm{vol}}(M)$ with respect to a Lagrangian given by the fluid's kinetic energy
\begin{equation}
\ell = \frac{1}{2} \int_M ||\mathbf{v}||^2 d\mathbf{x}.
\end{equation}
This formulation has deep consequences in the analysis of fluid dynamics~\cite{Marsden1972} and, as recently demonstrated by Pavlov and co-authors~\cite{Pavlov2009}, can provide a powerful framework for numerical discretizations of fluid flows.

For our purposes, the key pieces of insight deserving emphasis here are (1) that the configuration space of the ideal fluid is a Lie group, and (2) that the equations of motion~(\ref{fluid1a}-\ref{fluid2a}) on this group are variational.  The approach of this paper will be to construct a finite-dimensional approximation to the group $\mathrm{Diff}_{\mathrm{vol}}(M)$ in the manner of Pavlov and co-authors~\cite{Pavlov2009} and design variational Lie group integrators on the spatially discretized diffeomorphism group.  An extension to MHD and complex fluid flow will later be made through an appeal to the theory of Euler-Poincar\'{e} systems with advection, which provides a generalization of the variational formulation of ideal fluid flow to fluids with one or more advected parameters.

We devote this section to a study of the geometry of the diffeomorphism group, followed by the construction of a spatially discretized diffeomorphism group.

\subsection{The Continuous Diffeomorphism Group}

Let $M$ be a smooth manifold, hereafter referred to as the \emph{fluid domain}.  The \emph{volume-preserving diffeomorphism group} $\mathrm{Diff}_{\mathrm{vol}}(M)$ consists of smooth, bijective maps $\varphi : M \rightarrow M$ with smooth inverses.  The group multiplication in $\mathrm{Diff}_{\mathrm{vol}}(M)$ is given by function composition.

The Lie algebra of $\mathrm{Diff}_{\mathrm{vol}}(M)$ is $\mathfrak{X}_{\mathrm{div}}(M)$, the space of divergence-free vector fields tangent to the boundary of $M$.  Fixing a volume form $d\mathbf{x}$ on $M$, the space dual to $\mathfrak{X}_{\mathrm{div}}(M)$ may be identified with $\Omega^1(M)/d\Omega^0(M)$, the space of one-forms on $M$ modulo full differentials, under the pairing $\langle \cdot, \cdot \rangle : \mathfrak{X}_{\mathrm{div}}(M)^* \times \mathfrak{X}_{\mathrm{div}}(M) \rightarrow \mathbb{R}$ given by
\begin{equation}
\left\langle [\mathbf{w}^\flat], \mathbf{v} \right\rangle = \int_M \mathbf{w}^\flat(\mathbf{v}) \, d\mathbf{x} \label{pairing_diff}
\end{equation}
for any $\mathbf{w}^\flat \in \Omega^1(M), \mathbf{v} \in \mathfrak{X}_{\mathrm{div}}(M)$.  Here, $[\mathbf{w}^\flat]$ denotes the coset of one-forms in $\Omega^1(M)/d\Omega^0(M)$ with representative $\mathbf{w}^\flat$.  For easier reading, we will suppress the brackets when referring to cosets in $\Omega^1(M)/d\Omega^0(M)$ for the remainder of this paper.

\paragraph{Adjoint and Coadjoint Actions.}
Let $\varphi \in \mathrm{Diff}_{\mathrm{vol}}(M)$, $\mathbf{u},\mathbf{v} \in \mathfrak{X}_{\mathrm{div}}(M)$, and $\mathbf{w}^\flat \in \Omega^1(M)/d\Omega^0(M)$. The adjoint and coadjoint actions on $\mathfrak{X}_{\mathrm{div}}(M)$ and its dual are, respectively, the pushforward of vector fields and the pushforward of one-forms:
\begin{align}
\mathrm{Ad}_\varphi \mathbf{v} &= \varphi_* \mathbf{v} \label{Ad_diff} \\
\mathrm{Ad}_{\varphi^{-1}}^* \mathbf{w}^\flat &= \varphi_* \mathbf{w}^\flat. \label{coAd_diff}
\end{align}

The infinitesimal adjoint and coadjoint actions on $\mathfrak{X}_{\mathrm{div}}(M)$ and its dual are given by Lie differentiation:
\begin{align}
\mathrm{ad}_\mathbf{u} \mathbf{v} &= -\pounds_\mathbf{u} \mathbf{v} \label{ad_diff} \\
\mathrm{ad}_\mathbf{u}^* \mathbf{w}^\flat &= \pounds_\mathbf{u} \mathbf{w}^\flat. \label{coad_diff}
\end{align}
Note the sign of~(\ref{ad_diff}); it says that the Lie algebra bracket on $\mathfrak{X}_{\mathrm{div}}(M)$ is \emph{minus} the standard Jacobi-Lie bracket of vector fields.

\subsubsection{The Group Action on Scalar Fields}

The group $\mathrm{Diff}_{\mathrm{vol}}(M)$ acts naturally from the right on $\mathcal{F}(M) = \Omega^0(M)$, the space of scalar fields (zero-forms) on $M$, via the pullback:
\begin{equation}
f \cdot \varphi = \varphi^* f \label{action_diff}
\end{equation}
for any $f \in \mathcal{F}(M)$, $\varphi \in \mathrm{Diff}_{\mathrm{vol}}(M)$.

The induced infinitesimal action of an element $\mathbf{u}$ of $\mathfrak{X}_{\mathrm{div}}(M)$ on $\mathcal{F}(M)$ is given by Lie differentiation:
\begin{equation}
f \cdot \mathbf{u}  = \pounds_\mathbf{u} f. \label{inf_action_diff}
\end{equation}

We close this subsection with two remarks regarding the nature of the action of $\mathrm{Diff}_{\mathrm{vol}}(M)$ on scalar fields that shall motivate the definition of our discrete approximation to the volume-preserving diffeomorphism group.   First, any $\varphi \in \mathrm{Diff}_{\mathrm{vol}}(M)$ clearly preserves constant functions $f \in \mathcal{F}(M)$:
\begin{equation}
f \cdot \varphi = f \;\;\; \mathrm{if} \;\;\; f=\text{const}. \label{constant_preservation_diff}
\end{equation}
In addition, a theorem attributed to Koopman~\cite{Koopman1931} states that such diffeomorphisms preserve inner products of scalar functions $f,h$:
\begin{equation}
\langle f \cdot \varphi, h \cdot \varphi \rangle = \langle f, h \rangle \label{inner_product_preservation_diff}
\end{equation}
Here, the inner product is taken to be the standard $L_2$ inner product of scalar fields.  Note that the latter property relies on the fact that $\varphi$ is volume-preserving, while the former property does not.

\subsection{The Discrete Diffeomorphism Group}

Given a mesh $\mathbb{M}$ on the fluid domain $M$ with cells $\mathcal{C}_i$, $i=1,2,\dots,N$, define a diagonal $N \times N$ matrix $\Omega$ consisting of cell volumes: $\Omega_{ii} = \mathrm{vol}(\mathcal{C}_i)$.  To discretize the group $\mathrm{Diff}_{\mathrm{vol}}(M)$, define
\begin{equation}
\mathcal{D}(\mathbb{M}) = \{q \in GL(N)^+ \,|\, \displaystyle\sum_j q_{ij} = 1 \;\; \forall i, \,  q^T \Omega q = \Omega \}, \label{ddiff}
\end{equation}
the group of $\Omega$-orthogonal, signed stochastic matrices.  Elements of $\mathcal{D}(\mathbb{M})$ are referred to as \emph{discrete diffeomorphisms}.  Later we shall see that matrices in $\mathcal{D}(\mathbb{M})$ respect properties~(\ref{constant_preservation_diff}-\ref{inner_product_preservation_diff}) in a discrete sense.

The Lie algebra of $\mathcal{D}(\mathbb{M})$, denoted by $\mathfrak{d}(\mathbb{M})$, is the space of $\Omega$-antisymmetric, row-null matrices:
\begin{equation}
\mathfrak{d}(\mathbb{M}) = \{A \in \mathfrak{gl}(N) \,|\, \displaystyle\sum_j A_{ij} = 0 \;\; \forall i, \,  A^T \Omega + \Omega A = 0 \}.
\end{equation}
Elements of $\mathfrak{d}(\mathbb{M})$ are referred to as \emph{discrete vector fields}.

Remarkably, the space dual to $\mathfrak{d}(\mathbb{M})$ may be identified with a discrete analogue of $\Omega^1(M)/d\Omega^0(M)$ under a pairing given by a well-known matrix inner product.  The following definitions and the subsequent theorem make this statement precise.

\begin{definition}{\emph{(Discrete Zero-Forms.)}}
A \textbf{discrete zero-form (scalar field)} is a column vector $F \in \mathbb{R}^N$.  The components of such a vector $F$ are regarded as the cell averages of a continuous scalar field $f \in \mathcal{F}(M)$, i.e. $F_i = \int_{\mathcal{C}_i} \!f d\mathrm{x} \;/\; \mathrm{vol}(\mathcal{C}_i)$.  The space of discrete zero-forms is denoted $\Omega_d^0(\mathbb{M})$.
\end{definition}

\begin{definition}{\emph{(Discrete One-Forms.)}}
A \textbf{discrete one-form} is an antisymmetric matrix $C^\flat \in \mathfrak{so}(N)$.  The space of discrete one-forms is denoted $\Omega_d^1(\mathbb{M})$.
\end{definition}

\begin{definition}{\emph{(Discrete Exterior Derivative of Zero-Forms.)}}
The \textbf{discrete exterior derivative} is the map $d : \Omega_d^0(\mathbb{M}) \rightarrow \Omega_d^1(\mathbb{M})$ taking a discrete scalar field $F$ to the discrete one-form $dF$ whose entries are given by
\begin{equation}
(dF)_{ij} = F_i-F_j, \;\;\; i,j=1,2,\dots, N. \label{exterior_d}
\end{equation}
The image of $d$ is denoted $d\Omega_d^0(\mathbb{M})$ and its constituents are referred to as \textbf{discrete gradients} or \textbf{full discrete differentials}.
\end{definition}

\begin{theorem}{\emph{(The Space Dual to $\mathfrak{d}(\mathbb{M})$.)}}
The space dual to $\mathfrak{d}(\mathbb{M})$ may be identified with $\Omega_d^1(\mathbb{M})/d\Omega_d^0(\mathbb{M})$, the space of discrete one-forms on $\mathbb{M}$ modulo full discrete differentials, under the pairing $\langle \cdot, \cdot \rangle : \mathfrak{d}(\mathbb{M})^* \times \mathfrak{d}(\mathbb{M}) \rightarrow \mathbb{R}$ given by the $\Omega$-weighted Frobenius inner product
\begin{equation}
\left\langle [C^\flat], B \right\rangle = \mathrm{Tr}(C^{\flat T} \Omega B). \label{pairing_ddiff}
\end{equation}
Here, $B$ is a discrete vector field and $[C^\flat]$ denotes the coset of discrete one-forms in $\Omega_d^1(\mathbb{M})/d\Omega_d^0(\mathbb{M})$ with representative $C^\flat \in \Omega_d^1(\mathbb{M})$.
\end{theorem}
\begin{proof}
It is well known that the standard unweighted Frobenius inner product of matrices permits the identification of $\mathfrak{so}(N)$ with its dual.  Since the set $\Omega\mathfrak{d}(\mathbb{M}) := \{\Omega B | B \in \mathfrak{d}(\mathbb{M})\}$ constitutes the subspace of matrices in $\mathfrak{so}(N)$ that are row-null, it suffices to show that $d\Omega_d^0(\mathbb{M})$ is the orthogonal complement to $\Omega\mathfrak{d}(\mathbb{M})$ in $\mathfrak{so}(N)$ with respect to the standard unweighted Frobenius inner product of matrices.

To show this, observe that $\Omega\mathfrak{d}(\mathbb{M})$ has codimension $N-1$ in $\mathfrak{so}(N)$.  Indeed, any $D \in \Omega\mathfrak{d}(\mathbb{M})$ is constrained by a system of $N-1$ independent equations:
\begin{equation*}
\sum_{j} D_{ij} = 0, \;\;\; i=1,2,\dots, N-1.
\end{equation*}
(The nullity of the $N^{th}$ row of $D$ follows from the $N-1$ equalities above together with the antisymmetry of $D$.)

On the other hand, the space $d\Omega_d^0(\mathbb{M})$ of discrete gradients has dimension $N-1$, since any such quantity is defined by the $N$ components $F_1,F_2,\dots,F_N$ of a discrete zero-form $F$, modulo a constant additive factor.  Moreover, the standard unweighted Frobenius inner product $\mathrm{Tr}((dF)^T D)$ of any $D \in \Omega\mathfrak{d}(\mathbb{M})$ with a discrete gradient $dF \in d\Omega_d^0(\mathbb{M})$ is zero by the row-nullity and antisymmetry of $D$.  Hence, the orthogonal complement to $\Omega\mathfrak{d}(\mathbb{M})$ in $\mathfrak{so}(N)$ with respect to the standard unweighted Frobenius inner product coincides with the space of discrete gradients.
\end{proof}

As before, we will suppress the brackets when referring to cosets in $\Omega_d^1(\mathbb{M})/d\Omega_d^0(\mathbb{M})$ for the remainder of this paper.

\paragraph{Adjoint and Coadjoint Actions.}

Let $q \in \mathcal{D}(\mathbb{M})$, $A,B \in \mathfrak{d}(\mathbb{M})$, and $C^\flat \in \mathfrak{d}(\mathbb{M})^*$.  The adjoint and coadjoint actions on $\mathfrak{d}(\mathbb{M})$ and its dual are given by matrix conjugation:
\begin{align}
\mathrm{Ad}_q B &= q B q^{-1} \label{Ad_ddiff} \\
\mathrm{Ad}_{q^{-1}}^* C^\flat &= q C^\flat \Omega q^{-1} \Omega^{-1}. \label{coAd_ddiff}
\end{align}

The infinitesimal adjoint and coadjoint actions on $\mathfrak{d}(\mathbb{M})$ and its dual are given by matrix commutation:
\begin{align}
\mathrm{ad}_A B &= [A,B] \label{ad_ddiff} \\
\mathrm{ad}_A^* C^\flat &= -[A,C^\flat \Omega]\Omega^{-1}. \label{coad_ddiff}
\end{align}
Formulae~(\ref{Ad_ddiff}) and~(\ref{ad_ddiff}) are universal for matrix groups; formulae~(\ref{coAd_ddiff}) and~(\ref{coad_ddiff}) are dictated by the pairing~(\ref{pairing_ddiff}).

\subsubsection{The Group Action on Scalar Fields}

A right action of the group $\mathcal{D}(\mathbb{M})$ on $\Omega_d^0(\mathbb{M})$ can be constructed in the following manner:
\begin{equation}
F \cdot q = q^{-1} F \label{action_ddiff}
\end{equation}
for any $F \in \Omega_d^0(\mathbb{M})$, $q \in \mathcal{D}(\mathbb{M})$.  This action is merely the multiplication of a vector $F$ by a matrix $q^{-1}$; the inversion of $q$ ensures that the action is a right action rather than a left action.

The induced infinitesimal action of an element $A$ of $\mathfrak{d}(\mathbb{M})$ on $\Omega_d^0(\mathbb{M})$ is again given by multiplication:
\begin{equation}
F \cdot A  = -A F. \label{inf_action_ddiff}
\end{equation}

With respect to the $\Omega$-weighted Euclidean inner product $\langle F, H \rangle := F^T \Omega H$ on $\mathbb{R}^N$, the group action described above satisfies a pair of discrete analogues of properties~(\ref{constant_preservation_diff}-\ref{inner_product_preservation_diff}) of the continuous diffeomorphism group action on scalar fields.  Specifically, any $q \in \mathcal{D}(\mathbb{M})$ preserves constant functions:
\begin{equation}
F \cdot q = F \;\;\; \mathrm{if} \;\;\; F = (c,c,\dots,c)^T \in \Omega_d^0(\mathbb{M}), c \in \mathbb{R} \label{constant_preservation_ddiff}
\end{equation}
In addition, such discrete diffeomorphisms preserve inner products of discrete scalar fields $F,H$:
\begin{equation}
\langle F \cdot q, H \cdot q \rangle = \langle F, H \rangle. \label{inner_product_preservation_ddiff}
\end{equation}
The latter property relies on the fact that $q$ is $\Omega$-orthogonal, while the former property follows from the fact that $q$ is (signed) stochastic.

Properties~(\ref{constant_preservation_ddiff}-\ref{inner_product_preservation_ddiff}) illuminate our choice~(\ref{ddiff}) of a finite-dimensional approximation to the diffeomorphism group.

\paragraph{Correspondences.}

Table~\ref{tab:correspondences} summarizes the correspondences between the continuous and discrete diffeomorphism groups.  In light of these correspondences, we shall use the suggestive notation and terminology indicated in Table~\ref{tab:notation} for the various actions of the discrete diffeomorphism group on its Lie algebra, its Lie algebra's dual, and the space of discrete scalar fields.

\clearpage
\renewcommand\arraystretch{2.0}
\begin{table}[t]
	\centering
		\begin{tabular}{|c|c|c|}
		  \hline
			Abstract formulation & Continuous & Discrete \\ \hline
			$G$ & $\mathrm{Diff}_{\mathrm{vol}}(M)$ & $\mathcal{D}(\mathbb{M})$ \\ \hline
			$\mathfrak{g}$ & $\mathfrak{X}_{\mathrm{div}}(M)$ & $\mathfrak{d}(\mathbb{M})$ \\ \hline
			$\mathfrak{g}^*$ & $\Omega^1(M)/d\Omega^0(M)$ & $\Omega_d^1(\mathbb{M})/d\Omega_d^0(\mathbb{M})$ \\ \hline
			$\langle \cdot, \cdot \rangle : \mathfrak{g^*} \times \mathfrak{g} \rightarrow \mathbb{R}$ & $\displaystyle\int_M \mathbf{w}^\flat(\mathbf{v}) \, d\mathbf{x}$ & $\mathrm{Tr}(C^{\flat T} \Omega B)$ \\ \hline
		  $\mathrm{Ad}_g \eta$ & $\varphi_* \mathbf{v}$ & $qBq^{-1}$ \\ \hline
	    $\mathrm{Ad}_{g^{-1}}^* \mu$ & $\varphi_* \mathbf{w}^\flat$ & $q C^\flat \Omega q^{-1} \Omega^{-1}$ \\ \hline
	    $\mathrm{ad}_\xi \eta$ & $-\pounds_\mathbf{u} \mathbf{v}$ & $[A,B]$ \\ \hline
	    $\mathrm{ad}_\xi^* \mu$ & $\pounds_\mathbf{u} \mathbf{w}^\flat$ & $-[A,C^\flat \Omega]\Omega^{-1}$ \\ \hline
	    $G$-action on scalar fields & $\varphi^* f$ & $q^{-1}F$ \\ \hline
		\end{tabular}
		\caption{Table of correspondences between the continuous and discrete diffeomorphism groups.}
	\label{tab:correspondences}
\end{table}

\renewcommand\arraystretch{1.8}
\begin{table}[h]
	\centering
		\begin{tabular}{|c|c|c|}
		  \hline
			Notation & Meaning & Terminology \\ \hline
			$q_* B$ & $qBq^{-1}$ & Pushforward of a discrete vector field \\ \hline
			$q^* B$ & $q^{-1}Bq$ & Pullback of a discrete vector field \\ \hline
			$q_* C^\flat$ & $q C^\flat \Omega q^{-1} \Omega^{-1}$ & Pushforward of a discrete one-form \\ \hline
			$q^* C^\flat$ & $q^{-1} C^\flat \Omega q \Omega^{-1}$ & Pullback of a discrete one-form \\ \hline
			$\pounds_A B$ & $-[A,B]$ & Lie derivative of a discrete vector field \\ \hline
			$\pounds_A C^\flat$ & $-[A,C^\flat \Omega]\Omega^{-1}$ & Lie derivative of a discrete one-form \\ \hline
			$q_* F$ & $qF$ & Pushforward of a discrete zero-form \\ \hline
			$q^* F$ & $q^{-1}F$ & Pullback of a discrete zero-form \\ \hline
		\end{tabular}
		\caption{Table of notation and terminology for the actions of the discrete diffeomorphism group on its Lie algebra, its Lie algebra's dual, and the space of discrete scalar fields.}
	\label{tab:notation}
\end{table}

\clearpage
\subsection{Nonholonomic Constraints} \label{section:nonholonomic}

Define the \emph{constrained set} $\mathcal{S} \subset \mathfrak{d}(\mathbb{M})$ to be the set of matrices $A \in \mathfrak{d}(\mathbb{M})$ whose entries $A_{ij}$ are nonzero only if cells $\mathcal{C}_i$ and $\mathcal{C}_j$ are adjacent.  In addition to enhanced sparsity, the set $\mathcal{S}$ has the physically appealing feature that its members encode exchanges of fluid particles between \emph{adjacent cells only}.  As Pavlov and co-authors~\cite{Pavlov2009} show, a nonzero entry $A_{ij}$ of a matrix $A \in \mathcal{S}$ that discretizes a vector field $\mathbf{v} \in \mathfrak{X}_{\mathrm{div}}(M)$ represents (up to a constant multiplicative factor) the flux of the field $\mathbf{v}$ across the face $D_{ij}$ shared by adjacent cells $\mathcal{C}_i$ and $\mathcal{C}_j$:
\begin{equation}
A_{ij} \approx -\frac{1}{2\Omega_{ii}} \displaystyle\int_{D_{ij}} \mathbf{v} \cdot \hat{\mathbf{n}} \, dS \label{flux_relation}
\end{equation}
In applications, we will use matrices only in $\mathcal{S}$ to represent velocity fields, bearing in mind that this choice corresponds to the imposition of \emph{nonholonomic constraints} on the dynamics of the system; indeed, the commutator of a pair of matrices in $\mathcal{S}$ is not necessarily an element of $\mathcal{S}$.

The imposition of nonholonomic constraints will be effected by \emph{constraining the variations} in our derivations of variational integrators; this, in turn, will amount to replacing equalities on discrete vector fields and one-forms with \emph{weak equalities} in the sense laid forth below.  These notions will all become clearer in Section~\ref{section:EP} when we derive a family of integrators by taking variations of a discrete action and equating those variations to zero to obtain numerical update equations. These equations will later be replaced by weak equalities for the purposes of numerical implementation.

\subsubsection{The Discrete Flat Operator}

Prior to this section, we have employed the notation $C^\flat$ to refer to arbitrary elements of $\Omega_d^1(\mathbb{M})/d\Omega_d^0(\mathbb{M})$.  We shall now introduce an explicit meaning for the symbol $^\flat$ by defining a discrete flat operator taking discrete vector fields to discrete one-forms.  Designing a discrete flat operator in such a way that the discrete theory maintains its parallelism with the continuous theory is a nontrivial task and is perhaps one of the most important contributions of Pavlov and co-authors~\cite{Pavlov2009} toward the development of the discrete geometry of fluid flow.  We give the definition of the discrete flat operator here, as well as an example of a flat operator on a cartesian mesh, but we refer the reader to~\cite{Pavlov2009} for its derivation and a generalization to irregular meshes.

\begin{definition} {\emph{(Discrete Flat Operator.)}}
Choose a parameter $\epsilon$ that measures the resolution of a mesh $\mathbb{M}$ (e.g. the maximal length of a mesh edge). A \textbf{discrete flat operator} is a map $\flat: \mathcal{S} \rightarrow \Omega_d^1(\mathbb{M})/d\Omega_d^0(\mathbb{M})$ taking discrete vector fields (satisfying the nonholonomic constraints) to discrete one-forms that satisfies
\begin{align}
\langle C_\epsilon^{\flat_\epsilon}, B_\epsilon \rangle &\rightarrow \langle \mathbf{w}^\flat,\mathbf{v} \rangle \\
\langle C_\epsilon^{\flat_\epsilon}, \pounds_{A_\epsilon} B_\epsilon \rangle &\rightarrow \langle \mathbf{w}^\flat,\pounds_\mathbf{u}\mathbf{v} \rangle \\
\langle C^{\flat}, B \rangle &= \langle B^{\flat}, C \rangle
\end{align}
for any $A,B,C \in \mathcal{S}$ that approximate continuous vector fields $\mathbf{u},\mathbf{v},\mathbf{w}$, where the limits above are taken as the mesh resolution $\epsilon$ tends to zero and the parametric dependence of $A,B,C$, and the flat operator on $\epsilon$ have been denoted via subscripting.
\end{definition}

\begin{example*}
\emph{
On a regular two-dimensional regular cartesian grid with spacing $\epsilon$, the operator $\flat : \mathcal{S} \rightarrow \Omega_d^1(\mathbb{M})/d\Omega_d^0(\mathbb{M})$ defined by
\begin{equation}
C^\flat_{ij} = \left\{\begin{array}{rl}
2\epsilon^2 C_{ij} & \text{if } j \in N(i) \\
w_{ij} \epsilon^2 \displaystyle\sum_{k \in N(i)\cap N(j)} (C_{ik}+C_{kj}) & \text{if } j \in N(N(i))
\end{array} \right. \label{cartesian_flat_operator}
\end{equation}
is a discrete flat operator, where $w_{ij}=1$ if cells $\mathcal{C}_i$ and $\mathcal{C}_j$ are share a single vertex and $w_{ij}=2$ if cells $\mathcal{C}_i$ and $\mathcal{C}_j$ belong to the same row or column.
}
\end{example*}

\begin{remark*}
\emph{Throughout this paper, we continue to make no notational distinction between arbitrary elements of $\Omega_d^1(\mathbb{M})/d\Omega_d^0(\mathbb{M})$ and flattened elements of $\mathcal{S}$, even though the latter quantities constitute a proper subset of $\Omega_d^1(\mathbb{M})/d\Omega_d^0(\mathbb{M})$.}
\end{remark*}

\subsubsection{Weak Equalities} \label{section:weak_equalities}

Throughout this report, our use of variational principles for the derivation of numerical integrators will lead to \emph{weak equalities} on discrete forms and discrete vector fields. Such weak equalities will be denoted using the hat notation $\hat{=}$ to remind the reader of the absence of a strong equality. A discrete one-form $C^\flat \in \Omega_d^1(\mathbb{M})/d\Omega_d^0(\mathbb{M})$ will be called weakly null, i.e.,
\begin{equation}
C^\flat \;\hat{=}\; 0, \label{form_hat}
\end{equation}
iff $C^\flat$ is zero when paired with any vector field in the constraint space $\mathcal{S}$, i.e.,
\begin{equation}
\langle C^\flat, Z \rangle = 0 \quad \forall Z \in \mathcal{S}.\label{weak_form_equality}
\end{equation}
The equality~(\ref{weak_form_equality}) holds not for all $Z \in \mathfrak{d}(\mathbb{M})$ but for all $Z$ in a \emph{subset} of $\mathfrak{d}(\mathbb{M})$, the constraint space $\mathcal{S}$, hence the weak equality.
Similarly, a discrete vector field $B \in \mathfrak{d}(\mathbb{M})$ will be called weakly null, i.e.,
\begin{equation}
B \;\hat{=}\; 0, \label{field_hat}
\end{equation}
iff $B$ is orthogonal to every $Z$ in the constraint space $\mathcal{S}$, i.e.,
\begin{equation}
\langle Z^\flat, B \rangle = 0 \quad \forall Z \in \mathcal{S}.\label{weak_field_equality}
\end{equation}

\paragraph{Weak Equalities on Discrete One-Forms.}
The following lemma characterizes the nature of solutions to equalities of the form~(\ref{form_hat}).

\begin{lemma} \label{lemma:discrete_pressure}
Suppose $C^\flat$ is a discrete one-form satisfying $\langle C^\flat, Z \rangle = 0$ for every $Z \in \mathcal{S}$, i.e. $C^\flat \;\hat{=}\; 0$.  Then there exists a discrete zero-form $P$ for which
\begin{equation}
C^\flat_{ij} = P_i - P_j
\end{equation}
for every pair of neighboring cells $\mathcal{C}_i,\mathcal{C}_j$.
\end{lemma}
\begin{proof}
This follows directly from the quotient space structure of $\mathfrak{d}(\mathbb{M})^* = \Omega_d^1(\mathbb{M})/d\Omega_d^0(\mathbb{M})$ together with the sparsity structure of $Z$.
\end{proof}

\paragraph{Weak Equalities on Discrete Vector Fields.}
In a similar vein, equalities of the form~(\ref{field_hat}) require special care.  Most relevant to our studies will be the situation in which the quantity $B$ in~(\ref{field_hat}) has the form $B=X-A$ for a fixed matrix $A \in [\mathcal{S},\mathcal{S}]$ and an undetermined matrix $X \in \mathcal{D}(\mathbb{M})$.  In such a scenario, the solution $X \in \mathcal{D}(\mathbb{M})$ is in general not unique.  While choosing $X=A$ certainly ensures the satisfaction of the equality~(\ref{field_hat}), we are often interested in solutions $X$ that belong to $\mathcal{S}$, the physically meaningful space of discrete vector fields satisfying the nonholonomic constraints.

To achieve this goal, we define a \emph{sparsity operator} to be a map $^\downarrow : [\mathcal{S},\mathcal{S}] \rightarrow \mathcal{S}$ satisfying
\begin{equation}
\langle Z^\flat, A \rangle = \langle Z^\flat, A^\downarrow \rangle \label{sparsity}
\end{equation}
for any $A \in [\mathcal{S},\mathcal{S}], Z \in \mathcal{S}$. Given such a sparsity operator, a physically meaningful solution to the matrix equation $X-A \;\hat{=}\; 0$ is given by $X = A^\downarrow$.  Note that (up to a rescaling of nonzero matrix entries) the sparsity operator $^\downarrow$ is nothing more than the dual of the flat operator with respect to the $\Omega$-weighted Frobenius inner product.

\begin{figure}[t]
\centering
\includegraphics[width=0.5\textwidth]{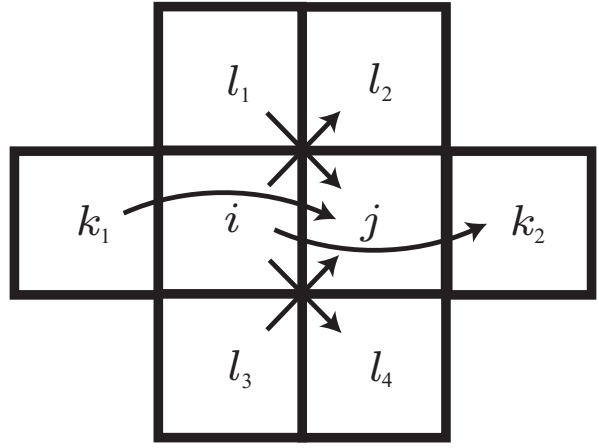}
\caption{Arrangement of cells with indices $i,j,k_1,k_2,l_1,l_2,l_3,l_4$ in definition~(\ref{cartesian_sparsity}) of the sparsity operator on a two-dimensional regular cartesian grid.  The corresponding arrangement of cells for a pair of vertically adjacent cells $\mathcal{C}_i$ and $\mathcal{C}_j$ is given by the 90-degree counterclockwise rotation of the diagram above.  Further rotations in increments of 90 degrees give the remaining two possible orientations of the pair of adjacent cells $\mathcal{C}_i$ and $\mathcal{C}_j$.}
\label{fig:cellsu}
\end{figure}

\begin{example*}
\emph{
On a regular two-dimensional regular cartesian grid with spacing $\epsilon$, the operator $^\downarrow : [\mathcal{S},\mathcal{S}] \rightarrow \mathcal{S}$ given by
\begin{equation}
(A^\downarrow)_{ij} = \left\{\begin{array}{rl}
A_{k_1j} + A_{ik_2} + \frac{1}{2}(A_{l_1j} + A_{il_2} + A_{l_3j} + A_{il_4}) & \text{if } j \in N(i) \\
0 & \text{otherwise}
\end{array} \right. \label{cartesian_sparsity}
\end{equation}
is a sparsity operator, where $k_1,k_2,l_1,l_2,l_3,l_4$ are the indices of the six cells of distance no more than two from both cell $i$ and cell $j$, as depicted in Fig.~\ref{fig:cellsu}.  Notice that the sparsity operator merely accumulates a weighted sum of all two-away transfers that pierce the interface between cells $\mathcal{C}_i$ and $\mathcal{C}_j$.
}
\end{example*}

\subsection{Discrete Loops}

In deriving conservation laws (namely Kelvin's circulation theorem and its variants) for fluid flows, a special role is played by the double dual $\mathfrak{X}_{\mathrm{div}}(M)^{**}$ of $\mathfrak{X}_{\mathrm{div}}(M)$ and its discrete counterpart $\mathfrak{d}(\mathbb{M})^{**}$.  In the continuous case, we will frequently identify $\mathfrak{X}_{\mathrm{div}}(M)^{**}$ with the space of closed loops in $M$ via the pairing
\begin{equation}
\langle \gamma, \mathbf{w}^\flat \rangle = \oint_\gamma \mathbf{w}^\flat,
\end{equation}
where $\gamma : S^1 \rightarrow M$ is a closed loop in $M$ and $\mathbf{w}^\flat \in \mathfrak{X}_{\mathrm{div}}(M)^* = \Omega^1(M)/d\Omega^0(M)$.

In the discrete case, we simply identify the double dual of $\mathfrak{d}(\mathbb{M})$ with itself.  This is consistent with Arnold's~\cite{Arnold1998} treatment of Kelvin's circulation theorem, which, roughly speaking, views the circulation of a one-form $\mathbf{w}^\flat$ around a closed loop $\gamma$ as the limit of the pairing of $\mathbf{w}^\flat$ with a family of ``narrow currents'' $\{\mathbf{v}_\varepsilon\} \subset \mathfrak{X}_{\mathrm{div}}(M)$ of width $\varepsilon$ as $\varepsilon \rightarrow 0$.  Here, $\mathbf{v}_\varepsilon$ denotes a vector field tangent to $\gamma$ that vanishes outside a strip of width $\varepsilon$ containing $\gamma$.

Throughout this paper, we use the term \emph{space of discrete loops} when referring to $\mathfrak{d}(\mathbb{M})^{**} \cong \mathfrak{d}(\mathbb{M})$.

\section{Euler Poincar\'{e} Equations with Advected Parameters} \label{section:EP}

Having discussed the geometry of the diffeomorphism group $\mathrm{Diff}_\mathrm{vol}(M)$ and its discrete counterpart, we now present the theory of Euler-Poincar\'{e} reduction.  This framework will later be used to precisely describe the manner in which solutions to the governing equations for ideal fluid flow may be realized as geodesics on $\mathrm{Diff}_\mathrm{vol}(M)$, or, equivalently, as solutions to the Euler-Poincar\'{e} equations on $\mathrm{Diff}_\mathrm{vol}(M)$ with respect to a kinetic energy Lagrangian $\ell = \frac{1}{2} \int_M ||\mathbf{v}||^2 d\mathbf{x}$.

In anticipation of our eventual extension to MHD and complex fluid flow, we present a generalization of Euler-Poincar\'{e} theory dealing with Euler-Poincar\'{e} systems with advected parameters.  We present this theory in its most general form, namely that in which one is dealing with a physical system whose configuration space is a Lie group $G$, whose advected variables belonging to the dual of a vector space $V$, and whose Lagrangian $L : TG \times V^* \rightarrow \mathbb{R}$ is left- or right-invariant.  We will see in Section~\ref{section:applications} that, with the appropriate choice of the group $G$, vector space $V$, and Lagrangian $L$, a variety of governing equations arising in continuum dynamics can be recast as Euler-Poincar\'{e} equations with advected parameters, including the governing equations for MHD, nematic liquid crystal flow, and microstretch continua.

In the latter half of this section, we derive a structure-preserving discretization of the Euler-Poincar\'{e} equations with advection.  Our derivation directly extends the work of Bou-Rabee and Marsden~\cite{BouRabee2009}, who derived variational Lie group integrators for ordinary Euler-Poincar\'{e} systems without advection.

To maintain consistency with Bou-Rabee and Marsden~\cite{BouRabee2009}, we shall perform derivations for systems with left-invariant Lagrangians whose configuration groups act from the left on the space of advected parameters.  Note, however, that most examples of semidirect products arising in continuum mechanics (including MHD and complex fluid flow) involve systems with right-invariant Lagrangians whose configuration groups act from the right on the space of advected parameters.  We shall take care to present both the left-left and the right-right versions of the theorems that follow, proving the left-left versions and leaving the proofs of the right-right versions as exercises for the reader.

\subsection{The Continuous Euler-Poincar\'{e} Equations with Advected Parameters} \label{section:continuousEP}

The following theorem is due to Holm and co-authors~\cite{Holm1998}:

\begin{theorem} \label{theorem:continuousEP}
Let $G$ be a Lie group which acts from the left on a vector space $V$. Assume that the function $L : TG \times V^* \rightarrow \mathbb{R}$ is left $G$-invariant, so that upon fixing $a_0 \in V^*$, the corresponding function $L_{a_0} : TG \rightarrow \mathbb{R}$ defined by $L_{a_0}(v_g) := L(v_g,a_0)$ is left $G_{a_0}$-invariant, where $G_{a_0} \subseteq G$ denotes the isotropy group of $a_0$.  Define $\ell : \mathfrak{g} \times V^* \rightarrow \mathbb{R}$ by
\begin{equation}
\ell(g^{-1}v_g,g^{-1}a_0) = L(v_g,a_0).
\end{equation}
Let $a(t)=g(t)^{-1}a_0$ and $\xi(t)=g(t)^{-1}\dot{g}(t)$.  Then the following are equivalent:
\begin{enumerate}
\item{Hamilton's variational principle
\begin{equation}
\delta \int_0^T L_{a_0}(g(t),\dot{g}(t)) dt = 0
\end{equation}
holds for variations of $g(t)$ vanishing at the endpoints.}
\item{$g(t)$ satisfies the Euler-Lagrange equations for $L_{a_0}$ on $G$.}
\item{The reduced action integral
\begin{equation}
\int_0^T \ell(\xi(t),a(t)) dt \label{continuous_reduced_action}
\end{equation}
is stationary under variations of $(\xi(t),a(t))$ of the form
\begin{align}
\delta \xi &= \dot{\eta} + [\xi,\eta] \label{xi_variations} \\
\delta a &= -\eta a \label{a_variations},
\end{align}
where $\eta(t)$ is an arbitrary curve in $\mathfrak{g}$ vanishing at the endpoints.}
\item{The Euler-Poincar\'{e} equations hold on $\mathfrak{g} \times V^*$:
\begin{align}
\frac{\partial}{\partial t}\frac{\delta \ell}{\delta \xi} &= \mathrm{ad}_\xi^* \frac{\delta \ell}{\delta \xi} + \frac{\delta \ell}{\delta a} \diamond a \label{EPcontinuous} \\
\frac{\partial a}{\partial t} &= -\xi a, \label{EPcontinuous2}
\end{align}
where $\diamond : V \times V^* \rightarrow \mathfrak{g}^*$ is the bilinear operator defined by
\begin{equation}
\langle \eta a, v \rangle_{V^* \times V} = -\langle v \diamond a, \eta \rangle_{\mathfrak{g}^* \times \mathfrak{g}} \label{diamond}
\end{equation}
for all $v \in V$, $a \in V^*$, and $\eta \in \mathfrak{g}$.}
\end{enumerate}
\end{theorem}

\begin{remarks*}
\emph{
\begin{enumerate}
\item{
If $L$ is right invariant and $G$ acts on $V$ from the right, then equations~(\ref{xi_variations}-\ref{EPcontinuous2}) read
\begin{align}
\delta \xi &= \dot{\eta} - [\xi,\eta] \label{xi_variations_right} \\
\delta a &= -a\eta \label{a_variations_right} \\
\frac{\partial}{\partial t}\frac{\delta \ell}{\delta \xi} &= -\mathrm{ad}_\xi^* \frac{\delta \ell}{\delta \xi} + \frac{\delta \ell}{\delta a} \diamond a \label{EPcontinuous_right} \\
\frac{\partial a}{\partial t} &= -a \xi \label{EPcontinuous2_right}
\end{align}
where $\xi(t)=\dot{g}(t)g(t)^{-1}$, $a(t)=a_0 g(t)^{-1}$, and $\diamond : V \times V^* \rightarrow \mathfrak{g}^*$ is the bilinear operator defined by
\begin{equation}
\langle a\eta, v \rangle_{V^* \times V} = -\langle v \diamond a, \eta \rangle_{\mathfrak{g}^* \times \mathfrak{g}} \label{diamond_right}
\end{equation}
for all $v \in V$, $a \in V^*$, and $\eta \in \mathfrak{g}$.
}
\item{There are also corresponding versions for right-invariant Lagrangians with $G$ acting on $V$ from the left, and left-invariant Lagrangians with $G$ acting on $V$ from the right; for the purposes of this study, only the left-left and right-right cases will be of interest.}
\item{As mentioned in Section~\ref{section:nonholonomic}, we will ultimately impose \emph{nonholonomic constraints} on the dynamics of discrete vector fields and discrete one-forms in our studies of fluid flow.  Such constraints can be effected via two equivalent means: (1) constraining the variations $\eta(t)$ to lie in the nonholonomic constraint space, or (2) replacing the equality~(\ref{EPcontinuous}) with a \emph{weak equality} in the sense laid forth in Section~\ref{section:weak_equalities}.  For a full development of nonholonomic variational principles, see the text of Bloch~\cite{Bloch2003}.}
\end{enumerate}
}
\end{remarks*}

\begin{proof}
We only sketch the proof here; see Holm and co-authors~\cite{Holm1998} for a full exposition.

The equivalence of (1) and (2) is well-known.  To prove the equivalence of (1) and (3), one uses the fact that generic variations of $g(t)$ induce non-generic variations at the level of the Lie algebra.  For matrix groups, this is seen as follows: Letting $\eta = g^{-1}(\delta g) $ shows that $\dot{\eta} = g^{-1}\dot{(\delta g)}  - g^{-1} \dot{g} g^{-1} (\delta g) = g^{-1} \dot{(\delta g)} - \xi \eta$ and hence $\delta \xi$ has the form $\delta \xi = g^{-1} \dot{(\delta g)} - g^{-1} (\delta g) g^{-1} \dot{g} = \dot{\eta} + \xi \eta - \eta \xi = \dot{\eta} + [\xi,\eta]$.

Lastly, to prove the equivalence of (3) and (4), take variations of~(\ref{continuous_reduced_action}), substitute the relations~(\ref{xi_variations}-\ref{a_variations}), and integrate by parts.
\end{proof}

\subsection{The Continuous Kelvin-Noether Theorem}

The following theorem, again due to Holm and co-authors~\cite{Holm1998}, describes a conservation law that may be viewed as an extension of Noether's theorem.

\begin{theorem} \label{theorem:continuousKN}
Let $G$ be a group which acts from the left on a manifold $\mathcal{C}$ and suppose $\mathcal{K} : \mathcal{C} \times V^* \rightarrow \mathfrak{g}^{**}$ is an equivariant map; that is,
\begin{equation}
\mathcal{K}(gc,ga) = \mathrm{Ad}_g^{**} \mathcal{K}(c,a)
\end{equation}
for any $g \in G$, $c \in \mathcal{C}$, and $a \in V^*$, where the actions of $G$ on $\mathcal{C}$ and $V^*$ have been denoted by concatenation.  Suppose the curve $(g(t),\xi(t),a(t))$ satisfies the (left-left) Euler-Poincar\'{e} equations~(\ref{EPcontinuous}-\ref{EPcontinuous2}) with an initial advected parameter value $a_0 \in V^*$.  Fix $c_0 \in \mathcal{C}$ and define $c(t)=g(t)^{-1}c_0$.  Define the (left-left) \textbf{Kelvin-Noether quantity} $I^{ll} : \mathcal{C} \times \mathfrak{g} \times V^* \rightarrow \mathbb{R}$ by
\begin{equation}
I^{ll}(c,\xi,a) = \left\langle \mathcal{K}(c,a), \frac{\delta\ell}{\delta\xi} \right\rangle.
\end{equation}
Then the quantity $I^{ll}(t) := I^{ll}(c(t),\xi(t),a(t))$ satisfies
\begin{equation}
\frac{d}{dt} I^{ll}(t) = \left\langle \mathcal{K}(c,a), \frac{\delta\ell}{\delta a} \diamond a \right\rangle. \label{Ievolution}
\end{equation}
Similarly, suppose $G$ acts from the right on a manifold $\mathcal{C}$ and the map $\mathcal{K}: \mathcal{C} \times V^* \rightarrow \mathfrak{g}$ is equivariant in the sense that
\begin{equation}
\mathcal{K}(cg,ag) = \mathrm{Ad}_{g^{-1}}^{**} \mathcal{K}(c,a)
\end{equation}
for all $g \in G$, $c \in \mathcal{C}$, and $a \in V^*$.  Suppose the curve $(g(t),\xi(t),a(t))$ satisfies the (right-right) Euler-Poincar\'{e} equations~(\ref{EPcontinuous_right}-\ref{EPcontinuous2_right}) with an initial advected parameter value $a_0 \in V^*$.  Fix $c_0 \in \mathcal{C}$ and define $c(t)=c_0 g(t)^{-1}$.  Define the (right-right) {Kelvin-Noether quantity} $I^{rr} : \mathcal{C} \times \mathfrak{g} \times V^* \rightarrow \mathbb{R}$ by
\begin{equation}
I^{rr}(c,\xi,a) = \left\langle \mathcal{K}(c,a), \frac{\delta\ell}{\delta\xi} \right\rangle.
\end{equation}
Then the quantity $I^{rr}(t) := I(c(t),\xi(t),a(t))$ satisfies
\begin{equation}
\frac{d}{dt} I^{rr}(t) = \left\langle \mathcal{K}(c,a), \frac{\delta\ell}{\delta a} \diamond a \right\rangle. \label{Ievolution_right}
\end{equation}
\end{theorem}

\begin{proof}
A proof of this theorem is given in the work of Holm and co-authors~\cite{Holm1998}; the proof uses the equivariance of $\mathcal{K}$ together with the Euler-Poincar\'{e} equations to differentiate $I^{ll}$ and manipulate $\frac{dI^{ll}}{dt}$ into the form of~(\ref{Ievolution}).
\end{proof}

\begin{corollary}
In the absence of the advected quantity $a(t)$, the Kelvin-Noether quantities $I^{ll} : \mathcal{C} \times \mathfrak{g} \rightarrow \mathbb{R}$ and $I^{rr} : \mathcal{C} \times \mathfrak{g} \rightarrow \mathbb{R}$ defined analogously as above are conserved by the Euler-Poincar\'{e} equations~(\ref{EPcontinuous}-\ref{EPcontinuous2}) and~(\ref{EPcontinuous_right}-\ref{EPcontinuous2_right}), respectively.
\end{corollary}

\begin{remarks*}
\emph{
\begin{enumerate}
\item{In the absence of advected parameters, the Kelvin-Noether theorem is an instance of Noether's theorem: it relates the symmetry of the Lagrangian under the left (respectively, right) action of the group $G$ on itself to the conservation of a momentum $\mathrm{Ad}_{g^{-1}}^* \frac{\delta\ell}{\delta\xi}$ (respectively, $\mathrm{Ad}_g^* \frac{\delta\ell}{\delta\xi}$).}
\item{In the context of ideal fluid flow, we shall see that the Kelvin-Noether theorem gives rise to \emph{Kelvin's circulation theorem}: the line integral of the velocity field along any closed loop moving passively with the fluid is constant in time.}
\end{enumerate}
}
\end{remarks*}

\subsection{The Discrete Euler-Poincar\'{e} Equations with Advected Parameters}

A discretization of the Euler-Poincar\'{e} equations with advected parameters may be obtained via a small modification of the discrete reduction procedure detailed by Bou-Rabee~\cite{BouRabee2007}.  Let $\tau : \mathfrak{g} \rightarrow G$ be a local approximant to the exponential map on $G$ (hereafter referred to as a \emph{group difference map}), let $h \in \mathbb{R}$ be a time step, and define the \emph{reduced action sum} $s_d^{a_0} : G^{K+1} \rightarrow \mathbb{R}$ with fixed $a_0 \in V^*$ by
\begin{equation}
s_d^{a_0}(\{g_k\}_{k=0}^K) =
\displaystyle\sum_{k=0}^{K-1} \ell(\xi_k,a_k)h \label{reduced_action}
\end{equation}
with
\begin{equation}
\xi_k = \tau^{-1}(g_k^{-1}g_{k+1})/h \label{xik}
\end{equation}
and
\begin{equation}
a_k = g_k^{-1}a_0. \label{advection}
\end{equation}
Notice that~(\ref{reduced_action}) provides a direct approximation to the action integral~(\ref{continuous_reduced_action}) via numerical quadrature.  Hamilton's principle in this discrete setting states that
\begin{equation}
\delta s_d^{a_0} = 0
\end{equation}
for arbitrary variations of $g_k \in G$ subject to $\delta g_0=\delta g_K=0$.

Before proceeding with a derivation of the discrete update equations arising from this principle of stationary action, let us recapitulate a few properties of group difference maps $\tau : \mathfrak{g} \rightarrow G$ derived in~\cite{BouRabee2007}.  For reference, the basic Lie group theoretic notations appearing below are fixed in Appendix~\ref{appendix:lie}.

\begin{definition} \emph{(Group Difference Map.)} \label{def:tau}
A local diffeomorphism $\tau : \mathfrak{g} \rightarrow G$ taking a neighborhood $\mathcal{N}$ of $0 \in \mathfrak{g}$ to a neighborhood of the identity $e \in G$ with $\tau(0)=e$ and $\tau(\xi)^{-1}=\tau(-\xi)$ for all $\xi \in \mathcal{N}$ is called a \textbf{group difference map}.
\end{definition}

\begin{definition} \emph{(Right-Trivialized Tangent.)}
The \textbf{right-trivialized tangent} $d\tau : \mathfrak{g} \times \mathfrak{g} \rightarrow \mathfrak{g}$ of a group difference map $\tau$ is defined by the relation
\begin{equation}
D\tau(\xi) \cdot \delta = TR_{\tau(\xi)} d\tau_\xi(\delta).
\end{equation}
This definition decomposes the differential of $\tau(\xi)$ into a map $d\tau_\xi$ on the Lie algebra and a translation from the tangent space at the identity of $G$ to the tangent space at $\tau(\xi)$.
The right-trivialized tangent satisfies the following useful identity:
\begin{equation}
d\tau_\xi(\delta) = \mathrm{Ad}_{\tau(\xi)} d\tau_{-\xi}(\delta).
\end{equation}
\end{definition}

\begin{definition} \emph{(Inverse Right-Trivialized Tangent.)}
The \textbf{inverse right-trivialized tangent} $d\tau^{-1} : \mathfrak{g} \times \mathfrak{g} \rightarrow \mathfrak{g}$ of a group difference map $\tau$ is defined by the relation
\begin{equation}
D\tau^{-1}(g) \cdot \delta = d\tau_\xi^{-1}(TR_{\tau(-\xi)}\delta) \label{dtauinverse}
\end{equation}
with $g = \tau(\xi)$.
This definition decomposes the differential of $\tau^{-1}(g)$ into a translation from the tangent space at $g$ to the tangent space at the identity of $G$ and a map $d\tau_\xi^{-1}$ on the Lie algebra.
The inverse right-trivialized tangent satisfies the following useful identity:
\begin{equation}
d\tau^{-1}_\xi(\delta) = d\tau^{-1}_{-\xi}(\mathrm{Ad}_{\tau(-\xi)}\delta). \label{inverse_identity}
\end{equation}
\end{definition}

A convenient formula for computing the inverse right-trivialized tangent of a given group difference map $\tau$ is given by
\begin{align}
d\tau_\xi^{-1}(\delta) = \left.\frac{d}{d\varepsilon}\right|_{\varepsilon=0} \tau^{-1}(\exp(\varepsilon\delta)\tau(\xi)). \label{dtauinv_formula}
\end{align}

For proofs of the properties enumerated above, the reader is referred to~\cite{BouRabee2007} and the references therein.  We now present the main theorem of this section, which provides a discrete analog of the continuous Euler-Poincar\'{e} principle with advection discussed in Section~\ref{section:continuousEP}.

\begin{theorem} \label{discreteHP}
Let $G$ act from the left on the vector space $V$ and assume that the function $L : TG \times V^* \rightarrow \mathbb{R}$ is left-invariant.  Let $\ell \, : \mathfrak{g} \times V^* \rightarrow \mathbb{R}$ be the left-trivialization of $L$ as in Theorem~\ref{theorem:continuousEP}.  Suppose $\{g_k,\xi_k,a_k\}_{k=0}^K$ is a discrete path for which~(\ref{reduced_action}) is stationary under variations of $g_k \in G$ with $\delta g_0=\delta g_K=0$ and with $\xi_k$ and $a_k$ given by~(\ref{xik}-\ref{advection}).  Then the sequence $\{g_k,\xi_k,a_k\}_{k=0}^K$ satisfies the (left-left) \textbf{discrete Euler-Poincar\'{e} equations with an advected parameter}:
\begin{align}
g_{k+1} &= g_k \tau(h\xi_k) \label{EP1} \\
a_{k+1} &= \tau(-h\xi_k) a_k \label{EP2} \\
(d\tau^{-1}_{h\xi_k})^*\frac{\delta\ell}{\delta\xi_k} &= (d\tau^{-1}_{-h\xi_{k-1}})^*\frac{\delta\ell}{\delta\xi_{k-1}} + h\frac{\delta\ell}{\delta a_k} \diamond a_k. \label{EP3}
\end{align}

Similarly, suppose $G$ acts on $V$ from the right, $L$ is right-invariant, and $\xi_k$ and $a_k$ given by
\begin{align}
\xi_k &= \tau^{-1}(g_{k+1}g_k^{-1})/h \\
a_k &= a_0 g_k^{-1}.
\end{align}
Then the stationarity of~(\ref{reduced_action}) implies that the sequence $\{g_k,\xi_k,a_k\}_{k=0}^K$ satisfies the (right-right) {discrete Euler-Poincar\'{e} equations with an advected parameter}:
\begin{align}
g_{k+1} &= \tau(h\xi_k) g_k \label{EP1_right} \\
a_{k+1} &= a_k \tau(-h\xi_k)  \label{EP2_right} \\
(d\tau^{-1}_{-h\xi_k})^*\frac{\delta\ell}{\delta\xi_k} &= (d\tau^{-1}_{h\xi_{k-1}})^*\frac{\delta\ell}{\delta\xi_{k-1}} + h\frac{\delta\ell}{\delta a_k} \diamond a_k, \label{EP3_right}
\end{align}
where $\ell$ is the right-trivialization of $L$.

Notice that~(\ref{EP2}-\ref{EP3}) and~(\ref{EP2_right}-\ref{EP3_right}) are decoupled from the group update laws~(\ref{EP1}) and~(\ref{EP1_right}), respectively.
\end{theorem}

\begin{proof}
We prove the left-left version of the theorem here; the corresponding proof for the right-right version is similar.

Let us first compute the variations $\delta \xi_k$ and $\delta a_k$ induced by variations of $g_k$, in accordance with relations~(\ref{xik}-\ref{advection}).  In terms of the quantity $\eta_k := g_k^{-1} \delta g_k$,
\begin{align}
\delta a_k
&= -g_k^{-1} \delta g_k g_k^{-1} a_0 \nonumber \\
&= -\eta_k a_k.
\end{align}
Similarly,
\begin{align}
\delta\xi_k
&= \delta \tau^{-1}(g_k^{-1}g_{k+1})/h \nonumber \\
&= D\tau^{-1}(\tau(h\xi_k)) \cdot (-TR_{\tau(h\xi_k)}\eta_k + TL_{\tau(h\xi_k)}\eta_{k+1})/h \nonumber \\
&= D\tau^{-1}(\tau(h\xi_k)) \cdot (-TR_{\tau(h\xi_k)}\eta_k + TR_{\tau(h\xi_k)}\mathrm{Ad}_{\tau(h\xi_k)}\eta_{k+1})/h.
\end{align}
Applying definition~(\ref{dtauinverse}) together with the identity~(\ref{inverse_identity}), the latter expression reduces to
\begin{align}
\delta\xi_k
&= -d\tau^{-1}_{h\xi_k}(\eta_k) + d\tau^{-1}_{h\xi_k}(\mathrm{Ad}_{\tau(h\xi_k)}\eta_{k+1}) \nonumber \\
&= -d\tau^{-1}_{h\xi_k}(\eta_k) + d\tau^{-1}_{-h\xi_k}(\eta_{k+1}).
\end{align}

We are now equipped to compute variations of the action~(\ref{reduced_action}) under variations of $\{g_k\}_{k=0}^K$ with fixed endpoints.  In terms of the diamond operator defined by~(\ref{diamond}),
\begin{align}
\delta s_d^{a_0}
&= \displaystyle\sum_{k=0}^{K-1}-\left\langle \eta_k a_k, h\frac{\delta\ell}{\delta a_k}\right\rangle + \displaystyle\sum_{k=0}^{K-1}\left\langle \frac{\delta\ell}{\delta\xi_k}, -d\tau^{-1}_{h\xi_k}(\eta_k) + d\tau^{-1}_{-h\xi_k}(\eta_{k+1})\right\rangle \nonumber \\
&= \displaystyle\sum_{k=0}^{K-1}\left\langle h\frac{\delta\ell}{\delta a_k} \diamond a_k, \eta_k \right\rangle + \displaystyle\sum_{k=0}^{K-1}\left\langle \frac{\delta\ell}{\delta\xi_k}, -d\tau^{-1}_{h\xi_k}(\eta_k) + d\tau^{-1}_{-h\xi_k}(\eta_{k+1})\right\rangle. \label{before_parts}
\end{align}
Using (discrete) integration by parts together with the boundary conditions $\eta_0=\eta_K=0$ gives
\begin{equation}
\delta s_d^{a_0} = \displaystyle\sum_{k=1}^{K-1}\left[\left\langle h\frac{\delta\ell}{\delta a_k} \diamond a_k, \eta_k \right\rangle + \left\langle -(d\tau^{-1}_{h\xi_k})^*\frac{\delta\ell}{\delta\xi_k} + (d\tau^{-1}_{-h\xi_{k-1}})^*\frac{\delta\ell}{\delta\xi_{k-1}}, \eta_k \right\rangle\right]. \label{after_parts}
\end{equation}
This implies that
\begin{equation}
(d\tau^{-1}_{h\xi_k})^*\frac{\delta\ell}{\delta\xi_k} = (d\tau^{-1}_{-h\xi_{k-1}})^*\frac{\delta\ell}{\delta\xi_{k-1}} + h\frac{\delta\ell}{\delta a_k} \diamond a_k
\end{equation}
must hold for each $k=1,2,\dots,K-1$ in order to ensure the stationarity of the action $s_d^{a_0}$.
This relation together with rearrangements of~(\ref{xik}-\ref{advection}) constitute the discrete Euler-Poincar\'{e} equations~(\ref{EP1}-\ref{EP3}) with an advected parameter.
\end{proof}

\begin{remark*}
\emph{This theorem holds for nonholonomically constrained systems, modulo the following amendment: we restict the variations $\{\eta_k\}_{k=0}^K$ to lie in the nonholonomic constraint space, leading to the replacement of~(\ref{EP3}) and~(\ref{EP3_right}) with \emph{weak equalities} in the sense of Section~\ref{section:weak_equalities}.}
\end{remark*}

\subsection{The Discrete Kelvin-Noether Theorem}

The following theorem describes a discrete analogue of the continuous Kelvin-Noether theorem~(\ref{theorem:continuousKN}).  We make a slight departure from the letter of the continuous theory by identifying $\mathfrak{g}^{**}$ with $\mathfrak{g}$. This is done in anticipation of our eventual discretizations of continuum theories, where we perform temporal discretizations \emph{after} performing spatial discretizations of the relevant configuration spaces; needless to say, the spatially discretized groups will be finite-dimensional, implying $\mathfrak{g}^{**} \cong \mathfrak{g}$.

\begin{theorem} \label{discreteKN}
Let $G$ act from the left on a manifold $\mathcal{C}$ and suppose $\mathcal{K}: \mathcal{C} \times V^* \rightarrow \mathfrak{g}$ is an equivariant map; that is,
\begin{equation}
\mathcal{K}(gc,ga) = \mathrm{Ad}_g \mathcal{K}(c,a)
\end{equation}
for any $g \in G$, $c \in \mathcal{C}$, and $a \in V^*$, where the actions of $G$ on $\mathcal{C}$ and $V^*$ have been denoted by concatenation.  Fix a group difference map $\tau : \mathfrak{g} \rightarrow G$ and suppose the sequence $\{g_k,\xi_k,a_k\}_{k=0}^K$ satisfies the (left-left) discrete Euler Poincar\'{e} equations~(\ref{EP1}-\ref{EP3}) with an initial advected parameter value $a_0 \in V^*$.  Fix $c_0 \in \mathcal{C}$ and define $c_k = g_k^{-1} c_0$.  Define the (left-left) \textbf{Kelvin-Noether quantity} $I : \mathcal{C} \times \mathfrak{g} \times V^* \rightarrow \mathbb{R}$ by
\begin{equation}
I^{ll}(c,\xi,a) = \left\langle (d\tau^{-1}_{h\xi})^*\frac{\delta\ell}{\delta\xi}, \mathcal{K}(c,a) \right\rangle.
\end{equation}
Then the quantity $I^{ll}_k := I^{ll}(c_k,\xi_k,a_k)$ satisfies
\begin{equation}
I^{ll}_k - I^{ll}_{k-1} = \left\langle h\frac{\delta\ell}{\delta a_k} \diamond a_k, \mathcal{K}(c_k,a_k) \right\rangle. \label{Ik_evolution_left}
\end{equation}

Similarly, suppose $G$ acts from the right on a manifold $\mathcal{C}$ and the map $\mathcal{K}: \mathcal{C} \times V^* \rightarrow \mathfrak{g}$ is equivariant in the sense that
\begin{equation}
\mathcal{K}(cg,ag) = \mathrm{Ad}_{g^{-1}} \mathcal{K}(c,a)
\end{equation}
for all $g \in G$, $c \in \mathcal{C}$, and $a \in V^*$.  Suppose the sequence $\{g_k,\xi_k,a_k\}_{k=0}^K$ satisfies the (right-right) discrete Euler Poincar\'{e} equations~(\ref{EP1_right}-\ref{EP3_right}) with an initial advected parameter value $a_0 \in V^*$.  Fix $c_0 \in \mathcal{C}$ and define $c_k = c_0 g_k^{-1}$.  Define the (right-right) {Kelvin-Noether quantity} $I : \mathcal{C} \times \mathfrak{g} \times V^* \rightarrow \mathbb{R}$ by
\begin{equation}
I^{rr}(c,\xi,a) = \left\langle (d\tau^{-1}_{-h\xi})^*\frac{\delta\ell}{\delta\xi}, \mathcal{K}(c,a) \right\rangle.
\end{equation}
Then the quantity $I^{rr}_k := I^{rr}(c_k,\xi_k,a_k)$ satisfies
\begin{equation}
I^{rr}_k - I^{rr}_{k-1} = \left\langle h\frac{\delta\ell}{\delta a_k} \diamond a_k, \mathcal{K}(c_k,a_k)\right\rangle.
\end{equation}
\end{theorem}

\begin{proof}
Again, we prove the left-left version of the theorem here and remark that the right-right version is nearly identical.  Since $\mathcal{K}$ is equivariant,
\begin{align}
I^{ll}_k - I^{ll}_{k-1}
&= \left\langle (d\tau^{-1}_{h\xi_k})^* \frac{\delta\ell}{\delta\xi_k}, \mathcal{K}(c_k,a_k)\right\rangle
- \left\langle (d\tau^{-1}_{h\xi_{k-1}})^* \frac{\delta\ell}{\delta\xi_{k-1}}, \mathrm{Ad}_{g_{k-1}^{-1}g_k}\mathcal{K}(c_k,a_k) \right\rangle \nonumber \\
&= \left\langle (d\tau^{-1}_{h\xi_k})^* \frac{\delta\ell}{\delta\xi_k}, \mathcal{K}(c_k,a_k) \right\rangle
- \left\langle \frac{\delta\ell}{\delta\xi_{k-1}}, d\tau^{-1}_{h\xi_{k-1}} (\mathrm{Ad}_{\tau(h\xi_{k-1})}\mathcal{K}(c_k,a_k)) \right\rangle
\end{align}
by~(\ref{EP1}).  Using the identity~(\ref{inverse_identity}) allows us to write
\begin{align}
I^{ll}_k - I^{ll}_{k-1}
&= \left\langle (d\tau^{-1}_{h\xi_k})^* \frac{\delta\ell}{\delta\xi_k}, \mathcal{K}(c_k,a_k) \right\rangle
- \left\langle \frac{\delta\ell}{\delta\xi_{k-1}}, d\tau^{-1}_{-h\xi_{k-1}} (\mathcal{K}(c_k,a_k)) \right\rangle \nonumber \\
&= \left\langle (d\tau^{-1}_{h\xi_k})^* \frac{\delta\ell}{\delta\xi_k}, \mathcal{K}(c_k,a_k) \right\rangle
- \left\langle (d\tau^{-1}_{-h\xi_{k-1}})^* \frac{\delta\ell}{\delta\xi_{k-1}}, \mathcal{K}(c_k,a_k) \right\rangle.
\end{align}
The result~(\ref{Ik_evolution_left}) then follows from~(\ref{EP3}).
\end{proof}

\begin{corollary}
In the absence of the advected quantity $a_k$, the Kelvin-Noether quantities $I^{ll} : \mathcal{C} \times \mathfrak{g} \rightarrow \mathbb{R}$ and $I^{rr} : \mathcal{C} \times \mathfrak{g} \rightarrow \mathbb{R}$ defined analogously as above are conserved by the discrete update equations~(\ref{EP1}-\ref{EP3}) and~(\ref{EP1_right}-\ref{EP3_right}), respectively.
\end{corollary}

\begin{remark*}
\emph{A careful reading of the above proof shows that the validity of Theorem~\ref{discreteKN} persists for nonholonomically constrained systems whenever $\mathcal{K}(c_k,a_k)$ belongs to the prescribed nonholonomic constraint space.  We shall invoke this observation tacitly throughout Section~\ref{section:applications}.}
\end{remark*}

\subsection{Symplecticity of the Discrete Flow}

The following paragraphs prove, via a small extension of the argument presented in~\cite{BouRabee2007}, the symplecticity of the flow of the discrete Euler-Poincar\'{e} equations~(\ref{EP1}-\ref{EP3}) with advected parameters.

For fixed $a_0 \in V^*$, let $\mathcal{A}_d$ denote the set of sequences $\{g_k,\xi_k,a_k\}_{k=0}^K$ satisfying the (left-left) discrete Euler Poincar\'{e} equations~(\ref{EP1}-\ref{EP3}).  Let $\hat{s}_d^{a_0} : \mathcal{A}_d \rightarrow \mathbb{R}$ denote the restriction of the reduced action sum~(\ref{reduced_action}) to sequences in $\mathcal{A}_d$, and define the \emph{initial condition space} to be the set
\begin{equation}
\mathcal{I} = \left\{(g_0,\xi_0) \mid g_0 \in G, \xi_0 \in \mathfrak{g} \right\}.
\end{equation}
Since the update equations~(\ref{EP1}-\ref{EP3}) define a discrete flow $F_d : \mathbb{Z} \times \mathcal{I} \rightarrow \mathcal{I}$, we may equivalently think of $\hat{s}_d^{a_0}$ as a real-valued function $\hat{s}_d^{a_0} : \mathcal{I} \rightarrow \mathbb{R}$ on the initial condition space.

Referring back to the proof of Theorem~\ref{discreteHP}, the differential of $\hat{s}_d^{a_0}$ then consists of the boundary terms that were dropped in passing from~(\ref{before_parts}) to~(\ref{after_parts}):
\begin{equation}
d\hat{s}_d^{a_0} \cdot (\delta g_0,\delta\xi_0) =  \left\langle (d\tau^{-1}_{-h\xi_{K-1}})^*\frac{\delta\ell}{\delta\xi_{K-1}}, g_K^{-1}\delta g_K \right\rangle - \left\langle (d\tau^{-1}_{h\xi_0})^*\frac{\delta\ell}{\delta\xi_0}, g_0^{-1}\delta g_0 \right\rangle + \left\langle h\frac{\delta\ell}{\delta a_0} \diamond a_0, g_0^{-1}\delta g_0 \right\rangle.
\end{equation}
Define the one-forms $\theta^+,\theta^- : T\mathcal{I} \rightarrow \mathbb{R}$ by
\begin{align}
\theta^+(g_k,\xi_k) \cdot (\delta g_k,\delta\xi_k) &= \left\langle (d\tau^{-1}_{h\xi_k})^*\frac{\delta\ell}{\delta\xi_k}, g_k^{-1}\delta g_k \right\rangle - \left\langle h\frac{\delta\ell}{\delta a_k} \diamond a_k, g_k^{-1}\delta g_k \right\rangle \\
\theta^-(g_k,\xi_k) \cdot (\delta g_k,\delta\xi_k) &= \left\langle (d\tau^{-1}_{-h\xi_k})^*\frac{\delta\ell}{\delta\xi_k}, \tau(-h\xi_k)g_k^{-1}(F_d^1)_*\delta g_k \right\rangle
\end{align}
so that
\begin{equation}
d\hat{s}_d^{a_0} \cdot (\delta g_0,\delta \xi_0) = \theta^-(g_{K-1},\xi_{K-1}) \cdot (\delta g_{K-1},\delta\xi_{K-1}) - \theta^+(g_0,\xi_0) \cdot (\delta g_0,\delta\xi_0).
\end{equation}
In terms of the discrete flow $F_d$,
\begin{equation}
d\hat{s}_d^{a_0} = (F_d^{K-1})^* \theta^- - \theta^+.
\end{equation}
Differentiating this relation and using the fact that the pullback commutes with the exterior derivative, we find
\begin{equation}
0 = d^2\hat{s}_d^{a_0} = (F_d^{K-1})^* d\theta^- - d\theta^+.
\end{equation}
Now notice that $d\theta^-$ and $d\theta^+$ are identical: Differentiating a single term of the discrete action sum and recalling the steps that led to~(\ref{before_parts}) in the proof of Theorem~\ref{discreteHP}, we have
\begin{equation}
d \left( \ell(\xi_k,a_k)h \right)
= -\theta^-(g_k,\xi_k)  + \theta^+(g_k,\xi_k)
\end{equation}
and hence $d\theta^- = d\theta^+$ since $d^2=0$.
Defining the two-form $\omega_d = d\theta^- = d\theta^+$ then shows that
\begin{equation}
(F_d^{K-1})^*\omega_d = \omega_d.
\end{equation}
In other words, \emph{the discrete flow preserves the discrete symplectic two-form $\omega_d$}.
A similar argument may be used to prove the symplecticity of the flow defined by the right-right discrete Euler-Poincar\'{e} equations with advected parameters.

\section{Applications to Continuum Theories} \label{section:applications}

This section leverages the theory presented in Sections~\ref{section:diff} and \ref{section:EP} in order to design structured discretizations of four continuum theories: ideal, incompressible fluid dynamics; ideal, incompressible magnetohydrodynamics; the dynamics of nematic liquid crystals; and the dynamics of microstretch continua~\footnote{As mentioned earlier, our numerical treatment of the last two examples of continuum theories will only be valid in 2D; extension to three and higher dimensions requires an alteration of our discretization of Lie-algebra-valued fields that will be addressed in a different paper.}. For each of these continuum theories, we present (1) the continuous equations of motion for the system, (2) the geometric, variational formulation of the governing equations, (3) an application of the Kelvin-Noether theorem to the system, (4) a spatial discretization of the governing equations, (5) a spatiotemporal discretization of the governing equations, and (6) an application of the discrete Kelvin-Noether theorem to the discrete-time, discrete-space system.

The continuous variational formulations that we present here for ideal fluid flow and MHD are well-documented; see references~\cite{Arnold1998,Holm1998,Marsden1984,Holm2005} for a comprehensive exposition.  The variational formulation of nematic liquid crystal flow and microstretch fluid flow is a more recent development due to Gay-Balmaz \& Ratiu~\cite{GayBalmaz2009}.

The reader will discover, upon completion of a first reading of this section, that the procedure employed for each of our discretizations adheres to a canonical prescription.  While the details of this prescription will be made clearer through examples, we summarize here for later reference our systematic approach to the design of structured discretizations of continuum theories:

\begin{enumerate}
	\item{Identify the configuration group and construct a finite-dimensional approximation $G$ to that group.  In the applications we present, this typically involves replacing the diffeomorphism group $\mathrm{Diff}_{\mathrm{vol}}(M)$ with the matrix group $\mathcal{D}(\mathbb{M})$ and replacing scalar fields with discrete zero-forms.}
	\item{Compute the infinitesimal adjoint and coadjoint actions on $\mathfrak{g}$ and $\mathfrak{g}^*$, respectively.}
	\item{Identify the space of advected parameters and construct a finite-dimensional approximation $V^*$ to that space.  Design an appropriate representation of $G$ on $V^*$.}
	\item{Compute the diamond operation associated with the action of $G$ on $V^*$.}
	\item{Write down the discrete-space Lagrangian $\ell: \mathfrak{g} \times V^* \rightarrow \mathbb{R}$.}
	\item{Compute the discrete-space, continuous-time Euler-Poincar\'{e} equations~(\ref{EPcontinuous_right}-\ref{EPcontinuous2_right}) with advected parameters.}
	    \newcounter{enumi_saved}
      \setcounter{enumi_saved}{\value{enumi}}
\end{enumerate}

If a variational temporal discretization is desired, one continues in the following manner:

\begin{enumerate}
			\setcounter{enumi}{\value{enumi_saved}}
	\item{Compute the exponential map $\exp: \mathfrak{g} \rightarrow G$.  If $G=\mathcal{D}(\mathbb{M})$, this is simply the usual matrix exponential.}
	\item{Construct a local approximant $\tau: \mathfrak{g} \rightarrow G$ to the exponential map.  It is often convenient to invoke the Cayley transform for this purpose.}
	\item{Compute the inverse right-trivialized tangent $d\tau^{-1} : \mathfrak{g} \times \mathfrak{g} \rightarrow \mathbb{R}$ of $\tau$, as well as its dual.}
	\item{Write down the discrete-space, discrete-time Euler-Poincar\'{e} equations~(\ref{EP1_right}-\ref{EP3_right}) with advected parameters.}
\end{enumerate}

For finite-dimensional systems, only the last four steps are relevant.  To illustrate the process of temporal discretization described in the last four steps, we will begin this section by presenting two finite-dimensional mechanical examples, the rigid body and the heavy top.  These mechanical examples are not merely pedagogical.  Quite remarkably, the discrete equations of motion for the rigid body will be seen to exhibit strong similarities to the discrete fluid equations expressed in our framework.  Another parallel between the heavy top and MHD will be highlighted. These analogies are manifestations of the fact that \emph{the governing equations for perfect fluid flow and those for the rigid body are both Euler-Poincar\'{e} equations associated with a quadratic Lagrangian, albeit on a different Lie group}.  The transparency of this correspondence in the discrete setting is one of the unifying themes of this section, and it serves as a strong testament to the scope and power of geometric mechanics in numerical applications.

\subsection{Finite-Dimensional Examples}

\subsubsection{The Rigid Body} \label{section:rigid_body}

Consider a rigid body moving freely in empty space.  In a center-of-mass frame, its configuration (relative to a reference configuration) is described by a rotation matrix $\mathbf{R} \in SO(3)$.  Its angular velocity in the body frame is related to its configuration according to the relation
\begin{equation}
\hat{\mathbf{\Omega}} = \mathbf{R}^{-1}\dot{\mathbf{R}},
\end{equation}
where $\mathbf{\Omega} = (\Omega_1,\Omega_2,\Omega_3) \in \mathbb{R}^3$ and $\;\hat{\cdot} : (\mathbb{R}^3,\times) \rightarrow (\mathfrak{so}(3), [\cdot,\cdot])$ is the Lie algebra isomorphism associating vectors in $\mathbb{R}^3$ to antisymmetric matrices:
\begin{equation}
\hat{\mathbf{\Omega}} =
\begin{pmatrix}
0 & -\Omega_3 & \Omega_2 \\
\Omega_3 & 0 & -\Omega_1 \\
-\Omega_2 & \Omega_1 & 0 \\
\end{pmatrix}. \label{hat_iso}
\end{equation}

The rigid body Lagrangian is
\begin{equation}
\ell(\hat{\mathbf{\Omega}}) = \frac{1}{2} \mathbb{I}\mathbf{\Omega} \cdot \mathbf{\Omega} = \frac{1}{2} \mathrm{Tr}(\hat{\mathbf{\Omega}}^T \mathbb{J} \hat{\mathbf{\Omega}}),
\end{equation}
where $\mathbb{I} \in Sym(3)^+$ is the inertia tensor of the body in a principal axis frame, and $\mathbb{J}$ is related to $\mathbb{I}$ via
\begin{equation}
\mathbb{J} = \mathrm{Tr}(\mathbb{I})I - \mathbb{I}. \label{modified_inertia}
\end{equation}
This Lagrangian is left-invariant since the map $\mathbf{R} \mapsto \mathbf{Q}\mathbf{R}$ leaves $\hat{\mathbf{\Omega}} = \mathbf{R}^{-1}\dot{\mathbf{R}}$ invariant for any $\mathbf{Q} \in SO(3)$.

\medskip

Identifying $\mathfrak{so}(3)$ with its dual via the pairing $\langle \hat{\mathbf{v}},\hat{\mathbf{w}} \rangle = \mathbf{v} \cdot \mathbf{w} = \frac{1}{2}\mathrm{Tr}(\hat{\mathbf{v}}^T \hat{\mathbf{w}})$, define
\begin{equation}
\hat{\mathbf{\Pi}} = \frac{\delta\ell}{\delta\hat{\mathbf{\Omega}}} = \hat{\mathbf{\Omega}}\mathbb{J} + \mathbb{J}\hat{\mathbf{\Omega}} = \widehat{\mathbb{I}\mathbf{\Omega}}.
\end{equation}

The Euler-Poincar\'{e} equations~(\ref{EP1}) (without advected parameters) on $\mathfrak{so}(3)^*$ read
\begin{equation}
\frac{d\hat{\mathbf{\Pi}}}{dt} - [\hat{\mathbf{\Pi}}, \hat{\mathbf{\Omega}}] = 0, \label{EP_rigid_body}
\end{equation}
or, equivalently,
\begin{equation}
\frac{d\mathbf{\Pi}}{dt} - \mathbf{\Pi} \times \mathbf{\Omega} = 0.
\end{equation}

\paragraph{Temporal Discretization.}

For a given group difference map $\tau: \mathfrak{so}(3) \rightarrow SO(3)$, the discrete-time Euler-Poincar\'{e} equations~(\ref{EP1}-\ref{EP3}) (again without advected parameters) read
\begin{align}
\mathbf{R}_{k+1} &= \mathbf{R}_k \tau(h \hat{\mathbf{\Omega}}_k) \label{ddEP1_rigid_body} \\
(d\tau^{-1}_{h\hat{\mathbf{\Omega}}_k})^*\hat{\mathbf{\Pi}}_k &= (d\tau^{-1}_{-h\hat{\mathbf{\Omega}}_{k-1}})^*\hat{\mathbf{\Pi}}_{k-1}. \label{ddEP2_rigid_body}
\end{align}

One choice for a group difference map $\tau$ is the Cayley transform
\begin{equation}
\mathrm{cay}(\hat{\mathbf{\Omega}}) =  \left(I-\frac{\hat{\mathbf{\Omega}}}{2}\right)^{-1}\left(I+\frac{\hat{\mathbf{\Omega}}}{2}\right), \label{cayley_rigid_body}
\end{equation}
whose inverse right-trivialized tangent takes the form (see~\cite{BouRabee2007})
\begin{equation}
d\mathrm{cay}^{-1}_{\hat{\mathbf{\Omega}}} (\hat{\mathbf{\Psi}}) =
\left(I-\frac{\hat{\mathbf{\Omega}}}{2}\right)\hat{\mathbf{\Psi}}\left(I+\frac{\hat{\mathbf{\Omega}}}{2}\right) = \hat{\mathbf{\Psi}} - \frac{1}{2}[\hat{\mathbf{\Omega}},\hat{\mathbf{\Psi}}] -\frac{1}{4}\hat{\mathbf{\Omega}}\hat{\mathbf{\Psi}}\hat{\mathbf{\Omega}}.
\end{equation}

Equation~(\ref{ddEP2_rigid_body}) with $\tau=\mathrm{cay}$ can then be written
\begin{equation}
\frac{\hat{\mathbf{\Pi}}_k-\hat{\mathbf{\Pi}}_{k-1}}{h} - \frac{[\hat{\mathbf{\Pi}}_{k-1},\hat{\mathbf{\Omega}}_{k-1}] +[\hat{\mathbf{\Pi}}_k,\hat{\mathbf{\Omega}}_k]}{2} + \frac{h}{4}(\hat{\mathbf{\Omega}}_{k-1}\hat{\mathbf{\Pi}}_{k-1} \hat{\mathbf{\Omega}}_{k-1} - \hat{\mathbf{\Omega}}_k \hat{\mathbf{\Pi}}_k \hat{\mathbf{\Omega}}_k ) = 0. \label{ddEP_rigid_body_rearranged}
\end{equation}

\paragraph{The Discrete Kelvin-Noether Theorem.}

To apply theorem~(\ref{discreteKN}), let $G = SO(3)$ act on $\mathcal{C} := \mathfrak{so}(3)$ via the adjoint representation and let $\mathcal{K} : \mathfrak{so}(3) \rightarrow \mathfrak{so}(3)$ be the equivariant map $\hat{\mathbf{W}} \mapsto \hat{\mathbf{W}}$.  The discrete Kelvin-Noether theorem (free of advected parameters) then implies that
\begin{equation}
\langle (d\tau^{-1}_{h\hat{\mathbf{\Omega}}_k})^*\hat{\mathbf{\Pi}}_k, \hat{\mathbf{W}}_k \rangle \label{pik}
\end{equation}
is independent of $k$ for any $\hat{\mathbf{W}}_0 \in \mathfrak{so}(3)$, where $\hat{\mathbf{W}}_k = \mathbf{R}_k^{-1} \hat{\mathbf{W}}_0 \mathbf{R}_k$.  Rearranging~(\ref{pik}) results in
\begin{equation}
\langle \mathbf{R}_k (d\tau^{-1}_{h\hat{\mathbf{\Omega}}_k})^*\hat{\mathbf{\Pi}}_k \mathbf{R}_k^{-1}, \hat{\mathbf{W}}_0 \rangle.
\end{equation}
Since $\mathbf{W}_0$ is arbitrary, this shows that the rigid body update equations preserve the quantity
\begin{equation}
\hat{\boldsymbol{\pi}}_k = \mathbf{R}_k (d\tau^{-1}_{h\hat{\mathbf{\Omega}}_k})^*\hat{\mathbf{\Pi}}_k \mathbf{R}_k^{-1},
\end{equation}
which is a discrete analogue of spatial angular momentum of the body.  In other words, the rigid body discrete update equations~(\ref{ddEP1_rigid_body}-\ref{ddEP2_rigid_body}) \emph{preserve a discrete spatial angular momentum exactly}.

\subsubsection{The Heavy Top}

The heavy top is a mechanical system consisting of a rigid body with a fixed point rotating in the presence of a uniform gravitational field in the $-\mathbf{e}_3$ direction.  Let $l\boldsymbol{\chi}$ denote the vector pointing from the top's point of fixture to its center of mass, where $\boldsymbol{\chi}$ is a unit vector and $l$ is a scalar.  Let $\mathbf{R} \in \mathrm{SO}(3)$ denote the configuration of the top relative to the upright configuration and let $\mathbf{\Omega}$ denote the top's body angular velocity, related to $\mathbf{R}$ via $\hat{\mathbf{\Omega}} = \mathbf{R}^{-1}\dot{\mathbf{R}}$.

If $\mathbf{\Gamma} \in \mathbb{R}^3$ is the direction of gravity as viewed in the body frame, then $\mathbf{\Gamma}$ is a \emph{purely advected quantity}:
\begin{equation}
\mathbf{\Gamma}(t) = \mathbf{R}(t)^{-1}\mathbf{\Gamma}_0 = \mathbf{R}(t)^{-1}\mathbf{e}_3.
\end{equation}
We therefore treat $\mathbf{\Gamma}$ as an element of a representation space $V^*:=\mathbb{R}^3$, on which $G:=SO(3)$ acts from the left by multiplication:
\begin{equation}
\mathbf{R} \cdot \mathbf{\Gamma} = \mathbf{R} \mathbf{\Gamma}.
\end{equation}
The infinitesimal action of $\mathfrak{so}(3)$ on $V^*$ is then
\begin{equation}
\hat{\mathbf{\Omega}} \cdot \mathbf{\Gamma} = \mathbf{\Omega} \times \mathbf{\Gamma}.
\end{equation}
Now note that for any $\mathbf{w} \in \mathbb{R}^3$,
\begin{equation}
(\mathbf{\Omega} \times \mathbf{\Gamma}) \cdot \mathbf{w} = -(\mathbf{w} \times \mathbf{\Gamma}) \cdot \mathbf{\Omega},
\end{equation}
showing that the diamond operation $\diamond : \mathbb{R}^3 \times \mathbb{R}^3 \rightarrow \mathfrak{so}(3)$ (see~\ref{diamond}) is given by
\begin{equation}
\mathbf{w} \diamond \mathbf{\Gamma} = \widehat{(\mathbf{w} \times \mathbf{\Gamma})} = [\hat{\mathbf{w}}, \hat{\mathbf{\Gamma}}].
\end{equation}

The Lagrangian $\ell : \mathfrak{so}(3) \times \mathbb{R}^3 \rightarrow \mathbb{R}$ for the heavy top is the kinetic energy of the top minus its potential energy:
\begin{equation}
\ell(\hat{\mathbf{\Omega}},\mathbf{\Gamma}) = \frac{1}{2}\langle \mathbb{I}\mathbf{\Omega}, \mathbf{\Omega} \rangle - Mgl\mathbf{\Gamma}\cdot\boldsymbol{\chi},
\end{equation}
where $M$ the mass of the top, $g$ is the acceleration due to gravity, and $\mathbb{I}$ is the top's inertia tensor.

The Euler-Poincar\'{e} equations~(\ref{EP1}-\ref{EP2}) with the advected parameter $\mathbf{\Gamma}$ now read
\begin{align}
\frac{d\hat{\mathbf{\Pi}}}{dt} - [\hat{\mathbf{\Pi}}, \hat{\mathbf{\Omega}}] &= Mgl [\hat{\mathbf{\Gamma}},\hat{\boldsymbol{\chi}}] \label{EP1_heavy_top} \\
\frac{d\hat{\mathbf{\Gamma}}}{dt} &= [\hat{\mathbf{\Gamma}}, \hat{\mathbf{\Omega}}], \label{EP2_heavy_top}
\end{align}
or, equivalently,
\begin{align}
\frac{d\mathbf{\Pi}}{dt} - \mathbf{\Pi} \times \mathbf{\Omega} &= Mgl \mathbf{\Gamma} \times \boldsymbol{\chi} \label{EP1_heavy_top_cross} \\
\frac{d\mathbf{\Gamma}}{dt} &= \mathbf{\Gamma} \times \mathbf{\Omega}. \label{EP2_heavy_top_cross}
\end{align}

\paragraph{Temporal Discretization.}

The discrete-time Euler-Poincar\'{e} equations~(\ref{EP1}-\ref{EP3}) for the heavy top read
\begin{align}
\mathbf{R}_{k+1} &= \mathbf{R}_k \tau(h \hat{\mathbf{\Omega}}_k) \label{ddEP1_heavy_top} \\
\mathbf{\Gamma}_{k+1} &= \tau(-h \hat{\mathbf{\Omega}}_k) \mathbf{\Gamma}_k \label{ddEP2_heavy_top} \\
(d\tau^{-1}_{h\hat{\mathbf{\Omega}}_k})^*\hat{\mathbf{\Pi}}_k &= (d\tau^{-1}_{-h\hat{\mathbf{\Omega}}_{k-1}})^*\hat{\mathbf{\Pi}}_{k-1} + hMgl[\hat{\mathbf{\Gamma}}_k, \hat{\boldsymbol{\chi}}]. \label{ddEP3_heavy_top}
\end{align}
To facilitate an eventual analogy with MHD, let us recast the second equation in terms of $\hat{\mathbf{\Gamma}} \in \mathfrak{so}(3)$:
\begin{equation}
\hat{\mathbf{\Gamma}}_{k+1} = \tau(-h \hat{\mathbf{\Omega}}_k) \hat{\mathbf{\Gamma}}_k \tau(h \hat{\mathbf{\Omega}}_k) \label{ddEP2_heavy_top_recast}
\end{equation}
Now use the Cayley group difference map~(\ref{cayley_rigid_body}) and rearrange~(\ref{ddEP3_heavy_top}-\ref{ddEP2_heavy_top_recast}) to obtain
\begin{align}
\frac{\hat{\mathbf{\Pi}}_k-\hat{\mathbf{\Pi}}_{k-1}}{h} - \frac{[\hat{\mathbf{\Pi}}_{k-1},\hat{\mathbf{\Omega}}_{k-1}]+[\hat{\mathbf{\Pi}}_k,\hat{\mathbf{\Omega}}_k]}{2} + \frac{h}{4}(\hat{\mathbf{\Omega}}_{k-1}\hat{\mathbf{\Pi}}_{k-1} \hat{\mathbf{\Omega}}_{k-1} - \hat{\mathbf{\Omega}}_k \hat{\mathbf{\Pi}}_k \hat{\mathbf{\Omega}}_k ) &= Mgl[\hat{\mathbf{\Gamma}}_k, \hat{\boldsymbol{\chi}}] \label{ddEP3_heavy_top_rearranged} \\
\frac{\hat{\mathbf{\Gamma}}_k - \hat{\mathbf{\Gamma}}_{k-1}}{h} + \left[\hat{\mathbf{\Omega}}_{k-1},\frac{\hat{\mathbf{\Gamma}}_{k-1}+\hat{\mathbf{\Gamma}}_k}{2}\right] + \frac{h}{4}(\hat{\mathbf{\Omega}}_{k-1} \hat{\mathbf{\Gamma}}_{k-1} \hat{\mathbf{\Omega}}_{k-1} - \hat{\mathbf{\Omega}}_{k-1}\hat{\mathbf{\Gamma}}_k \hat{\mathbf{\Omega}}_{k-1}) &= 0. \label{ddEP2_heavy_top_rearranged}
\end{align}

\paragraph{The Discrete Kelvin-Noether Theorem.}

To apply theorem~(\ref{discreteKN}), again let $G = SO(3)$ act on $\mathcal{C} := \mathfrak{so}(3)$ via the adjoint representation and let $\mathcal{K} : \mathfrak{so}(3) \times \mathbb{R}^3 \rightarrow \mathfrak{so}(3)$ be the equivariant map $(\hat{\mathbf{W}}, \mathbf{\Gamma}) \mapsto \hat{\mathbf{W}}$.  The discrete Kelvin-Noether theorem then says that
\begin{equation}
\langle (d\tau^{-1}_{h\hat{\mathbf{\Omega}}_k})^*\hat{\mathbf{\Pi}}_k, \hat{\mathbf{W}}_k \rangle - \langle (d\tau^{-1}_{h\hat{\mathbf{\Omega}}_{k-1}})^*\hat{\mathbf{\Pi}}_{k-1}, \hat{\mathbf{W}}_{k-1} \rangle
= hMgl \langle [\hat{\mathbf{\Gamma}}_k, \hat{\boldsymbol{\chi}}], \hat{\mathbf{W}}_k \rangle, \label{KN_heavy_top}
\end{equation}
where $\hat{\mathbf{W}}_k = \mathbf{R}_k^{-1} \hat{\mathbf{W}}_0 \mathbf{R}_k = \widehat{\mathbf{R}_k^{-1} \mathbf{W}_0}$.  That is, $\mathbf{W}_k \in \mathbb{R}^3$ is the body frame representation of a fixed spatial vector $\mathbf{W}_0 \in \mathbb{R}^3$.  Choosing $\mathbf{W}_0 = \mathbf{\Gamma}_0$ so that $\mathbf{W}_k = \mathbf{\Gamma}_k$ causes the right-hand side of~(\ref{KN_heavy_top}) to vanish, showing that $\langle (d\tau^{-1}_{h\hat{\mathbf{\Omega}}_k})^*\hat{\mathbf{\Pi}}_k, \hat{\mathbf{\Gamma}}_k \rangle = \langle \mathbf{R}_k (d\tau^{-1}_{h\hat{\mathbf{\Omega}}_k})^*\hat{\mathbf{\Pi}}_k \mathbf{R}_k^{-1}, \hat{\mathbf{e}}_3 \rangle$ is constant.  Thus, \emph{the discrete update equations~(\ref{ddEP1_heavy_top}-\ref{ddEP3_heavy_top}) preserve the $\mathbf{e}_3$-component of the heavy top's spatial angular momentum}
\begin{equation}
\hat{\boldsymbol{\pi}}_k = \mathbf{R}_k (d\tau^{-1}_{h\hat{\mathbf{\Omega}}_k})^*\hat{\mathbf{\Pi}}_k \mathbf{R}_k^{-1}.
\end{equation}

\subsection{Discretization of Continuum Theories} \label{section:continuum_discretization}

We now proceed with our discretization of four continuum theories: ideal, incompressible fluid dynamics; ideal, incompressible magnetohydrodynamics; the dynamics of nematic liquid crystals; and the dynamics of microstretch continua.

\subsubsection{Ideal, Incompressible Fluid Flow}

\paragraph{The Continuous Equations of Motion.}

The governing equations for the dynamics of an ideal, incompressible fluid occupying a domain $M \subseteq \mathbb{R}^3$ are the Euler equations:
\begin{align}
\frac{\partial \mathbf{v}}{\partial t} + (\mathbf{v} \cdot \nabla) \mathbf{v} &= - \nabla p \label{fluid1} \\
\nabla \cdot \mathbf{v} &= 0 \label{fluid2}
\end{align}
Equation~(\ref{fluid1}) describes the temporal evolution of the fluid velocity field $\mathbf{v}$ in terms of the coevolving scalar pressure field $p$.  The latter quantity is uniquely determined (up to a constant) from the constraint~(\ref{fluid2}), which expresses mathematically the incompressibility of the fluid.

\paragraph{Geometric Formulation.}

The configuration space of the ideal, incompressible fluid is the group $G=\mathrm{Diff}_{\mathrm{vol}}(M)$.  In terms of material particle labels $X$ and fixed spatial locations $x \in M$, a passive fluid particle traces a path $x(X,t) = \varphi_t(X)$ under a motion $\varphi_t$ in $G$.  The spatial velocity field $\mathbf{v}(x,t)$ is related to $\varphi_t$ via
\begin{equation}
\mathbf{v}(x,t)=\dot{\varphi}_t(\varphi_t^{-1}(x)). \label{v_relation}
\end{equation}
Recognizing~(\ref{v_relation}) as the right translation of a tangent vector $\dot{\varphi}_t \in T_{\varphi_t} G$ to the identity, we regard $\mathbf{v}(x,t)$ as an element of $\mathfrak{g}=\mathfrak{X}_{\mathrm{div}}(M)$, the Lie algebra of $G$.

The fluid Lagrangian $\ell : \mathfrak{g} \rightarrow \mathbb{R}$, regarded as the right-trivialization of a Lagrangian $L : TG \rightarrow \mathbb{R}$, is the fluid's total kinetic energy
\begin{equation}
\ell(\mathbf{v}) = \frac{1}{2} \langle \mathbf{v}^\flat, \mathbf{v} \rangle,
\end{equation}
where the pairing above is that given by~(\ref{pairing_diff}).

The Euler-Poincar\'{e} equations~(\ref{EPcontinuous_right}) (without advected parameters) read
\begin{equation}
\frac{\partial \mathbf{v}^\flat}{\partial t} + \pounds_\mathbf{v} \mathbf{v}^\flat = -d\tilde{p}. \label{EPfluid}
\end{equation}
The pressure differential appearing here is an explicit reminder that the Euler-Poincar\'{e} equations formally govern motion in $\mathfrak{g}^*=\Omega^1(M)/d\Omega^0(M)$, a quotient space in this case.  In the language of vector calculus,~(\ref{EPfluid}) has the form of equation~(\ref{fluid1}), with $p$ related to $\tilde{p}$ by $p = \tilde{p} + \frac{1}{2}\mathbf{v}\cdot\mathbf{v}$.

\paragraph{The Continuous Kelvin-Noether Theorem.}

The ideal, incompressible fluid provides a canonical illustration of the utility of the Kelvin-Noether Theorem~(\ref{theorem:continuousKN}) in the absence of advected parameters.  Take $\mathcal{C}$ to be the space of loops (closed curves) in the fluid domain $M$, acted upon by $G$ from the right via the pullback, and take $\mathcal{K} : \mathcal{C} \rightarrow \mathfrak{g}^{**}$ to be the equivariant map defined by
\begin{equation}
\langle \mathcal{K}(\gamma), \mathbf{w}^\flat \rangle = \oint_\gamma \mathbf{w}^\flat,
\end{equation}
where $\gamma : S^1 \rightarrow M$ is a loop and $\mathbf{w}^\flat$ is a one-form.  The Kelvin-Noether theorem then says that
\begin{equation}
\frac{d}{dt} \oint_{\gamma(t)} \mathbf{v}^\flat(t) = 0
\end{equation}
where $\mathbf{v}(t)$ is the fluid velocity field and $\gamma(t)$ is a loop advected with the flow.  This is known classically as \emph{Kelvin's circulation theorem}: the line integral of the velocity field along any closed loop moving passively with the fluid is constant in time.

\paragraph{Spatial Discretization.}

To discretize~(\ref{EPfluid}), cover $M$ with an mesh $\mathbb{M}$ and replace $\mathrm{Diff}_{\mathrm{vol}}(M)$ with the finite-dimensional matrix group $G=\mathcal{D}(\mathbb{M})$.  Denote the discrete counterpart of $\varphi$ by $y$ and that of $\mathbf{v}$ by $Y$; the quantities $y$ and $Y$ are elements of the finite-dimensional diffeomorphism group and its Lie algebra $\mathfrak{g} = \mathfrak{d}(\mathbb{M})$, respectively.  As such, $y$ is an $\Omega$-orthogonal, signed stochastic matrix, and $Y$ is an $\Omega$-antisymmetric, row-null matrix, where $\Omega$ is the diagonal matrix of cell volumes described in Section~\ref{section:diff}.

The spatially discretized fluid Lagrangian $\ell : \mathfrak{g} \rightarrow \mathbb{R}$ is the fluid's total kinetic energy
\begin{equation}
\ell(Y) = \frac{1}{2} \langle Y^\flat, Y \rangle, \label{dLagrangian_fluid}
\end{equation}
where the pairing above is that given by~(\ref{pairing_ddiff}).  This Lagrangian is \emph{right invariant}: in analogy with~(\ref{v_relation}), the relation between $y$ and $Y$ is given explicitly by
\begin{equation}
Y = \dot{y}y^{-1},
\end{equation}
showing that the map $y \mapsto yq$ with $q \in \mathcal{D}(\mathbb{M})$ leaves~(\ref{dLagrangian_fluid}) invariant.

The discrete-space, continuous-time Euler-Poincar\'{e} equations~(\ref{EPcontinuous_right}) (without advected parameters) read
\begin{equation}
\frac{\partial Y^\flat}{\partial t} + \pounds_Y Y^\flat \,\hat{=}\, 0, \label{dEPfluid}
\end{equation}
where $\pounds_Y Y^\flat$ is the $\Omega$-weighted matrix commutator~(\ref{coad_ddiff}) and the hatted equality denotes a weak equality in the sense of Section~\ref{section:weak_equalities}.  Namely, there exists a discrete scalar field $P$ for which
\begin{equation}
\left( \frac{\partial Y^\flat}{\partial t} + \pounds_Y Y^\flat \right)_{ij} = -(dP)_{ij}
\end{equation}
for every pair of neighboring cells $\mathcal{C}_i$ and $\mathcal{C}_j$.

Note the similarity between~(\ref{dEPfluid}) and the rigid body equation~(\ref{EP_rigid_body}) upon writing $\pounds_Y Y^\flat$ in its explicit form $[Y^\flat\Omega,Y]\Omega^{-1}$.

\paragraph{Temporal Discretization.}

The discrete-space, discrete-time Euler-Poincar\'{e} equations~(\ref{EP1_right}-\ref{EP3_right}) (again without advected parameters) read
\begin{align}
y_{k+1} &= \tau(h Y_k) y_k \\
(d\tau^{-1}_{-hY_k})^*Y_k^\flat &\,\hat{=}\, (d\tau^{-1}_{hY_{k-1}})^*Y_{k-1}^\flat. \label{ddEPfluid}
\end{align}

A convenient choice for a group difference map $\tau : \mathfrak{d}(\mathbb{M}) \rightarrow \mathcal{D}(\mathbb{M})$ is once again the Cayley transform
\begin{equation}
\tau(Y) = \mathrm{cay}(Y) = \left(I-\frac{Y}{2}\right)^{-1} \left(I+\frac{Y}{2}\right). \label{cayley}
\end{equation}
It is well-known that the Cayley transform is a local approximant to the matrix exponential and preserves structure for quadratic Lie groups; that is, if $Y$ is $\Omega$-antisymmetric, then $\mathrm{cay}(Y)$ is $\Omega$-orthogonal.  Conveniently, the Cayley transform maps row-null matrices to signed stochastic matrices, making it a genuine map from the Lie algebra $\mathfrak{d}(\mathbb{M})$ to the Lie group $\mathcal{D}(\mathbb{M})$.  To see this, note that if $Y$ is row-null, then $\left(I-\frac{Y}{2}\right)$ and $\left(I+\frac{Y}{2}\right)$ are each stochastic; now use the fact that the set of signed stochastic matrices is closed under multiplication and inversion.

As we saw with the rigid body, the inverse right-trivialized tangent of the Cayley transform takes the form (see~\cite{BouRabee2007})
\begin{equation}
d\mathrm{cay}^{-1}_Y(Z) = \left(I-\frac{Y}{2}\right)Z\left(I+\frac{Y}{2}\right) = \left(I+\frac{1}{2}\pounds_Y\right)Z -\frac{1}{4}YZY. \label{dcayinv}
\end{equation}
The dual of $d\mathrm{cay}^{-1}$ with respect to the pairing~(\ref{pairing_ddiff}) can be computed as follows:  For any $Z \in \mathfrak{d}(\mathbb{M}), X^\flat \in \mathfrak{d}(\mathbb{M})^*$,
\begin{align}
\langle X^\flat,d\mathrm{cay}^{-1}_Y(Z) \rangle
&= -\mathrm{Tr}\left(\Omega \left(I-\frac{Y}{2}\right)Z\left(I+\frac{Y}{2}\right) X^\flat\right) \nonumber \\
&= -\mathrm{Tr}\left(\Omega Z \left(I+\frac{Y}{2}\right) X^\flat \Omega \left(I-\frac{Y}{2}\right) \Omega^{-1}\right) \nonumber \\
&= \left\langle \left(I+\frac{Y}{2}\right) X^\flat \Omega \left(I-\frac{Y}{2}\right) \Omega^{-1}, Z \right\rangle.
\end{align}
This shows that
\begin{equation}
(d\mathrm{cay}^{-1}_Y)^*(X^\flat) = \left(I+\frac{Y}{2}\right) X^\flat \Omega \left(I-\frac{Y}{2}\right) \Omega^{-1} = \left(I-\frac{1}{2}\pounds_Y\right)X^\flat - \frac{1}{4}YX^\flat\Omega Y\Omega^{-1}. \label{dcayinvstar}
\end{equation}

Substituting into~(\ref{ddEPfluid}) and rearranging leads to the following discrete update equation for ideal, incompressible fluid flow:
\begin{equation}
\frac{Y^\flat_k-Y^\flat_{k-1}}{h} + \frac{\pounds_{Y_{k-1}}Y^\flat_{k-1}+\pounds_{Y_k}Y^\flat_k}{2} + \frac{h}{4}(Y_{k-1}Y_{k-1}^\flat\Omega Y_{k-1}\Omega^{-1} - Y_k Y^\flat_k \Omega Y_k \Omega^{-1}) \,\hat{=}\, 0. \label{ddEPfluid_rearranged}
\end{equation}
Notice once again the evident parallel between~(\ref{ddEPfluid_rearranged}) and the rigid body update equation~(\ref{ddEP_rigid_body_rearranged}).

\paragraph{The Discrete Kelvin-Noether Theorem.}

Let $\mathcal{C}$ be the space of discrete loops in $\mathbb{M}$ and let $G=\mathcal{D}(\mathbb{M})$ act on $\mathcal{C} \cong \mathfrak{g}$ via the discrete pullback: $\Gamma \cdot y := y^* \Gamma = y^{-1} \Gamma y$ for any $\Gamma \in \mathcal{C}$, $y \in G$.  Let $\mathcal{K} : \mathcal{C} \rightarrow \mathfrak{g}$ be the equivariant map $\Gamma \mapsto \Gamma$. The discrete Kelvin-Noether theorem then says that the quantity
\begin{equation}
M_k
= \left\langle (d\tau^{-1}_{-hY_k})^*Y^\flat_k, \Gamma_k \right\rangle \label{circulation1}
\end{equation}
satisfies
\begin{equation}
M_k - M_{k-1} = 0 \label{circulation_evolution}
\end{equation}
with $\Gamma_k = \Gamma_0 \cdot y_k^{-1} = (y_k)_*\Gamma_0$, i.e. $\Gamma_k$ is a discrete loop advected passively by the fluid flow. Identifying~(\ref{circulation1}) as the circulation of the one-form $(d\tau^{-1}_{-hY_k})^*Y_k^\flat$ along the curve $\Gamma_k$, we see that~(\ref{circulation_evolution}) gives the discrete analogue of Kelvin's circulation theorem.  In other words, \emph{a discrete Kelvin's circulation theorem holds for the discrete-space, discrete-time fluid equations} laid forth by Pavlov and co-authors~\cite{Pavlov2009}.
Notice the beautiful parallel between the discrete Kelvin's circulation theorem for the ideal fluid and the discrete angular momentum conservation law derived earlier (in Section~\ref{section:rigid_body}) for the rigid body.

\subsubsection{Ideal, Incompressible Magnetohydrodynamics}

\paragraph{The Continuous Equations of Motion.}

The governing equations for ideal, incompressible MHD, which studies the motion of a perfectly conducting incompressible fluid occupying a domain $M \subseteq \mathbb{R}^3$ in the presence of a coevolving magnetic field $\mathbf{B}$, are the \emph{ideal MHD equations}~\cite{Goedbloed2004}:
\begin{align}
\frac{\partial \mathbf{v}}{\partial t} + (\mathbf{v} \cdot \nabla) \mathbf{v} &= (\nabla \times \mathbf{B}) \times \mathbf{B} - \nabla p \label{MHD1} \\
\frac{\partial \mathbf{B}}{\partial t} - \nabla \times (\mathbf{v} \times \mathbf{B}) &= 0 \label{MHD2} \\
\nabla \cdot \mathbf{B} &= 0 \label{MHD3} \\
\nabla \cdot \mathbf{v} &= 0 \label{MHD4}
\end{align}
Equation~(\ref{MHD1}) is the classic Euler equation for the fluid velocity field $\mathbf{v}$ with an additional term $(\nabla \times \mathbf{B}) \times \mathbf{B}$ corresponding to the Lorentz force $\mathbf{j} \times \mathbf{B}$ acting on mobile charges in the fluid. The second equation~(\ref{MHD2}) describes the evolution of the magnetic field $\mathbf{B}$ and has a particularly simple interpretation when recast in the form
\begin{equation}
\frac{\partial \mathbf{B}}{\partial t} + \pounds_\mathbf{v} \mathbf{B}= 0; \label{magnetic_advection}
\end{equation}
that is, \emph{the magnetic field is advected with the fluid flow}.  The constraints~(\ref{MHD3}-\ref{MHD4}) represent the absence of magnetic charge and the incompressibility of the fluid flow.  The latter constraint uniquely determines the pressure $p$ appearing in~(\ref{MHD1}), while the former constraint automatically holds for all time if it holds initially, as can be seen by taking the divergence of equation~(\ref{MHD2}) and using the fact that $\nabla \cdot (\nabla \times \mathbf{u}) = 0$ for any vector field $\mathbf{u}$.

One may check through differentiation that the quantities
\begin{equation}
E = \int_M \left(\frac{1}{2} \mathbf{v}\cdot\mathbf{v} + \frac{1}{2} \mathbf{B}\cdot\mathbf{B}\right) \, d \mathbf{x}
\end{equation}
and
\begin{equation}
J = \int_M \mathbf{v}\cdot\mathbf{B} \, d \mathbf{x}
\end{equation}
are constants of motion for the ideal MHD equations~(\ref{MHD1}-\ref{MHD4}).  The former quantity is the total energy of the system, being the volume-integrated sum of the kinetic energy density $\frac{1}{2} \mathbf{v} \cdot \mathbf{v}$ of the fluid and the potential energy density $\frac{1}{2} \mathbf{B} \cdot \mathbf{B}$ of the magnetic field.  The quantity $J$ is known as the \emph{cross-helicity} of the fluid, which may be shown to bear a relation to the topological linking of the magnetic field $\mathbf{B}$ and the fluid vorticity $\nabla \times \mathbf{v}$~\cite{Arnold1998}.

\paragraph{Geometric Formulation.}

As with the ideal, incompressible fluid, the configuration space for the ideal, incompressible magnetofluid is the group $G=\mathrm{Diff}_{\mathrm{vol}}(M)$.  Curves $\varphi_t$ in $G$ encode fluid motions as usual, and the fluid's spatial velocity field $\mathbf{v}$ belongs to $\mathfrak{g}=\mathfrak{X}_{\mathrm{div}}(M)$.

According to~(\ref{magnetic_advection}), the magnetic field $\mathbf{B}$ is an advected parameter.  With this in mind, we choose the representation space $V=\Omega^1(M)/d\Omega^0(M)$ and identify $V^*$ with $\mathfrak{X}_{\mathrm{div}}(M)$ under the usual pairing.  Elements $\varphi$ of $G$ act on $V$ from the right via the pullback, so that the induced right action of $G$ on $V^*$ is again given by the pullback:
\begin{equation}
\mathbf{B} \cdot \varphi = \varphi^* \mathbf{B}. \label{Gaction_MHD}
\end{equation}
The infinitesimal action of $\mathfrak{g}$ on $V^*$ is via the Lie derivative:
\begin{equation}
\mathbf{B} \cdot \mathbf{v} = \pounds_\mathbf{v} \mathbf{B}. \label{gaction_MHD}
\end{equation}
The diamond operation~(\ref{diamond_right}) may be computed as follows: For any $\mathbf{C}^\flat \in V$, $\mathbf{B} \in V^*$, and $\mathbf{v} \in \mathfrak{g}$,
\begin{align}
\langle \mathbf{B} \cdot \mathbf{v}, \mathbf{C}^\flat \rangle_{V^* \times V}
&= \langle \mathbf{C}^\flat, \pounds_\mathbf{v} \mathbf{B} \rangle_{\mathfrak{g}^* \times \mathfrak{g}} \nonumber \\
&= -\langle \mathbf{C}^\flat, \pounds_\mathbf{B} \mathbf{v} \rangle_{\mathfrak{g}^* \times \mathfrak{g}} \nonumber \\
&= \langle \pounds_\mathbf{B} \mathbf{C}^\flat, \mathbf{v} \rangle_{\mathfrak{g}^* \times \mathfrak{g}} \nonumber \\
&= -\langle \mathbf{C}^\flat \diamond \mathbf{B}, \mathbf{v} \rangle_{\mathfrak{g}^* \times \mathfrak{g}}.
\end{align}
Hence,
\begin{equation}
\mathbf{C}^\flat \diamond \mathbf{B} = -\pounds_\mathbf{B} \mathbf{C}^\flat. \label{diamond_MHD}
\end{equation}

The fluid Lagrangian $\ell : \mathfrak{g} \times V^* \rightarrow \mathbb{R}$ is the fluid's total kinetic energy minus the potential energy stored in the magnetic field:
\begin{equation}
\ell(\mathbf{v},\mathbf{B}) = \frac{1}{2} \langle \mathbf{v}^\flat, \mathbf{v} \rangle - \frac{1}{2} \langle \mathbf{B}^\flat, \mathbf{B} \rangle,
\end{equation}
where the pairing above is that given by~(\ref{pairing_diff}).

The Euler-Poincar\'{e} equations~(\ref{EPcontinuous_right}) with the advected parameter $\mathbf{B}$ read
\begin{align}
\frac{\partial \mathbf{v}^\flat}{\partial t} + \pounds_\mathbf{v} \mathbf{v}^\flat &= \pounds_\mathbf{B} \mathbf{B}^\flat - d\tilde{p} \label{EP1_MHD} \\
\frac{\partial \mathbf{B}}{\partial t} + \pounds_\mathbf{v} \mathbf{B} &= 0. \label{EP2_MHD}
\end{align}
In the language of vector calculus,~(\ref{EP1_MHD}-\ref{EP2_MHD}) have the form of equations~(\ref{MHD1}-\ref{MHD2}), with $p$ related to $\tilde{p}$ by $p = \tilde{p} + \frac{1}{2}\mathbf{v}\cdot\mathbf{v} - \mathbf{B}\cdot\mathbf{B}$.

\paragraph{The Continuous Kelvin-Noether Theorem.}

As with the ideal fluid, take $\mathcal{C}$ to be the space of loops in $M$ and let $G=\mathrm{Diff}_\mathrm{vol}(M)$ act on $\mathcal{C}$ via the pullback.  Let $\mathcal{K} : \mathcal{C} \times V^* \rightarrow \mathfrak{g}^{**}$ be the equivariant map defined by
\begin{equation}
\langle \mathcal{K}(\gamma,\mathbf{B}), \mathbf{w}^\flat \rangle = \oint_\gamma \mathbf{w}^\flat,
\end{equation}
where $\gamma : S^1 \rightarrow M$ is a loop, $\mathbf{B} \in V^*$, and $\mathbf{w}^\flat$ is a one-form.  The Kelvin-Noether theorem then gives the statement
\begin{equation}
\frac{d}{dt} \oint_{\gamma(t)} \mathbf{v}^\flat(t) = \oint_{\gamma(t)} \pounds_\mathbf{B} \mathbf{B}^\flat(t)
\end{equation}
where $\mathbf{v}(t)$ is the fluid velocity field, $\mathbf{B}(t)$ is the magnetic field, and $\gamma(t)$ is a loop advected with the flow.

A second momentum evolution law is obtained from the Kelvin-Noether theorem if one identifies $\mathfrak{g}^{**}$ with $\mathfrak{g} = \mathfrak{X}_{\mathrm{div}}(M)$ and takes $\mathcal{C} = \mathfrak{X}_{\mathrm{div}}(M)$ and $\mathcal{K}(\mathbf{u},\mathbf{B})=\mathbf{u}$, with $G$ acting on $\mathfrak{X}_{\mathrm{div}}(M)$ via the pullback.  The Kelvin-Noether theorem then gives
\begin{equation}
\frac{d}{dt} \langle \mathbf{v}^\flat(t), \mathbf{u}(t) \rangle = \langle \pounds_\mathbf{B} \mathbf{B}^\flat(t), \mathbf{u}(t)  \rangle,
\end{equation}
where $\mathbf{u}(t) = \varphi(t)_* \mathbf{u}(0)$ is a vector field advected with the flow.  Choosing $\mathbf{u}(0)=\mathbf{B}(0)$ leads to the conclusion
\begin{align}
\frac{d}{dt} \langle \mathbf{v}^\flat(t), \mathbf{B}(t) \rangle
&= \langle \pounds_\mathbf{B} \mathbf{B}^\flat(t), \mathbf{B}(t)  \rangle \nonumber \\
&= -\langle \mathbf{B}^\flat(t), \pounds_\mathbf{B} \mathbf{B}(t)  \rangle \nonumber \\
&=0;
\end{align}
that is, the quantity
\begin{equation}
J(t) = \langle \mathbf{v}^\flat(t), \mathbf{B}(t) \rangle
\end{equation}
is conserved along solutions to the governing equations for ideal MHD.  As mentioned in this section's introduction, the quantity $J$ is known as the \emph{cross-helicity} in MHD, and can be shown to bear a relation to the topological linking of the vorticity field $\nabla \times \mathbf{v}$ and the magnetic field $\mathbf{B}$~\cite{Arnold1998}.

\paragraph{Spatial Discretization.}

A spatial discretization of the ideal, incompressible MHD equations is obtained by replacing $\mathrm{Diff}_{\mathrm{vol}}(M)$ with the finite dimensional matrix group $G=\mathcal{D}(\mathbb{M})$.  As before, denote the discrete counterpart of $\varphi$ by $y$ and that of $\mathbf{v}$ by $Y$.  Replace the representation space $\Omega^1(M)/d\Omega^0(M)$ and its dual $\mathfrak{X}_{\mathrm{div}}(M)$ with $V=\Omega_d^1(\mathbb{M})/d\Omega_d^0(\mathbb{M})$ and $V^*=\mathfrak{d}(\mathbb{M})$, respectively.  Denote the discrete counterpart of $\mathbf{B}$ by $R$; the quantity $\!R$ is an $\Omega$-antisymmetric, row-null matrix belonging to $V^*$.

Elements $y$ of $G$ act on $V$ from the right via the discrete pullback, so that the induced right action of $G$ on $V^*$ is again given by the discrete pullback:
\begin{equation}
R \cdot y = y^* R = y^{-1} R y. \label{dGaction_MHD}
\end{equation}
The infinitesimal action of $\mathfrak{g}$ on $V^*$ is via the discrete Lie derivative:
\begin{equation}
R \cdot Y = \pounds_Y R = -[Y,R]. \label{dgaction_MHD}
\end{equation}
The derivation of the continuous diamond operation~(\ref{diamond_MHD}) carries over seamlessly to the discrete setting, giving
\begin{equation}
S^\flat \diamond R = -\pounds_R S^\flat.
\end{equation}

The spatially discretized MHD Lagrangian $\ell : \mathfrak{g} \times V^* \rightarrow \mathbb{R}$ is the fluid's total kinetic energy minus the potential energy stored in the magnetic field:
\begin{equation}
\ell(Y,R) = \frac{1}{2} \langle Y^\flat, Y \rangle - \frac{1}{2} \langle R^\flat, R \rangle,
\end{equation}
where the pairing above is that given by~(\ref{pairing_ddiff}).

The discrete-space, continuous-time Euler-Poincar\'{e} equations~(\ref{EPcontinuous_right}) with the advected parameter $R$ read
\begin{align}
\frac{\partial Y^\flat}{\partial t} + \pounds_Y Y^\flat &\,\hat{=}\, \pounds_R R^\flat \label{dEP1_MHD} \\
\frac{\partial R}{\partial t} + \pounds_Y R &\,\hat{=}\, 0, \label{dEP2_MHD}
\end{align}
where $P$ is discrete scalar field.

\paragraph{Temporal Discretization.}

The discrete-space, discrete-time Euler-Poincar\'{e} equations~(\ref{EP1_right}-\ref{EP3_right}) with the advected parameter $R$ read
\begin{align}
y_{k+1} &= \tau(h Y_k) y_k \\
R_{k+1} &\,\hat{=}\, \tau(h Y_k) R_k \tau(-h Y_k) \label{ddEP_MHD1} \\
(d\tau^{-1}_{-hY_k})^*Y_k^\flat &\,\hat{=}\, (d\tau^{-1}_{hY_{k-1}})^*Y_{k-1}^\flat + h\pounds_{R_k} R_k^\flat. \label{ddEP_MHD2}
\end{align}

Using the Cayley transform~(\ref{cayley}) as a group difference map, the update equations~(\ref{ddEP_MHD1}-\ref{ddEP_MHD2}) become, upon rearrangement,
\begin{align}
\frac{Y^\flat_k-Y^\flat_{k-1}}{h} + \frac{\pounds_{Y_{k-1}}Y^\flat_{k-1}+\pounds_{Y_k}Y^\flat_k}{2} + \frac{h}{4}(Y_{k-1}Y_{k-1}^\flat\Omega Y_{k-1}\Omega^{-1} - Y_k Y^\flat_k \Omega Y_k \Omega^{-1}) &\,\hat{=}\, \pounds_{R_k}R^\flat_k \label{ddEP_MHD1_rearranged} \\
\frac{R_k - R_{k-1}}{h} + \pounds_{Y_{k-1}}\left(\frac{R_{k-1}+R_k}{2}\right) + \frac{h}{4}(Y_{k-1} R_{k-1} Y_{k-1} - Y_{k-1}R_k Y_{k-1}) &\,\hat{=}\, 0. \label{ddEP_MHD2_rearranged}
\end{align}
Referring back to equations~(\ref{ddEP3_heavy_top_rearranged}-\ref{ddEP2_heavy_top_rearranged}), the reader can note the striking similarity between the heavy top update equations and the MHD update equations above.

\paragraph{The Discrete Kelvin-Noether Theorem.}

Take $\mathcal{C}$ to be the space of discrete loops in $\mathbb{M}$ and let $G=\mathcal{D}(\mathbb{M})$ act on $\mathcal{C} \cong \mathfrak{d}(\mathbb{M})$ via the pullback.  Let $\mathcal{K} : \mathcal{C} \times V^* \rightarrow \mathfrak{g}$ be the equivariant map $(\Gamma,A) \mapsto \Gamma$. The discrete Kelvin-Noether theorem then says that the quantity
\begin{equation}
J_k
= \left\langle (d\tau^{-1}_{-hY_k})^*Y_k^\flat, \Gamma_k \right\rangle \label{discrete_cross_helicity}
\end{equation}
satisfies
\begin{equation}
J_k - J_{k-1} = h \langle \pounds_{R_k}R^\flat_k, \Gamma_k \rangle
\end{equation}
with $\Gamma_k = \Gamma_0 \cdot y_k^{-1} = (y_k)_*\Gamma_0$.  If $\Gamma_0 = R_0$, then $\Gamma_k=R_k$ and we find
\begin{align}
J_k - J_{k-1}
&= h \langle \pounds_{R_k}R^\flat_k, R_k \rangle \nonumber \\
&= -h \langle R^\flat_k, \pounds_{R_k} R_k \rangle \nonumber \\
&= 0.
\end{align}
Notice that~(\ref{discrete_cross_helicity}) serves as the discrete analogue of the cross-helicity $J=\langle \mathbf{v}^\flat,\mathbf{B}\rangle$ arising in MHD.  Thus, we have proven that \emph{the cross-helicity~(\ref{discrete_cross_helicity}) is preserved exactly along solutions to the discrete-space, discrete-time MHD update equations~(\ref{ddEP_MHD1}-\ref{ddEP_MHD2})}.

\subsubsection{Nematic Liquid Crystals} \label{section:nematic}

The dynamics of \emph{complex fluids} differ from those of ordinary fluids through their dependence upon \emph{micromotions}, i.e., internal changes in the shape and orientation of individual fluid particles that couple with their ordinary translational motion.  In the particular case of \emph{nematic liquid crystals}, the fluid particles are rod-like and the internal motions are purely rotational.  In a three-dimensional domain $M' \subseteq \mathbb{R}^3$ the variables of interest are the fluid velocity field $\mathbf{v} \in \mathfrak{X}_{\mathrm{div}}(M')$, the \emph{local angular velocity field} $\boldsymbol{\nu} \in \mathcal{F}(M',\mathbb{R}^3)$, and the \emph{director} $\mathbf{n} \in \mathcal{F}(M',\mathbb{R}^3)$.  The latter variable describes the local orientation of fluid particles and its value at any point $\mathbf{x} \in M'$ is restricted to have unit length, so that $\mathbf{n}$ may equivalently be viewed as a map from the domain to $S^2$.

\paragraph{The Continuous Equations of Motion.}

In terms of these variables, the equations of motion for three-dimensional, incompressible, homogeneous nematic liquid crystal flow are given by
\begin{align}
\frac{\partial \mathbf{v}}{\partial t} + (\mathbf{v} \cdot \nabla) \mathbf{v} &= -\partial_i\left(\frac{\partial F}{\partial\mathbf{n}_{,i}} \cdot \nabla \mathbf{n} \right)  - \nabla p \label{nematic1a} \\
\frac{\partial \boldsymbol{\nu}}{\partial t} + (\mathbf{v} \cdot \nabla) \boldsymbol{\nu} &= \mathbf{h}\times\mathbf{n} \label{nematic2a} \\
\frac{\partial \mathbf{\mathbf{n}}}{\partial t} + (\nabla\mathbf{n})\mathbf{v} + \mathbf{n}\times\boldsymbol{\nu} &= 0 \label{nematic3a} \\
\nabla \cdot \mathbf{v} &= 0, \label{nematic4a}
\end{align}
where $p$ is a scalar pressure field,
\begin{equation}
\mathbf{h} = \frac{\partial F}{\partial\mathbf{n}} - \partial_i\left(\frac{\partial F}{\partial\mathbf{n}_{,i}} \right),
\end{equation}
and $F(\mathbf{n},\nabla\mathbf{n})$ is a scalar function describing the potential energy stored by irregularities in the fluid particles' alignment, hereafter referred to as the \emph{free energy}.  The notation $\nabla\mathbf{n}$ denotes the $3 \times 3$ matrix whose $i^{th}$ row is $\nabla\mathbf{n}_i$.

The governing equations for nematic liquid crystal flow have the following interpretation: Equation~(\ref{nematic1a}) is the classic Euler equation for the fluid velocity $\mathbf{v}$ with an additional term accounting for irregularities in the director field that react back on the flow; equations~(\ref{nematic2a}-\ref{nematic3a}) describe the advection of the local angular velocity field $\boldsymbol{\nu}$ and director field $\mathbf{n}$; and~(\ref{nematic4a}) expresses the incompressibility of the flow.

We will simplify the analysis considerably by restricting the rest of our discussion to \emph{two-dimensional flows} with the director everywhere tangent to the plane. For concreteness, choose $M \subset M'$ to be the $x$-$y$ plane and define the variables $\omega\in \mathcal{F}(M)$, $\alpha \in \mathcal{F}(M,S^1)$ via
\begin{align}
\boldsymbol{\nu} &= (0,0,\omega)^T \\
\mathbf{n} &= (\cos\alpha,\sin\alpha,0)^T,
\end{align}
where we identify $S^1$ with the real numbers modulo $2\pi$.

In terms of these variables, the equations of motion for \emph{two-dimensional}, incompressible, homogeneous nematic liquid crystal flow are given by
\begin{align}
\frac{\partial \mathbf{v}}{\partial t} + (\mathbf{v} \cdot \nabla) \mathbf{v} &= -\partial_i\left(\frac{\partial F}{\partial\alpha_{,i}}\right)\nabla\alpha  - \nabla p \label{nematic1} \\
\frac{\partial \omega}{\partial t} + \mathbf{v} \cdot \nabla \omega &= \partial_i\left(\frac{\partial F}{\partial\alpha_{,i}}\right) - \frac{\partial F}{\partial\alpha} \label{nematic2} \\
\frac{\partial \alpha}{\partial t} + \mathbf{v} \cdot \nabla \alpha &= \omega \label{nematic3} \\
\nabla \cdot \mathbf{v} &= 0, \label{nematic4}
\end{align}

\begin{remark*}
\emph{We will deal only with the two-dimensional case for the remainder of this section.
Note that for higher dimensions, Lie algebra valued functions such as $\boldsymbol{\nu}$ will require a different discretization than the one used here in order to respect the group product involved in this model; we will explore this extension in a future paper. }
\end{remark*}

\paragraph{Geometric Formulation.}

Rather remarkably, the governing equations~(\ref{nematic1a}-\ref{nematic4a}) for the dynamics of nematic liquid crystals arise from a variational principle on a certain Lie group with respect to a standard Lagrangian given by the difference between kinetic and potential energies.  Gay-Balmaz \& Ratiu~\cite{GayBalmaz2009} show, in particular, that the appropriate configuration group for three-dimensional nematic liquid cyrstal flow is a continuum analogue of the special Euclidean group, namely the semidirect product $\mathrm{Diff}_{\mathrm{vol}}(M') \, \circledS \, \mathcal{F}(M',SO(3))$, where $\mathcal{F}(M',SO(3))$ is the space of maps from the fluid domain $M' \subseteq \mathbb{R}^3$ to $SO(3)$.

The configuration space for two-dimensional nematic liquid crystal flow is the semidirect product
\begin{equation}
G = \mathrm{Diff}_{\mathrm{vol}}(M) \, \circledS \, \mathcal{F}(M,S^1),
\end{equation}
where $\mathcal{F}(M,S^1)$ is the space of maps from the two-dimensional domain to $S^1$.  The group product is given by
\begin{equation}
(\varphi_1,\theta_1)(\varphi_2,\theta_2) = (\varphi_1\circ\varphi_2, \varphi_2^*\theta_1+\theta_2)
\quad \text{for } (\varphi_1,\theta_1),(\varphi_2,\theta_2) \in G. \label{G_multiplication_nematic}
\end{equation}

The Lie algebra of $G$ is
\begin{equation}
\mathfrak{g} = \mathfrak{X}_{\mathrm{div}}(M) \, \circledS \, \mathcal{F}(M).
\end{equation}
We identify the space dual to $\mathfrak{g}$ with
\begin{equation}
\mathfrak{g}^* = \Omega^1(M)/d\Omega^0(M) \times \mathcal{F}(M)
\end{equation}
through the pairing
\begin{equation}
\left\langle (\mathbf{w}^\flat,\pi), (\mathbf{v},\psi) \right\rangle = \langle \mathbf{w}^\flat,\mathbf{v} \rangle + \langle \pi, \psi \rangle
\end{equation}
for $\mathbf{w}^\flat \in \Omega^1(M)/d\Omega^0(M)$, $\mathbf{v}\in \mathfrak{X}_{\mathrm{div}}(M)$, and $\pi,\psi \in \mathcal{F}(M)$, where
\begin{equation}
\langle \mathbf{w}^\flat,\mathbf{v} \rangle = \int_M \mathbf{w}^\flat(\mathbf{v}) \, d\mathbf{x}
\end{equation}
and
\begin{equation}
\langle \pi,\psi \rangle = \int_M \pi(\mathbf{x})\psi(\mathbf{x}) \, d\mathbf{x}.
\end{equation}

The bracket on $\mathfrak{g}$ is computed to be
\begin{equation}
\mathrm{ad}_{(\mathbf{u},\omega)} (\mathbf{v},\psi) = (-\pounds_\mathbf{u}\mathbf{v}, \pounds_\mathbf{v}\omega - \pounds_\mathbf{u}\psi),
\end{equation}
with dual
\begin{equation}
\mathrm{ad}^*_{(\mathbf{u},\omega)} (\mathbf{w}^\flat,\pi) = (\pounds_\mathbf{u}\mathbf{w}^\flat + \pi d\omega, \pounds_\mathbf{u}\pi).
\end{equation}

The space of advected parameters (not to be confused with the second factor of $G$) is $\mathcal{F}(M,S^1)$, whose elements $\alpha$ are acted upon from the right by elements $(\varphi,\theta)$ of $G$ in the following manner:
\begin{equation}
\alpha \cdot (\varphi,\theta) = \varphi^* \alpha - \theta. \label{Gaction_nematic}
\end{equation}
One checks that~(\ref{Gaction_nematic}) is a well-defined right action that is consistent with the group multiplication law~(\ref{G_multiplication_nematic}).

Note that the theory presented in Section~\ref{section:EP} does not, strictly speaking, apply to this situation, as $\mathcal{F}(M,S^1)$ is not a vector space.  Rather than introducing a flurry of new notation and terminology, we will simply regard members of $\mathcal{F}(M,S^1)$ as real-valued functions on $M$ when necessary.  Those readers interested in the theoretical framework underpinning Euler-Poincar\'{e} systems with nonlinear advected parameter spaces are encouraged to consult the work of Gay-Balmaz \& Tronci~\cite{GayBalmaz2010}.

The infinitesimal action of $\mathfrak{g}$ on $\mathcal{F}(M,S^1)$ is given by
\begin{equation}
\alpha \cdot (\mathbf{u},\omega) = \pounds_\mathbf{u} \alpha - \omega
\quad \text{for } (\mathbf{u},\omega) \in \mathfrak{X}_{\mathrm{div}}(M) \, \circledS \, \mathcal{F}(M).
\end{equation}

The diamond operation~(\ref{diamond_right}) is computed as follows: For any $\alpha,\beta \in \mathcal{F}(M,S^1)$ and $(\mathbf{u},\omega) \in \mathfrak{g}$,
\begin{align}
\langle \alpha \cdot (\mathbf{u},\omega), \beta \rangle
&= \langle \pounds_\mathbf{u} \alpha, \beta \rangle - \langle \omega, \beta \rangle \nonumber \\
&= \langle \beta d\alpha, \mathbf{u} \rangle - \langle \beta, \omega \rangle \nonumber \\
&= -\left\langle (-\beta d\alpha, \beta), (\mathbf{u},\omega) \right\rangle \label{diamond_calculation_nematic}
\end{align}
implying that
\begin{equation}
\beta \diamond \alpha = (-\beta d\alpha, \beta).
\end{equation}
The reader may find it helpful at this point to refer to~\cite{Holm1998} for more examples of calculations of this type.

The Lagrangian $\ell : [\mathfrak{X}_{\mathrm{div}}(M) \, \circledS \, \mathcal{F}(M)] \times \mathcal{F}(M,S^1) \rightarrow \mathbb{R}$ for nematic liquid crystals is
\begin{equation}
\ell(\mathbf{v},\omega,\alpha) = \frac{1}{2}\langle\mathbf{v}^\flat,\mathbf{v}\rangle + \frac{1}{2}\langle\omega,\omega\rangle - \int_M F(\alpha,d\alpha) d\mathbf{x},
\end{equation}
where $F(\alpha,d\alpha)$ is the free energy.  The first two terms in the Lagrangian may be recognized as the kinetic energy of the fluid due to translational motion and internal rotations, respectively.

The Euler-Poincar\'{e} equations~(\ref{EPcontinuous_right}) with the advected parameter $\alpha$ read
\begin{align}
\frac{\partial \mathbf{v}^\flat}{\partial t} + \pounds_\mathbf{v} \mathbf{v}^\flat &= -\frac{\delta\ell}{\delta\alpha}d\alpha - d\tilde{p} \label{EP1_nematic} \\
\frac{\partial \omega}{\partial t} + \pounds_\mathbf{v} \omega &= \frac{\delta\ell}{\delta\alpha} \label{EP2_nematic} \\
\frac{\partial \alpha}{\partial t} + \pounds_\mathbf{v} \alpha &= \omega, \label{EP3_nematic}
\end{align}
where the term $\omega d\omega = d(\omega^2/2)$ in~(\ref{EP1_nematic}) has been absorbed into the pressure differential $d\tilde{p}$.

If $(\varphi,\theta) \in G$ is used to denote the fluid configuration, then the evolution of $\alpha$ is given explicitly by
\begin{equation}
\alpha = \alpha_0 \cdot (\varphi,\theta)^{-1} = \alpha_0 \cdot (\varphi^{-1},-\varphi_*\theta) = \varphi_*(\alpha_0+\theta).
\end{equation}
In particular, if $\alpha_0=0$, $\alpha$ and $\theta$ are related via a transformation between spatial and material coordinates:
\begin{equation}
\alpha \circ \varphi = \theta.
\end{equation}

A common choice for the free energy function $F$ is
\begin{equation}
F(\alpha,d\alpha) = \frac{1}{2} || d\alpha ||^2, \label{free_energy}
\end{equation}
which corresponds to the so-named ``one-constant approximation'' of the Oseen-Z$\mathrm{\ddot{o}}$cher-Frank free energy~(\cite{GayBalmaz2009}). The variational derivative $\frac{\delta\ell}{\delta\alpha}$ is then
\begin{equation}
\frac{\delta\ell}{\delta\alpha} = *d*d\alpha = \Delta\alpha,
\end{equation}
i.e., simply the Laplacian of $\alpha$.

\paragraph{The Continuous Kelvin-Noether Theorem.}

Let $\mathcal{C}$ be a direct sum consisting of the space of loops in $M$ together with the space of scalar functions on $M$. Let $G = \mathrm{Diff}_\mathrm{vol}(M) \, \circledS \, \mathcal{F}(M,S^1)$ act on $\mathcal{C}$ via the pullback on each factor: $(\gamma,\phi) \cdot (\varphi,\theta) = (\varphi^*\gamma,\varphi^*\phi)$ for any loop $\gamma$ in $M$, any scalar function $\phi \in \mathcal{F}(M)$, any diffeomorphism $\varphi \in \mathrm{Diff}_\mathrm{vol}(M)$, and any $\theta \in \mathcal{F}(M,S^1)$.  Let $\mathcal{K} : \mathcal{C} \times \mathcal{F}(M,S^1) \rightarrow \mathfrak{g}^{**}$ be the equivariant map defined by
\begin{equation}
\langle \mathcal{K}((\gamma,\phi),\alpha), (\mathbf{w}^\flat,\psi) \rangle = \oint_\gamma \mathbf{w}^\flat + \langle \psi, \phi \rangle,
\end{equation}
where $\mathbf{w}^\flat$ is a one-form and $\psi \in \mathcal{F}(M)$.  The Kelvin-Noether theorem then gives the statement
\begin{equation}
\frac{d}{dt} \left(\oint_{\gamma(t)} \mathbf{v}^\flat(t) + \langle \omega(t),\phi(t) \rangle \right) = -\oint_{\gamma(t)} \frac{\delta\ell}{\delta\alpha}(t)d\alpha(t) + \left\langle \frac{\delta\ell}{\delta\alpha}(t), \phi(t) \right\rangle
\end{equation}
where $\mathbf{v}(t)$ is the fluid velocity field, $\omega(t)$ is the local angular velocity field, $\gamma(t)$ is a loop advected with the flow, and $\phi(t) = \varphi(t)_*\phi(0)$ is a scalar function advected with the flow.

To obtain a more specific conservation law, employ the free energy function~(\ref{free_energy}) so that $\frac{\delta\ell}{\delta\alpha} = \Delta \alpha$ and impose the boundary condition $d\alpha|_{\partial M}=0$.  Let $\phi(0)\equiv 1$ and take the limit as the loop $\gamma$ contracts to a point.  Using Stoke's theorem, we obtain
\begin{equation}
\frac{d}{dt} \int_M \omega(t) d\mathbf{x} = \int_M \Delta\alpha(t) d\mathbf{x} = 0.
\end{equation}
In other words, the \emph{total angular momentum due to micromotions} is preserved along solutions to the governing equations for nematic liquid crystal flow.  (This is easier to see directly from~(\ref{EP2_nematic}), although the Kelvin-Noether approach illuminates the connection between this conservation law and the symmetry of the nematic liquid crystal Lagrangian under the action of $\mathrm{Diff}_\mathrm{vol}(M)$ on the second factor $\mathcal{F}(M,S^1)$ of $G = \mathrm{Diff}_\mathrm{vol}(M) \, \circledS \, \mathcal{F}(M,S^1)$.)

\paragraph{Spatial Discretization.}

A spatial discretization of nematic liquid crystal flow is obtained by replacing the continuous configuration space with
\begin{equation}
G = \mathcal{D}(\mathbb{M}) \, \circledS \, \Omega_d^0(\mathbb{M},S^1),
\end{equation}
where $\Omega_d^1(\mathbb{M},S^1)$ denotes the space of $S^1$-valued discrete zero-forms on the mesh $\mathbb{M}$.  Elements of $\Omega_d^1(\mathbb{M},S^1)$ are merely vectors in $(S^1)^N$, where $N$ is the number of cells in the mesh.  In the notation of Table~\ref{tab:notation}, the group product is given by
\begin{equation}
(q_1,\theta_1)(q_2,\theta_2) = (q_1 q_2, q_2^*\theta_1+\theta_2) \quad \text{for } (q_1,\theta_1),(q_2,\theta_2) \in G. \label{dG_multiplication_nematic}
\end{equation}

The Lie algebra of $G$ is
\begin{equation}
\mathfrak{g} = \mathfrak{d}(\mathbb{M}) \, \circledS \, \Omega_d^0(\mathbb{M}).
\end{equation}
We identify the space dual to $\mathfrak{g}$ with
\begin{equation}
\mathfrak{g}^* = \Omega_d^1(\mathbb{M})/d\Omega_d^0(\mathbb{M}) \times \Omega_d^0(\mathbb{M})
\end{equation}
through the pairing
\begin{equation}
\left\langle (C^\flat,\pi), (B,\psi) \right\rangle = \langle C^\flat,B \rangle + \langle \pi, \psi \rangle
\end{equation}
for $C^\flat \in \Omega_d^1(\mathbb{M})/d\Omega_d^0(\mathbb{M})$, $B \in \mathfrak{d}(\mathbb{M})$, and $\pi,\psi \in \Omega_d^0(\mathbb{M})$, where
\begin{equation}
\langle C^\flat,B \rangle = \mathrm{Tr}(C^{\flat T} \Omega B)
\end{equation}
and
\begin{equation}
\langle \pi,\psi \rangle = \pi^T \Omega \psi.
\end{equation}

The bracket on $\mathfrak{g}$ is computed to be
\begin{equation}
\mathrm{ad}_{(A,\omega)} (B,\psi) = (-\pounds_A B, -B\omega + A\psi),
\end{equation}
with dual
\begin{equation}
\mathrm{ad}^*_{(A,\omega)} (C^\flat,\pi) = (\pounds_A C^\flat + \mathrm{skew}(\omega \pi^T), -A \pi).
\end{equation}

The space of advected parameters is $\Omega_d^0(\mathbb{M},S^1)$, whose elements $\alpha$ are acted upon from the right by elements $(q,\theta)$ of $G$ in the following manner:
\begin{equation}
\alpha \cdot (q,\theta) = q^* \alpha - \theta. \label{dGaction_nematic}
\end{equation}

Note once again that $\Omega_d^0(\mathbb{M},S^1)$ is not a vector space, but we will regard its members as real-valued discrete zero-forms on $\mathbb{M}$ when necessary in order to circumvent the need to introduce new notation and terminology.

The infinitesimal action of $\mathfrak{g}$ on $\Omega_d^0(\mathbb{M},S^1)$ is given by
\begin{equation}
\alpha \cdot (A,\omega) = -A \alpha - \omega \quad \text{for } (A,\omega) \in \mathfrak{d}(\mathbb{M}) \, \circledS \, \Omega_d^0(\mathbb{M}).
\end{equation}

The diamond operation~(\ref{diamond_right}) is computed as follows: For any $\alpha,\beta \in \Omega_d^0(\mathbb{M},S^1)$ and $(A,\omega) \in \mathfrak{g}$,
\begin{align}
\langle \alpha \cdot (A,\omega), \beta \rangle
&= -\langle A \alpha, \beta \rangle - \langle \omega, \beta \rangle \nonumber \\
&= -\alpha^T A^T \Omega \beta - \langle \beta, \omega \rangle \nonumber \\
&= -\mathrm{Tr}(A^T \Omega \beta \alpha^T) - \langle \beta, \omega \rangle \nonumber \\
&= \mathrm{Tr}(\Omega A \beta \alpha^T) - \langle \beta, \omega \rangle \nonumber \\
&= -\langle \mathrm{skew}(\beta \alpha^T), A \rangle - \langle \beta, \omega \rangle \nonumber \\
&= -\left\langle (\mathrm{skew}(\beta \alpha^T), \beta), (A,\omega) \right\rangle,
\end{align}
implying that
\begin{equation}
\beta \diamond \alpha = (\mathrm{skew}(\beta \alpha^T), \beta).
\end{equation}

The spatially discretized nematic liquid crystal Lagrangian $\ell : [\mathfrak{d}(\mathbb{M}) \, \circledS \, \Omega_d^0(\mathbb{M})] \times \Omega_d^0(\mathbb{M},S^1) \rightarrow \mathbb{R}$ is
\begin{equation}
\ell(Y,\omega,\alpha) = \frac{1}{2}\langle Y^\flat,Y \rangle + \frac{1}{2}\langle\omega,\omega\rangle - F_d(\alpha),
\end{equation}
where $F_d : \Omega_d^0(\mathbb{M},S^1) \rightarrow \mathbb{R}$ is a discrete approximation to the volume-integrated free energy $F$.  Here, $Y \in \mathfrak{d}(\mathbb{M})$ and $\omega \in \Omega_d^0(\mathbb{M})$ are the discrete counterparts to the continuous velocity field $\mathbf{v} \in \mathfrak{X}_{\mathrm{div}}(M)$ and the continuous angular velocity field $\omega \in \mathcal{F}(M)$, respectively.

The discrete-space, continuous-time Euler-Poincar\'{e} equations~(\ref{EPcontinuous_right}) with the advected parameter $\alpha$ read
\begin{align}
\frac{\partial Y^\flat}{\partial t} + \pounds_Y Y^\flat &\,\hat{=}\, \mathrm{skew}(\frac{\delta\ell}{\delta\alpha}\alpha^T) \label{dEP1_nematic} \\
\frac{\partial \omega}{\partial t} - Y \omega &= \frac{\delta\ell}{\delta\alpha} \label{dEP2_nematic} \\
\frac{\partial \alpha}{\partial t} - Y \alpha &= \omega, \label{dEP3_nematic}
\end{align}
where the term $\mathrm{skew}(\omega\omega^T)$ in~(\ref{dEP1_nematic}) vanishes by the symmetry of $\omega\omega^T$.

If $(q,\theta) \in G$ is used to denote the configuration of the fluid, then the evolution of $\alpha$ is given explicitly by
\begin{equation}
\alpha = \alpha_0 \cdot (q,\theta)^{-1} = \alpha_0 \cdot (q^{-1},-q_*\theta) = q_*(\alpha_0+\theta).
\end{equation}

A discretization of the free energy function~(\ref{free_energy}) and its variational derivative is obtained by employing the tools of classical discrete exterior calculus.  (See~\cite{Desbrun2005} for details.)  Define
\begin{equation}
F_d(\alpha) = \frac{1}{2} ||\mathbf{d}\alpha||_2^2, \label{dfree_energy}
\end{equation}
where $\mathbf{d}$ denotes the classical discrete exterior derivative that takes real-valued functions on $k$-simplices to real-valued functions on $(k+1)$-simplices, and $||\cdot||_2$ is the discrete $L^2$ norm of discrete forms introduced by Desbrun and co-authors~\cite{Desbrun2005}.
The variational derivative $\frac{\delta\ell}{\delta\alpha}$ is then
\begin{equation}
\frac{\delta\ell}{\delta\alpha} = *\mathbf{d}*\mathbf{d}\alpha = \mathbf{\Delta}\alpha,
\end{equation}
the discrete Laplacian of $\alpha$~\cite{Desbrun2005}.

\paragraph{Temporal Discretization.}

To perform a temporal discretization of~(\ref{dEP1_nematic}-\ref{dEP3_nematic}), we require a group difference map $\tau : \mathfrak{g} \rightarrow G$ to approximate the exponential.  For the semidirect product $G = \mathcal{D}(\mathbb{M}) \, \circledS \, \Omega_d^0(\mathbb{M},S^1)$, the exponential turns out to be a nontrivial extension of the exponential on $\mathcal{D}(\mathbb{M})$.  The following lemma, whose proof is presented in Appendix~\ref{appendix:nematic}, gives a formula for the exponential on $G$.

\begin{lemma} \label{lemma:nematic1}
The exponential map $\exp : \mathfrak{g} \rightarrow G$ for the group $G = \mathcal{D}(\mathbb{M}) \, \circledS \, \Omega_d^0(\mathbb{M},S^1)$ is given by
\begin{equation}
\exp(t(A,\omega)) = (e^{tA}, A^{-1}(I-e^{-tA})\omega), \label{exp_nematic}
\end{equation}
where $t \in \mathbb{R}$, $A \in \mathfrak{d}(\mathbb{M})$, $\omega \in \Omega_d^0(\mathbb{M})$, $e^{tA}$ is the usual matrix exponential, and $A^{-1}(I-e^{-tA})$ is to be regarded as a power series
\begin{equation}
A^{-1}(I-e^{-tA}) =  tI - \frac{t^2 A}{2} + \frac{t^3 A^2}{6} - \dots
\end{equation}
(so that it is defined even for $A$ not invertible).
\end{lemma}

In the following three lemmas, we derive a structure-preserving approximant $\tau$ to the exponential on $G$ and compute its inverse right-trivialized tangent $d\tau^{-1}$, as well as the dual of $d\tau^{-1}$.  Proofs of these lemmas are presented in Appendix~\ref{appendix:nematic}.

\begin{lemma} \label{lemma:nematic2}
The map $\tau : \mathfrak{g} \rightarrow G$ given by
\begin{equation}
\tau(A,\omega) = (\mathrm{cay}(A),\mathrm{cay}(-A/2)\omega) \label{tau_nematic}
\end{equation}
is a group difference map for the group $G = \mathcal{D}(\mathbb{M}) \, \circledS \, \Omega_d^0(\mathbb{M},S^1)$.  That is, $\tau$ is a local approximant to $\exp$ with $\tau(0,0)=(I,0)$, and $\tau$ satisfies
\begin{equation}
\tau(A,\omega)^{-1} = \tau(-A,-\omega) \quad \forall (A,\omega) \in \mathfrak{g}.\label{tau_inv_nematic}
\end{equation}
\end{lemma}

\begin{lemma} \label{lemma:nematic3}
The inverse right-trivialized tangent $d\tau^{-1} : \mathfrak{g} \times \mathfrak{g} \rightarrow \mathfrak{g}$ of the map~(\ref{tau_nematic}) is given by
\begin{equation}
d\tau^{-1}_{(A,\omega)} (B,\psi) = \left( d\mathrm{cay}^{-1}_A B, \: \mathrm{cay}(-A/2)\psi+\frac{1}{2}d\mathrm{cay}_{A/2}(d\mathrm{cay}^{-1}_A B)\omega \right). \label{dtau_inv_nematic}
\end{equation}
\end{lemma}

\begin{lemma} \label{lemma:nematic4}
For the map $\tau$ given by~(\ref{tau_nematic}), the dual of the operator $d\tau^{-1}$ is given by
\begin{equation}
(d\tau^{-1}_{(A,\omega)})^* (C^\flat,\pi) = \left( (d\mathrm{cay}^{-1}_A)^*C^\flat + \frac{1}{2}(d\mathrm{cay}^{-1}_A)^*(d\mathrm{cay}_{A/2})^*\mathrm{skew}(\pi\omega^T), \: \mathrm{cay}(A/2)\pi \right).
\end{equation}
Note that if $\pi$ is a scalar multiple of $\omega$, then this reduces to
\begin{equation}
(d\tau^{-1}_{(A,\omega)})^* (C^\flat,\pi) = \left( (d\mathrm{cay}^{-1}_A)^*C^\flat, \: \mathrm{cay}(A/2)\pi \right). \label{dtauinvstar_nematic_simplified}
\end{equation}
\end{lemma}

We are now equipped to write down the discrete-space, discrete-time Euler-Poincar\'{e} equations~(\ref{EP2_right}-\ref{EP3_right}) for nematic liquid crystal flow:  Using the fact that $\frac{\delta\ell}{\delta\omega}=\omega$, we use~(\ref{dtauinvstar_nematic_simplified}) obtain
\begin{align}
(d\mathrm{cay}^{-1}_{-hY_k})^*Y_k^\flat &\,\hat{=}\, (d\mathrm{cay}^{-1}_{hY_{k-1}})^*Y_{k-1}^\flat + h\mathrm{skew}\left(\frac{\delta\ell}{\delta\alpha_k}\alpha_k^T\right) \\
\mathrm{cay}(-hY_k/2)\omega_k &= \mathrm{cay}(hY_{k-1}/2)\omega_{k-1} + h\frac{\delta\ell}{\delta\alpha_k} \label{ddEPnematic2_pre} \\
\alpha_k &= \mathrm{cay}(hY_{k-1}) (\alpha_{k-1} + h\mathrm{cay}(-hY_{k-1}/2)\omega_{k-1}).
\end{align}
Upon rearrangement, these become
\begin{align}
\frac{Y^\flat_k-Y^\flat_{k-1}}{h} + &\frac{\pounds_{Y_{k-1}}Y^\flat_{k-1}+\pounds_{Y_k}Y^\flat_k}{2} + \frac{h}{4}(Y_{k-1}Y_{k-1}^\flat\Omega Y_{k-1}\Omega^{-1} - Y_k Y^\flat_k \Omega Y_k \Omega^{-1}) \,\hat{=}\, \mathrm{skew}\left(\frac{\delta\ell}{\delta\alpha_k}\alpha_k^T\right) \label{ddEPnematic1} \\
&\omega_k = \mathrm{cay}(hY_k/2)\left[\mathrm{cay}(hY_{k-1}/2)\omega_{k-1} + h\frac{\delta\ell}{\delta\alpha_k}\right] \label{ddEPnematic2} \\
&\alpha_k = \mathrm{cay}(hY_{k-1})\alpha_{k-1} + h\mathrm{cay}(hY_{k-1}/2)\omega_{k-1}. \label{ddEPnematic3}
\end{align}

\paragraph{The Discrete Kelvin-Noether Theorem.}

Take $\mathcal{C}$ to be the Lie algebra of $G=\mathcal{D}(\mathbb{M}) \, \circledS \, \Omega_d^0(\mathbb{M},S^1)$ and let $G$ act on $\mathcal{C} = \mathfrak{d}(\mathbb{M}) \, \circledS \, \Omega_d^0(\mathbb{M})$ via the adjoint representation.  Let $\mathcal{K} : \mathcal{C} \times \Omega_d^0(\mathbb{M},S^1) \rightarrow \mathfrak{g}$ be the equivariant map $((\Gamma,\psi),\alpha) \mapsto (\Gamma,\psi)$. The discrete Kelvin-Noether theorem then says that the quantity
\begin{equation}
M_k
= \left\langle (d\tau^{-1}_{-h(Y_k,\omega_k)})^*(Y_k^\flat,\omega_k), (\Gamma_k,\psi_k) \right\rangle
\end{equation}
satisfies
\begin{align}
M_k - M_{k-1}
&= h \left\langle \left(\mathrm{skew}\left(\frac{\delta\ell}{\delta\alpha_k}\alpha_k^T\right),\frac{\delta\ell}{\delta\alpha_k}\right), (\Gamma_k,\psi_k) \right\rangle \nonumber \\
&= h\left\langle \mathrm{skew}\left(\frac{\delta\ell}{\delta\alpha_k}\alpha_k^T\right),\Gamma_k \right\rangle
+ h\left\langle \frac{\delta\ell}{\delta\alpha_k}, \psi_k \right\rangle
\end{align}
with
\begin{align}
(\Gamma_k,\psi_k)
&= \mathrm{Ad}_{(y_k,\theta_k)} (\Gamma_0,\psi_0) \nonumber \\
&= \left((y_k)_*\Gamma_0, y_k(-\Gamma_0\theta_k+\psi_0)\right).
\end{align}

To obtain a more specific conservation law, consider the group difference map~(\ref{tau_nematic}) together with the free energy function~(\ref{dfree_energy}).  Choose $\Gamma_0 = 0$ and $\psi_0=\mathbf{1}:=(1,1,\cdots,1)^T$ so that $\Gamma_k=0$ and $\psi_k = y_k \mathbf{1} = \mathbf{1}$ by the stochasticity of $y_k$.  It then follows that
\begin{align}
\left\langle \mathrm{cay}(-hY_k/2) \omega_k, \mathbf{1} \right\rangle - \left\langle \mathrm{cay}(-hY_{k-1}/2) \omega_{k-1}, \mathbf{1} \right\rangle
&= h \langle \mathbf{\Delta}\alpha_k , \mathbf{1} \rangle \nonumber \\
&= 0, \label{nematic_angular_momentum}
\end{align}
where the last equality follows from an application of discrete Stoke's theorem to the discrete zero-form $\mathbf{\Delta}\alpha_k$.  Indeed, pairing such a form with the quantity $\mathbf{1}$ returns, in the framework of classical discrete exterior calculus, the integral of $\mathbf{d}*\mathbf{d}\alpha_k$ over the domain.  Recasting this area integral as a line integral of $*\mathbf{d}\alpha_k$ along the boundary of the domain, this quantity vanishes provided the appropriate boundary condition $\mathbf{d}\alpha_k=0$ is enforced along the domain's boundary.

Using the stochasticity of $\mathrm{cay}(-hY/2)$,~(\ref{nematic_angular_momentum}) can be rewritten as
\begin{equation}
\langle \omega_k, \mathbf{1} \rangle = \langle \omega_{k-1}, \mathbf{1} \rangle
\end{equation}
In other words, \emph{the total angular momentum due to micromotions is preserved exactly} along solutions to the discrete-space, discrete-time nematic liquid crystal equations~(\ref{ddEPnematic1}-\ref{ddEPnematic3}).

\begin{remark*}
\emph{
Alternatively, just as we observed in the continuous case, the angular momentum conservation law~(\ref{nematic_angular_momentum}) can be proven directly by pairing the left- and right-hand sides of~(\ref{ddEPnematic2_pre}) with the vector $\mathbf{1}$.  The Kelvin-Noether approach connects this conservation law with the symmetry of the discrete Lagrangian under the action of $G$ on the second factor $\Omega_d^0(\mathbb{M},S^1)$ of $G=\mathcal{D}(\mathbb{M}) \, \circledS \, \Omega_d^0(\mathbb{M},S^1)$.
}
\end{remark*}

\subsubsection{Microstretch Fluids} \label{section:microstretch}

A \emph{microstretch fluid} is a complex fluid whose constituent particles undergo three types of motion: translation, rotation, and stretch.  The variables of interest include the fluid velocity $\mathbf{v} \in \mathfrak{X}_{\mathrm{div}}(M')$, the local angular velocity field $\boldsymbol{\nu} \in \mathcal{F}(M',\mathbb{R}^3)$, the director $\mathbf{n} \in \mathcal{F}(M',\mathbb{R}^3)$, and the \emph{local stretch rate} $\nu_0 \in \mathcal{F}(M',\mathbb{R})$.  The first three of these variables have the same physical interpretation as with nematic liquid crystals.  The new variable $\nu_0$ measures the local rate of expansion of fluid particles, taking positive values at locations of expansion and negative values at locations of contraction.

A final variable required for a complete description of microstretch fluid flow is the \emph{microinertia tensor} $i \in \mathcal{F}(M',Sym(3)^+)$, which assigns a symmetric positive-definite inertia tensor to each point in the domain.  (The classic director approach to nematic liquid crystal flow presented in the previous section implicitly assumes $i$ to be uniformly the identity matrix, which fortunately holds for all time if it holds at $t=0$ in the absence of stretching.)

In practice, it is common to work with the \emph{modified microinertia tensor} $j$ and its trace $j_0$, defined as follows:
\begin{align}
j &= \mathrm{Tr}(i)I - i \\
j_0 &= \mathrm{Tr}(j)
\end{align}

\paragraph{The Continuous Equations of Motion.}

In terms of the variables defined above, the equations of motion for three-dimensional, incompressible, homogeneous microstretch fluid flow are given by
\begin{align}
\frac{\partial \mathbf{v}}{\partial t} + (\mathbf{v} \cdot \nabla) \mathbf{v} &= -\partial_i\left(\frac{\partial F}{\partial\mathbf{n}_{,i}} \cdot \nabla \mathbf{n} \right) - \nabla p \label{microstretch1_3D} \\
\frac{j_0}{2}\left(\frac{\partial \nu_0}{\partial t} + \mathbf{v} \cdot \nabla \nu_0 - \nu_0^2 \right) + (j\boldsymbol{\nu})\cdot\boldsymbol{\nu} &= -\mathbf{h}\cdot\mathbf{n} \label{microstretch2_3D} \\
j\left(\frac{\partial \boldsymbol{\nu}}{\partial t} + (\mathbf{v} \cdot \nabla) \boldsymbol{\nu}\right) - 2\nu_0 j \boldsymbol{\nu} - (j \boldsymbol{\nu}) \times \boldsymbol{\nu} &= \mathbf{h}\times\mathbf{n} \label{microstretch3_3D} \\
\frac{\partial \mathbf{n}}{\partial t} + (\nabla\mathbf{n})\mathbf{v} + \mathbf{n}\times\boldsymbol{\nu} - \nu_0 \mathbf{n} &= 0 \label{microstretch4_3D} \\
\frac{\partial j}{\partial t} + (\mathbf{v} \cdot \nabla) j + 2\nu_0 j + [j,\hat{\boldsymbol{\nu}}] &= 0 \label{microstretch5_3D} \\
\nabla \cdot \mathbf{v} &= 0, \label{microstretch6_3D}
\end{align}
where $p$ is a scalar pressure field,
\begin{equation}
\mathbf{h} = \frac{\partial F}{\partial\mathbf{n}} - \partial_i\left(\frac{\partial F}{\partial\mathbf{n}_{,i}} \right),
\end{equation}
$F(\mathbf{n},\nabla\mathbf{n})$ is the free energy, and $\hat{\boldsymbol{\nu}} \in \mathcal{F}(M',\mathfrak{so}(3))$ is the skew-symmetric matrix-valued map related to $\boldsymbol{\nu} \in \mathcal{F}(M',\mathbb{R}^3)$ via the hat isomorphism~(\ref{hat_iso}).

The equations of motion for microstretch fluid flow have an interpretation similar that of the governing equations for nematic liquid crystal flow: Equation~(\ref{microstretch1_3D}) is Euler's fluid equation with additional terms arising from micromotions; equations~(\ref{microstretch2_3D}-\ref{microstretch5_3D}) describe the advection of the local stretch rate $\nu_0$, the local angular velocity field $\boldsymbol{\nu}$, the director field $\mathbf{n}$, and the modified microinertia tensor $j$, respectively; and equation~(\ref{microstretch6_3D}) expresses the incompressibility of the flow.

Once again, we will specialize to the case of \emph{two-dimensional microstretch fluid flow} for simplicity.  Assume that the motion is restricted to the $x$-$y$ plane so that the director $\mathbf{n}$ is everywhere orthogonal to the $z$-axis and the modified microinertia tensor $j$ has the form
\begin{equation}
j =
\begin{pmatrix}
j_{11} & j_{12} & 0      \\
j_{21} & j_{22} & 0      \\
0      & 0      & j_{33} \\
\end{pmatrix}. \label{jform_microstretch}
\end{equation}
Define the variables $(\omega, R) \in \mathcal{F}(M,\mathbb{R}\oplus\mathbb{R}), (\alpha,\lambda) \in \mathcal{F}(M,S^1\oplus\mathbb{R}), (j_r, j_s) \in \mathcal{F}(M,\mathbb{R}\oplus\mathbb{R})$ via
\begin{align}
\boldsymbol{\nu} &= (0,0,\omega)^T \\
\nu_0 &= R \\
\mathbf{n} &= e^\lambda(\cos\alpha,\sin\alpha,0)^T \label{iso3_microstretch} \\
\frac{1}{2}j_0 &= e^{j_s} \label{iso4_microstretch} \\
j_{33} &= e^{j_r}. \label{iso5_microstretch}
\end{align}
The motivation behind this choice of variables will be made clear in the next section.

The governing equations for \emph{two-dimensional}, incompressible, homogeneous microstretch fluid flow are then given by
\begin{align}
\frac{\partial \mathbf{v}}{\partial t} + (\mathbf{v} \cdot \nabla) \mathbf{v} &= -\partial_i\left(\frac{\partial F}{\partial\alpha_{,i}}\right)\nabla\alpha  -\partial_i\left(\frac{\partial F}{\partial\lambda_{,i}}\right)\nabla\lambda - \nabla p \label{microstretch1} \\
\frac{\partial \pi}{\partial t} + \mathbf{v} \cdot \nabla \pi &= \partial_i\left(\frac{\partial F}{\partial\alpha_{,i}}\right) - \frac{\partial F}{\partial\alpha} \label{microstretch2} \\
\frac{\partial Q}{\partial t} + \mathbf{v} \cdot \nabla Q &= -2i_r - 2i_s + \partial_i\left(\frac{\partial F}{\partial\lambda_{,i}}\right) - \frac{\partial F}{\partial\lambda} \label{microstretch3} \\
\frac{\partial \alpha}{\partial t} + \mathbf{v} \cdot \nabla \alpha &= \omega \label{microstretch4} \\
\frac{\partial \lambda}{\partial t} + \mathbf{v} \cdot \nabla \lambda &= R \label{microstretch5} \\
\frac{\partial j_r}{\partial t} + \mathbf{v} \cdot \nabla j_r &= -2R \label{microstretch6} \\
\frac{\partial j_s}{\partial t} + \mathbf{v} \cdot \nabla j_s &= -2R, \label{microstretch7}
\end{align}
where
\begin{align}
\pi &= e^{j_r}\omega \\
Q &= e^{j_s} R \\
i_r &= \frac{1}{2}e^{j_r}\omega^2 \\
i_s &= \frac{1}{2}e^{j_s}R^2.
\end{align}

\paragraph{Geometric Formulation.}

In yet another remarkable fashion, the governing equations for microstretch fluid flow can be cast as Euler-Poincar\'{e} equations on a certain Lie group with respect to a natural Lagrangian.  In this case, as Gay-Balmaz \& Ratiu~\cite{GayBalmaz2009} show, the appropriate configuration group (on a three dimensional domain $M'$) is the semidirect product $\mathrm{Diff}_{\mathrm{vol}}(M') \, \circledS \, \mathcal{F}(M',CSO(3))$, where $CSO(3)$ is the \emph{conformal special orthogonal group}: the group of (positive) scalar multiples of rotation matrices.

In two dimensions, the configuration space for microstretch fluid flow is the semidirect product
\begin{equation}
G = \mathrm{Diff}_{\mathrm{vol}}(M) \, \circledS \, \mathcal{F}(M,S^1 \oplus \mathbb{R}), \label{G_microstretch}
\end{equation}
where $\mathcal{F}(M,S^1 \oplus \mathbb{R})$ is the space of maps from the domain to the direct sum $S^1 \oplus \mathbb{R}$.  The group product is given by
\begin{equation}
(\varphi_1,(\theta_1,r_1))(\varphi_2,(\theta_2,r_2)) = (\varphi_1\circ\varphi_2, (\varphi_2^*\theta_1+\theta_2,\varphi_2^*r_1+r_2)) \quad \text{for } (\varphi_1,(\theta_1,r_1)),(\varphi_2,(\theta_2,r_2)) \in G. \label{G_multiplication_microstretch}
\end{equation}

We regard $G$ as the two-dimensional realization of the group
\begin{equation}
G' = \mathrm{Diff}_{\mathrm{vol}}(M') \, \circledS \, \mathcal{F}(M',CSO(3))
\end{equation}
with group product given by
\begin{equation}
(\varphi_1,\chi_1)(\varphi_2,\chi_2) = (\varphi_1\circ\varphi_2, (\varphi_2^*\chi_1)\chi_2),
\end{equation}
where $M'$ is an open subset of $\mathbb{R}^3$ containing an embedding of $M$ in the $x$-$y$ plane and any $\chi \in \mathcal{F}(M',CSO(3))$ is related to a quantity $(\theta,r) \in \mathcal{F}(M,S^1 \oplus \mathbb{R})$ via
\begin{equation}
\chi = e^r
\begin{pmatrix}
\cos\theta & -\sin\theta & 0 \\
\sin\theta &  \cos\theta & 0 \\
0 & 0 & 1
\end{pmatrix}. \label{chi_theta_relation}
\end{equation}

The Lie algebra of $G$ is
\begin{equation}
\mathfrak{g} = \mathfrak{X}_{\mathrm{div}}(M) \, \circledS \, \mathcal{F}(M,\mathbb{R}\oplus\mathbb{R}).
\end{equation}
We identify the space dual to $\mathfrak{g}$ with
\begin{equation}
\mathfrak{g}^* = \Omega^1(M)/d\Omega^0(M) \times \mathcal{F}(M,\mathbb{R}\oplus\mathbb{R})
\end{equation}
through the pairing
\begin{equation}
\left\langle (\mathbf{w}^\flat,(\pi,S)), (\mathbf{v},(\psi,R)) \right\rangle = \langle \mathbf{w}^\flat,\mathbf{v} \rangle + \langle \pi, \psi \rangle + \langle S,R \rangle
\end{equation}
for $\mathbf{w}^\flat \in \Omega^1(M)/d\Omega^0(M)$, $\mathbf{v}\in \mathfrak{X}_{\mathrm{div}}(M)$, and $(\pi,S),(\psi,R) \in \mathcal{F}(M,\mathbb{R}\oplus\mathbb{R})$, where
\begin{align}
\langle \mathbf{w}^\flat,\mathbf{v} \rangle &= \int_M \mathbf{w}^\flat(\mathbf{v}) \, d\mathbf{x} \\
\langle \pi,\psi \rangle &= \int_M \pi(\mathbf{x})\psi(\mathbf{x}) \, d\mathbf{x} \\
\langle S,R \rangle &= \int_M S(\mathbf{x})R(\mathbf{x}) \, d\mathbf{x}.
\end{align}

The bracket on $\mathfrak{g}$ is computed to be
\begin{equation}
\mathrm{ad}_{(\mathbf{u},(\omega,Q))} (\mathbf{v},(\psi,R)) = (-\pounds_\mathbf{u}\mathbf{v}, (\pounds_\mathbf{v}\omega - \pounds_\mathbf{u}\psi, \pounds_\mathbf{v}Q - \pounds_\mathbf{u}R)),
\end{equation}
with dual
\begin{equation}
\mathrm{ad}^*_{(\mathbf{u},(\omega,Q))} (\mathbf{w}^\flat,(\pi,S)) = (\pounds_\mathbf{u}\mathbf{w}^\flat + \pi d\omega + SdQ, ( \pounds_\mathbf{u}\pi,\pounds_\mathbf{u}S)).
\end{equation}

The advected parameters are $(\alpha,\lambda) \in \mathcal{F}(\mathbb{M},S^1\oplus\mathbb{R})$ and $(j_r,j_s) \in \mathcal{F}(\mathbb{M},\mathbb{R}\oplus\mathbb{R})$, which are acted upon from the right by elements $(\varphi,(\theta,r))$ of $G$ via
\begin{align}
(\alpha,\lambda)\cdot(\varphi,(\theta,r)) = (\varphi^*\alpha-\theta, \varphi^*\lambda-r) \label{Gaction1_microstretch} \\
(j_r,j_s)\cdot(\varphi,(\theta,r)) = (\varphi^*j_r+2r, \varphi^*j_s+2r). \label{Gaction2_microstretch}
\end{align}

The choice of such a set of advected parameters and the corresponding group actions~(\ref{Gaction1_microstretch}-\ref{Gaction2_microstretch}) is motivated by the ambient geometric structure of three-dimensional microstretch fluid flow.  In three dimensions, the advected parameters are the director $\mathbf{n} \in \mathcal{F}(M',\mathbb{R}^3)$ and the microinertia tensor $i \in \mathcal{F}(M',Sym(3)^+)$, related to $(\alpha,\lambda)$ and $(j_r,j_s)$ via~(\ref{iso3_microstretch}-\ref{iso5_microstretch}).  These parameters are acted upon from the right by elements $(\varphi,\chi)$ of $\mathrm{Diff}_{\mathrm{vol}}(M') \, \circledS \, \mathcal{F}(M',CSO(3))$ via
\begin{align}
\mathbf{n} \cdot (\varphi,\chi) &= \chi^{-1}(\varphi^*\mathbf{n}) \label{Gaction1_microstretch3D} \\
i \cdot (\varphi,\chi) &= \chi^T(\varphi^*i)\chi. \label{Gaction2_microstretch3D}
\end{align}
One checks that for $\chi$ of the form~(\ref{chi_theta_relation}), the three-dimensional actions~(\ref{Gaction1_microstretch3D}-\ref{Gaction2_microstretch3D}) on $\mathbf{n}$ and $i$ induce the two-dimensional actions~(\ref{Gaction1_microstretch}-\ref{Gaction2_microstretch}) on $(\theta,r)$ and $(j_r,j_s)$.  One furthermore checks that, in a rather convenient manner, the quantities $j_r$ and $j_s$ are sufficient to reconstruct the products $j\boldsymbol{\nu}$ and $j_0\nu_0$ appearing in~(\ref{microstretch1_3D}-\ref{microstretch6_3D}), provided the modified microinertia tensor $j=\mathrm{Tr}(i)I-i$ has the form~(\ref{jform_microstretch}).

Note that the exponentials appearing in relations~(\ref{iso3_microstretch}-\ref{iso5_microstretch}) play an essential role in ensuring the feasibility of an eventual discretization of the configuration space and its action on the advected variables.  They endow the space of advected parameters with an additive, rather than multiplicative, structure, thereby guaranteeing the sensibleness of replacing continuous pullbacks of scalar fields by matrix-vector products for the purposes of discretization.

Having made note of these observations, let us proceed with a derivation of the Euler-Poincar\'{e} equations associated with the group G given by~(\ref{G_microstretch}).  The infinitesimal actions of $\mathfrak{g} = \mathfrak{X}_{\mathrm{div}}(M) \, \circledS \, \mathcal{F}(M,\mathbb{R}\oplus\mathbb{R})$ on the advected variables $(\alpha,\lambda) \in \mathcal{F}(\mathbb{M},S^1\oplus\mathbb{R})$ and $(j_r,j_s) \in \mathcal{F}(\mathbb{M},\mathbb{R}\oplus\mathbb{R})$ are given by
\begin{align}
(\alpha,\lambda)\cdot(\mathbf{u},(\omega,R)) &= (\pounds_\mathbf{u}\alpha-\omega, \pounds_\mathbf{u}\lambda-R) \label{gaction1_microstretch} \\
(j_r,j_s)\cdot(\mathbf{u},(\omega,R)) &= (\pounds_\mathbf{u}j_r+2R, \pounds_\mathbf{u}j_s+2R). \label{gaction2_microstretch}
\end{align}
These give rise to the following diamond operations, which are easily derived through analogy with~(\ref{diamond_calculation_nematic}):
\begin{align}
(\beta,\sigma)\diamond(\alpha,\lambda) &= (-\beta d\alpha - \sigma d\lambda, (\beta,\sigma)) \label{diamond1_microstretch} \\
(i_r,i_s)\diamond(j_r,j_s) &= (-i_r dj_r - i_s dj_s, (0,-2i_r-2i_s)). \label{diamond2_microstretch}
\end{align}

The Lagrangian $\ell : [\mathfrak{X}_{\mathrm{div}}(M) \, \circledS \, \mathcal{F}(M,\mathbb{R}\oplus\mathbb{R})] \times [\mathcal{F}(\mathbb{M},S^1\oplus\mathbb{R}) \oplus \mathcal{F}(\mathbb{M},\mathbb{R}\oplus\mathbb{R})] \rightarrow \mathbb{R}$ for two-dimensional microstretch fluid flow is given by
\begin{equation}
\ell(\mathbf{v},\omega,R,\alpha,\lambda,j_r,j_s) = \frac{1}{2}\langle \mathbf{v}^\flat,\mathbf{v} \rangle + \frac{1}{2}\langle e^{j_r}\omega,\omega \rangle + \frac{1}{2}\langle e^{j_s}R,R \rangle - \int_M F(\alpha,d\alpha,\lambda,d\lambda)d\mathbf{x}, \label{lagrangian_microstretch}
\end{equation}
where $F$ is the microstretch free energy function.  The first three terms in this Lagrangian correspond to, respectively, the kinetic energy due to translational, rotational, and stretching motion.

Make the following definitions for the variational derivatives of $\ell$:
\begin{align}
\pi = \frac{\delta\ell}{\delta\omega} &= e^{j_r}\omega \\
Q = \frac{\delta\ell}{\delta R} &= e^{j_s} R \\
i_r = \frac{\delta\ell}{\delta j_r} &= \frac{1}{2}e^{j_r}\omega^2 \\
i_s = \frac{\delta\ell}{\delta j_s} &= \frac{1}{2}e^{j_s}R^2.
\end{align}
The Euler-Poincar\'{e} equations~(\ref{EPcontinuous_right}) associated with the Lagrangian~(\ref{lagrangian_microstretch}) then read
\begin{align}
\frac{\partial \mathbf{v}^\flat}{\partial t} + \pounds_\mathbf{v} \mathbf{v}^\flat &= -\frac{\delta\ell}{\delta\alpha}d\alpha  -\frac{\delta\ell}{\delta\lambda}d\lambda - d\tilde{p} \label{EP1_microstretch} \\
\frac{\partial \pi}{\partial t} + \pounds_\mathbf{v} \pi &= \frac{\delta\ell}{\delta\alpha} \label{EP2_microstretch} \\
\frac{\partial Q}{\partial t} + \pounds_\mathbf{v} Q &= -2i_r - 2i_s + \frac{\delta\ell}{\delta\lambda} \label{EP3_microstretch} \\
\frac{\partial \alpha}{\partial t} + \pounds_\mathbf{v} \alpha &= \omega \label{EP4_microstretch} \\
\frac{\partial \lambda}{\partial t} + \pounds_\mathbf{v} \lambda &= R \label{EP5_microstretch} \\
\frac{\partial j_r}{\partial t} + \pounds_\mathbf{v} j_r &= -2R \label{EP6_microstretch} \\
\frac{\partial j_s}{\partial t} + \pounds_\mathbf{v} j_s &= -2R, \label{EP7_microstretch}
\end{align}
where four of the terms in~(\ref{EP1_microstretch}) have conspired to produce a full differential, which we have absorbed into the pressure differential $d\tilde{p}$.

A convenient choice for the free energy $F$ is
\begin{equation}
F(\alpha,d\alpha,\lambda,d\lambda) = \frac{1}{2} || d\alpha ||^2 + \frac{1}{2} \lambda^2 , \label{free_energy_microstretch}
\end{equation}
which corresponds to a modification of~(\ref{free_energy}), obtained through the addition of a potential energy due to internal stretching in accordance with a leading order approximation to Hooke's Law.  (Recall that the length of the director $\mathbf{n}$ is related to $\lambda$ via $||\mathbf{n}||=e^\lambda$.) The variational derivatives $\frac{\delta\ell}{\delta\alpha}$ and $\frac{\delta\ell}{\delta\lambda}$ are then
\begin{align}
\frac{\delta\ell}{\delta\alpha} &= *d*d\alpha = \Delta\alpha \\
\frac{\delta\ell}{\delta\lambda} &= -\lambda.
\end{align}

\paragraph{The Continuous Kelvin-Noether Theorem.}

In analogy with nematic liquid crystal flow, the Kelvin-Noether theorem applied to microstretch fluid flow with the free energy~(\ref{free_energy_microstretch}) and the boundary condition $d\alpha|_{\partial M}=0$ gives, as one corollary, the conservation law
\begin{equation}
\frac{d}{dt} \int_M \pi(t) d\mathbf{x} = 0.
\end{equation}
That is, the \emph{total angular momentum due to micromotions} is preserved along solutions to the governing equations for microstretch fluid flow.

\paragraph{Spatial Discretization.}

A spatial discretization of two-dimensional microstretch fluid flow is obtained by replacing the continuous configuration space with
\begin{equation}
G = \mathcal{D}(\mathbb{M}) \, \circledS \, \Omega_d^0(\mathbb{M},S^1\oplus\mathbb{R}),
\end{equation}
where $\Omega_d^0(\mathbb{M},S^1\oplus\mathbb{R})$ denotes the space of $(S^1\oplus\mathbb{R})$-valued discrete zero-forms on the mesh $\mathbb{M}$.  For a mesh with $N$ cells, elements of $\Omega_d^0(\mathbb{M},S^1\oplus\mathbb{R})$ are merely pairs $(\theta,r)$ of real-valued $N\times 1$ vectors $\theta$ and $r$, with the entries of $\theta$ taken modulo $2\pi$.  The group product is given by
\begin{equation}
(q_1,(\theta_1,r_1))(q_2,(\theta_2,r_2)) = (q_1 q_2, (q_2^{-1}\theta_1+\theta_2,q_2^{-1}r_1+r_2)) \quad \text{for } (q_1,(\theta_1,r_1)),(q_2,(\theta_2,r_2)) \in G. \label{dG_multiplication_microstretch}
\end{equation}

The Lie algebra of $G$ is
\begin{equation}
\mathfrak{g} = \mathfrak{d}(\mathbb{M}) \, \circledS \, \Omega_d^0(\mathbb{M},\mathbb{R}\oplus\mathbb{R}).
\end{equation}
We identify the space dual to $\mathfrak{g}$ with
\begin{equation}
\mathfrak{g}^* = \Omega_d^1(\mathbb{M})/d\Omega_d^0(\mathbb{M}) \times \Omega_d^0(\mathbb{M},\mathbb{R}\oplus\mathbb{R})
\end{equation}
through the pairing
\begin{equation}
\left\langle (C^\flat,(\pi,S)), (B,(\psi,R)) \right\rangle = \langle C^\flat,B \rangle + \langle \pi, \psi \rangle + \langle S, R \rangle
\end{equation}
for $C^\flat \in \Omega_d^1(\mathbb{M})/d\Omega_d^0(\mathbb{M})$, $B \in \mathfrak{d}(\mathbb{M})$, and $\pi,\psi, S,R \in \Omega_d^0(\mathbb{M})$, where
\begin{align}
\langle C^\flat,B \rangle &= \mathrm{Tr}(C^{\flat T} \Omega B) \\
\langle \pi,\psi \rangle &= \pi^T \Omega \psi \\
\langle S, R \rangle &= S^T \Omega R.
\end{align}

The bracket on $\mathfrak{g}$ is computed to be
\begin{equation}
\mathrm{ad}_{(A,(\omega,Q))} (B,(\psi,R)) = (-\pounds_A B, (-B\omega + A\psi, -BQ + AR)),
\end{equation}
with dual
\begin{equation}
\mathrm{ad}^*_{(A,(\omega,Q))} (C^\flat,(\pi,S)) = (\pounds_A C^\flat + \mathrm{skew}(\omega\pi^T) + \mathrm{skew}(QS^T), ( -A\pi,-AS)).
\end{equation}

The advected parameters are $(\alpha,\lambda) \in \Omega_d^0(\mathbb{M},S^1\oplus\mathbb{R})$ and $(j_r,j_s) \in \Omega_d^0(\mathbb{M},\mathbb{R}\oplus\mathbb{R})$, which are acted upon from the right by elements $(q,(\theta,r))$ of $G$ via
\begin{align}
(\alpha,\lambda)\cdot(q,(\theta,r)) = (q^{-1}\alpha-\theta, q^{-1}\lambda-r) \label{dGaction1_microstretch} \\
(j_r,j_s)\cdot(q,(\theta,r)) = (q^{-1}j_r+2r, q^{-1}j_s+2r). \label{dGaction2_microstretch}
\end{align}

The induced infinitesimal actions of $\mathfrak{g} = \mathfrak{d}(\mathbb{M}) \, \circledS \, \Omega_d^0(\mathbb{M},\mathbb{R}\oplus\mathbb{R})$ on the advected variables $(\alpha,\lambda) \in \Omega_d^0(\mathbb{M},S^1\oplus\mathbb{R})$ and $(j_r,j_s) \in \Omega_d^0(\mathbb{M},\mathbb{R}\oplus\mathbb{R})$ are given by
\begin{align}
(\alpha,\lambda)\cdot(A,(\omega,R)) &= (-A\alpha-\omega, -A\lambda-R) \label{dgaction1_microstretch} \\
(j_r,j_s)\cdot(A,(\omega,R)) &= (-Aj_r+2R, -Aj_s+2R). \label{dgaction2_microstretch}
\end{align}
The astute reader will have recognized the pattern relating continuous and discrete diamond operations on scalar fields discussed in Section~\ref{section:nematic}, thereby permitting an easy calculation of the discrete diamond operations via inspection of their continuous counterparts~(\ref{diamond1_microstretch}-\ref{diamond2_microstretch}):
\begin{align}
(\beta,\sigma)\diamond(\alpha,\lambda) &= (\mathrm{skew}(\beta\alpha^T) + \mathrm{skew}(\sigma\lambda^T), (\beta,\sigma)) \\
(i_r,i_s)\diamond(j_r,j_s) &= (\mathrm{skew}(i_r j_r^T) + \mathrm{skew}(i_s j_s^T), (0,-2i_r-2i_s)).
\end{align}

The spatially discretized Lagrangian  $\ell : [\mathfrak{d}(\mathbb{M}) \, \circledS \, \Omega_d^0(\mathbb{M},\mathbb{R}\oplus\mathbb{R})] \, \times \, [\Omega_d^0(\mathbb{M},S^1\oplus\mathbb{R}) \oplus \Omega_d^0(\mathbb{M},\mathbb{R}\oplus\mathbb{R})] \rightarrow \mathbb{R}$ for two-dimensional microstretch fluid flow is given by
\begin{equation}
\ell(Y,\omega,R,\alpha,\lambda,j_r,j_s) = \frac{1}{2}\langle Y^\flat,Y \rangle + \frac{1}{2}\langle e^{j_r}\odot\omega,\omega \rangle + \frac{1}{2}\langle e^{j_s}\odot R,R \rangle - F_d(\alpha,\lambda), \label{dlagrangian_microstretch}
\end{equation}
where $F_d : \Omega_d^0(\mathbb{M},S^1\oplus\mathbb{R}) \rightarrow \mathbb{R}$ is a discrete approximation to the volume-integrated microstretch free energy.  The notation $\odot$ here refers to the Hadamard (entry-wise) product of vectors.  Likewise, the exponentials appearing here are entry-wise exponentials.

In analogy with the continuous definitions from before, make the following definitions for the variational derivatives of $\ell$:
\begin{align}
\pi = \frac{\delta\ell}{\delta\omega} &= e^{j_r}\odot\omega \\
Q = \frac{\delta\ell}{\delta R} &= e^{j_s}\odot R \\
i_r = \frac{\delta\ell}{\delta j_r} &= \frac{1}{2}e^{j_r}\odot\omega\odot\omega \\
i_s = \frac{\delta\ell}{\delta j_s} &= \frac{1}{2}e^{j_s} \odot R \odot R.
\end{align}

The discrete-space, continuous-time Euler-Poincar\'{e} equations~(\ref{EPcontinuous_right}-\ref{EPcontinuous2_right}) associated with the Lagrangian defined in equation~(\ref{dlagrangian_microstretch}) then read
{\allowdisplaybreaks
\begin{align}
\frac{\partial Y^\flat}{\partial t} + \pounds_Y Y^\flat &\,\hat{=}\, \mathrm{skew}(\pi\omega^T) + \mathrm{skew}(QR^T) + \mathrm{skew}(i_r j_r^T) \nonumber \\
&\:\:\:\:\:+ \mathrm{skew}(i_s j_s^T) + \mathrm{skew}\left(\frac{\delta\ell}{\delta\alpha} \alpha^T\right) + \mathrm{skew}\left(\frac{\delta\ell}{\delta\lambda} \lambda^T\right) \label{dEP1_microstretch} \\
\frac{\partial \pi}{\partial t} - Y\pi &= \frac{\delta\ell}{\delta\alpha} \label{dEP2_microstretch} \\
\frac{\partial Q}{\partial t} - YQ &= -2i_r - 2i_s + \frac{\delta\ell}{\delta\lambda} \label{dEP3_microstretch} \\
\frac{\partial \alpha}{\partial t} - Y\alpha &= \omega \label{dEP4_microstretch} \\
\frac{\partial \lambda}{\partial t} - Y\lambda &= R \label{dEP5_microstretch} \\
\frac{\partial j_r}{\partial t} - Yj_r &= -2R \label{dEP6_microstretch} \\
\frac{\partial j_s}{\partial t} - Yj_s &= -2R. \label{dEP7_microstretch}
\end{align}
}
Note the presence of the four extra terms in~(\ref{dEP1_microstretch}) relative to~(\ref{EP1_microstretch}); these terms only conspire to produce a full discrete differential in the continuous limit.

A discretization of the free energy~(\ref{free_energy_microstretch}) is given by
\begin{equation}
F_d(\alpha,\lambda) = \frac{1}{2} || \mathbf{d}\alpha ||_2^2 + \frac{1}{2} \langle \lambda, \lambda \rangle. \label{dfree_energy_microstretch}
\end{equation}
The variational derivatives $\frac{\delta\ell}{\delta\alpha}$ and $\frac{\delta\ell}{\delta\lambda}$ are then
\begin{align}
\frac{\delta\ell}{\delta\alpha} &= *\mathbf{d}*\mathbf{d}\alpha = \mathbf{\Delta}\alpha \\
\frac{\delta\ell}{\delta\lambda} &= -\lambda.
\end{align}

\paragraph{Temporal Discretization.}

A temporal discretization of~(\ref{dEP1_microstretch}-\ref{dEP7_microstretch}) is obtained most easily by choosing
\begin{equation}
\tau(A,\omega) = (\mathrm{cay}(A),\omega), \label{tau_microstretch}
\end{equation}
which is not a group difference map in the sense of Definition~(\ref{def:tau}) (since $\tau(A,\omega)^{-1} \neq \tau(-A,-\omega)$), but it nonetheless maps $\mathfrak{g}$ into $G$ and locally approximates the exponential.   While this choice provides a lower order approximation to $\exp$ than the genuine group difference map~(\ref{tau_nematic}), it has the appealing feature that its inverse right-trivialized tangent has a simpler form than that shown in~(\ref{dtau_inv_nematic}).

Note that in general, the use of a map $\tau$ with $\tau(\xi)^{-1} \neq \tau(-\xi)$ for $\xi \in \mathfrak{g}$ leads to the following generalization of the (right-right) discrete Euler-Poincar\'{e} equations with an advected parameter~(\ref{EP1_right}-\ref{EP3_right}):
\begin{align}
g_{k+1} &= \tau(h\xi_k)g_k \\
a_{k+1} &= a_k\tau(h\xi_k)^{-1} \label{EP2_right_badtau} \\
(D\tau^{-1}(\tau(h\xi_k)) \cdot TL_{\tau(h\xi_k)})^* \frac{\delta\ell}{\delta\xi_k} &= (D\tau^{-1}(\tau(h\xi_{k-1})) \cdot TR_{\tau(h\xi_{k-1})})^* \frac{\delta\ell}{\delta\xi_{k-1}} + h\frac{\delta\ell}{\delta a_k} \diamond a_k, \label{EP3_right_badtau}
\end{align}
which are easily derived through an argument that parallels the proof of Theorem~\ref{discreteHP}.

For the map~(\ref{tau_microstretch}), one computes that for $(A,\omega) \in \mathfrak{g}, (C^\flat,\pi) \in \mathfrak{g}^*$,
\begin{align}
(D\tau^{-1}(\tau(A,\omega))) \cdot TL_{\tau(A,\omega))})^* (C^\flat,\pi) &= \left((d\mathrm{cay}^{-1}_{-A})^* C^\flat - \mathrm{skew}(\pi\omega^T), \pi \right) \\
(D\tau^{-1}(\tau(A,\omega))) \cdot TR_{\tau(A,\omega))})^* (C^\flat,\pi) &= \left((d\mathrm{cay}^{-1}_{A})^* C^\flat, \mathrm{cay}(A)\pi \right)
\end{align}

Substituting into~(\ref{EP2_right_badtau}-\ref{EP3_right_badtau}) gives the discrete-space, discrete-time Euler-Poincar\'{e} equations for microstretch fluid flow:

\begin{align}
\frac{Y^\flat_k-Y^\flat_{k-1}}{h} + &\frac{\pounds_{Y_{k-1}}Y^\flat_{k-1}+\pounds_{Y_k}Y^\flat_k}{2} + \frac{h}{4}(Y_{k-1}Y_{k-1}^\flat\Omega Y_{k-1}\Omega^{-1} - Y_k Y^\flat_k \Omega Y_k \Omega^{-1}) \nonumber \\
\,\hat{=}\,&\; \mathrm{skew}(\pi_k\omega_k^T)
+\mathrm{skew}(Q_k R_k^T)
+\mathrm{skew}(i_{r,k} j_{r,k}^T)
+\mathrm{skew}(i_{s,k} j_{s,k}^T) \nonumber \\
&\; +\mathrm{skew}\left(\frac{\delta\ell}{\delta\alpha_k}\alpha_k^T\right)
+\mathrm{skew}\left(\frac{\delta\ell}{\delta\lambda_k}\lambda_k^T\right) \label{ddEP1_microstretch} \\
\pi_k =&\; \mathrm{cay}(hY_{k-1})\pi_{k-1} + h\frac{\delta\ell}{\delta\alpha_k} \label{ddEP2_microstretch} \\
Q_k =&\; \mathrm{cay}(hY_{k-1})Q_{k-1} - 2hi_{r,k} - 2hi_{s,k} + h\frac{\delta\ell}{\delta\lambda_k} \label{ddEP3_microstretch} \\
\alpha_k =&\; \mathrm{cay}(hY_{k-1})(\alpha_{k-1} + h\omega_{k-1}). \label{ddEP4_microstretch} \\
\lambda_k =&\; \mathrm{cay}(hY_{k-1})(\lambda_{k-1} + hR_{k-1}). \label{ddEP5_microstretch} \\
j_{r,k} =&\; \mathrm{cay}(hY_{k-1})(j_{r,k-1} - 2hR_{k-1}). \label{ddEP6_microstretch} \\
j_{s,k} =&\; \mathrm{cay}(hY_{k-1})(j_{s,k-1} - 2hR_{k-1}). \label{ddEP7_microstretch}
\end{align}

\paragraph{The Discrete Kelvin-Noether Theorem.}

The Kelvin-Noether theorem applied to microstretch fluid flow with the discrete free energy~(\ref{dfree_energy_microstretch}) and the boundary condition $d\alpha|_{\partial M}=0$ gives, as one corollary, the conservation law
\begin{equation}
\langle \pi_k, \mathbf{1} \rangle = \langle \pi_{k-1}, \mathbf{1} \rangle.
\end{equation}
That is, \emph{the total angular momentum due to micromotions is preserved exactly} along solutions to the discrete-space, discrete-time microstretch continua equations~(\ref{ddEP1_microstretch}-\ref{ddEP7_microstretch}).

\section{Update Equations on a Cartesian Grid} \label{section:cartesian}

We now specialize to the case in which the mesh $\mathbb{M}$ is a two-dimensional cartesian grid and derive the cartesian realizations of the discrete update equations~(\ref{ddEPfluid_rearranged}),~(\ref{ddEP_MHD1_rearranged}-\ref{ddEP_MHD2_rearranged}),~(\ref{ddEPnematic1}-\ref{ddEPnematic3}), and~(\ref{ddEP1_microstretch}-\ref{ddEP7_microstretch}) for ideal fluid flow, MHD, nematic liquid crystal flow, and microstretch fluid flow, respectively.  While the matrix forms of the update equations are readily implementable (on regular or irregular meshes), the reader will find that upon specializing to a cartesian grid and analyzing the matrix equations in an entry-wise fashion, easily implementable finite difference schemes emerge with a more familiar form.

In fact, a comparison with the literature will reveal that our MHD scheme, in particular, employs a \emph{spatial} discretization (although not its temporal discretization) that is identical to that of the MAC-Yee scheme on a two-dimensional cartesian grid~\cite{Liu2001}.  As a method that integrates two popular structure-preserving finite difference schemes -- the MAC scheme from fluid dynamics and the staggered mesh-based Yee scheme from electrodynamics -- the MAC-Yee scheme debuted in 2001 as a promising integrator for ideal incompressible magnetohydrodynamics that maintains the $\nabla \cdot \mathbf{B} = 0$ constraint exactly and respects energy and cross-helicity conservation in the following sense: the spatially discretized, continuous-time equations presented by Liu and Wang~\cite{Liu2001} conserve a discrete energy and a discrete cross-helicity.

Our MHD integrator surpasses these milestones in several key respects. Firstly, while our spatial discretization coincides with that of the MAC-Yee scheme, our temporal discretization differs markedly.  As a consequence of our careful choice of temporal discretization:
\begin{enumerate}
\item{Our integrator is variational, suggesting that our integrator should exhibit good energy behavior \emph{even after temporal discretization}, not merely in the context of the spatially discretized, continuous-time equations.}
\item{Through the discrete Kelvin-Noether theorem, our integrator preserves a discrete cross-helicity \emph{exactly} after temporal discretization.}
\end{enumerate}
Finally, our derivation of an MHD integrator was geometric, providing physical insight into the success of staggered grid schemes and generalizing the MAC-Yee scheme to unstructured meshes.

\paragraph{The Discrete Diffeomorphism Group on Cartesian Grids.}

In the case of a two-dimensional cartesian grid $\mathbb{M}$ with uniform spacing $\epsilon$, the diagonal matrix of cell volumes $\Omega$ coincides with a multiple of the identity matrix:
\begin{equation}
\Omega = \epsilon^2 I.
\end{equation}
Consequently, the group $\mathcal{D}(\mathbb{M})$  reduces to the group of orthogonal, signed stochastic matrices:
\begin{equation}
\mathcal{D}(\mathbb{M}) = \{q \in GL(N)^+ \,|\, \displaystyle\sum_j q_{ij} = 1 \;\; \forall i, \,  q^T q = I \}, \label{ddiff_cartesian}
\end{equation}
and its Lie algebra $\mathfrak{d}(\mathbb{M})$ reduces to the space of antisymmetric, row-null matrices:
\begin{equation}
\mathfrak{d}(\mathbb{M}) = \{A \in \mathfrak{gl}(N) \,|\, \displaystyle\sum_j A_{ij} = 0 \;\; \forall i, \,  A^T  +  A = 0 \}. \label{dM_cartesian}
\end{equation}

Matrices $A$ in the nonholonomically constrained space $\mathcal{S} \subset \mathfrak{d}(\mathbb{M})$ that approximate vector fields $\mathbf{v}=(u,v) \in \mathfrak{X}_{\mathrm{div}}(M)$, $M \subset \mathbb{R}^2$, have nonzero entries given by
\begin{align}
A_{mn} &\approx -\frac{1}{2\epsilon^2} \displaystyle\int_{D_{mn}} \mathbf{v} \cdot \hat{\mathbf{n}} \, dS \nonumber \\
&= \left\{\begin{array}{l}
-\frac{1}{2\epsilon}u(\mathbf{x}_{mn}) \text{ if cells $\mathcal{C}_m$ and $\mathcal{C}_n$ are horizontally adjacent} \\
-\frac{1}{2\epsilon}v(\mathbf{x}_{mn}) \text{ if cells $\mathcal{C}_m$ and $\mathcal{C}_n$ are vertically adjacent},
\end{array}\right.
 \label{flux_relation_cartesian}
\end{align}
where $D_{mn}$ denotes the edge shared by cells $\mathcal{C}_m$ and $\mathcal{C}_n$ and $\mathbf{x}_{mn} \in M$ is the position of its midpoint.

The constraints~(\ref{dM_cartesian}) appearing in the definition of $\mathfrak{d}(\mathbb{M})$ have the following implication for matrices $A \in \mathfrak{d}(\mathbb{M})$ approximating continuous vector fields: the fluxes of the velocity field across the edges bounding any given cell $\mathcal{C}_n$ must sum to zero.  This constitutes the discrete analogue of the constraint $\nabla \cdot \mathbf{v}$ on the velocity field in incompressible fluid flow, and we will record this constraint in every finite-difference scheme presented below for completeness.

We should emphasize, on the other hand, that the MHD cartesian update equations~(\ref{Bxupdate_MHD}-\ref{Byupdate_MHD}) (to be derived in this section) \emph{automatically preserve the divergence-freeness of the magnetic field} at the discrete level since the ambient matrix update equation~(\ref{ddEP_MHD2}) is an equation on the Lie algebra of antisymmetric, row-null matrices, a space that is closed under addition and commutation.  This is consistent with the fact that in the continuous world, $\nabla \cdot \mathbf{B}$ holds automatically if it holds at $t=0$, whereas $\nabla \cdot \mathbf{v} = 0$ can be viewed as a constraint that uniquely determines the pressure field $p$.

\paragraph{Nonholonomic Constraints and Cubic Terms.}

As discussed in Section~\ref{section:nonholonomic}, we have imposed nonholonomic constraints throughout Section~\ref{section:continuum_discretization} to guarantee the desired sparsity of the matrices used to represent vector fields and one-forms.  When dealing with the resulting weak equalities that appear in~(\ref{ddEPfluid_rearranged}),~(\ref{ddEP_MHD1_rearranged}-\ref{ddEP_MHD2_rearranged}),~(\ref{ddEPnematic1}), and~(\ref{ddEP1_microstretch}), the following two procedures will be employed, in accordance with the remarks in Section~\ref{section:nonholonomic}:
\begin{enumerate}
\item{When working with Lie derivatives $\pounds_A C^\flat$ of discrete one-forms $C^\flat \in \Omega_d^1(\mathbb{M})/d\Omega_d^0(\mathbb{M})$, we concern ourselves only with those components $(\pounds_A C^\flat)_{mn}$ of $\pounds_A C^\flat$ corresponding to adjacent cells $\mathcal{C}_m$ and $\mathcal{C}_n$.}
\item{When working with discrete vector fields in $\mathfrak{d}(\mathbb{M})$, we apply the sparsity operator~(\ref{cartesian_sparsity}) to any commutators that arise.}
\end{enumerate}

Lastly, for simplicity, we shall drop any cubic terms (matrix products involving three elements of $\mathfrak{d}(\mathbb{M}) \cup \Omega_d^1(\mathbb{M})/d\Omega_d^0(\mathbb{M})$) appearing in the update equations~(\ref{ddEPfluid_rearranged}),~(\ref{ddEP_MHD1_rearranged}-\ref{ddEP_MHD2_rearranged}),~(\ref{ddEPnematic1}-\ref{ddEPnematic3}), and~(\ref{ddEP1_microstretch}-\ref{ddEP7_microstretch}).  This practice is advocated in the context of rigid body simulations by Kobilarov and co-authors~\cite{Kobilarov2009}, who point out that the dropping of such terms does not reduce the second-order of accuracy of the discrete Euler-Poincar\'{e} equations arising from the Cayley group difference map.  Note also that \emph{the discrete Kelvin-Noether theorem still holds exactly} if $d\mathrm{cay}_Y^{-1}(Z)$ and its dual $(d\mathrm{cay}_Y^{-1})^*X^\flat$ are replaced by
\begin{equation}
\left(I+\frac{1}{2}\pounds_Y\right)Z
\end{equation}
and
\begin{equation}
\left(I-\frac{1}{2}\pounds_Y\right)X^\flat,
\end{equation}
respectively -- which is equivalent to dropping the cubic terms.  (Compare with~(\ref{dcayinv}) and~(\ref{dcayinvstar}).)

\paragraph{Notation.}  In what follows, we parameterize the vertices of our coordinate grid with integer pairs $(i,j) \in \mathbb{Z} \times \mathbb{Z}$.  Spatial locations of quantities will be encoded with superscript notation.  Thus, vertex-based quantities are denoted $q^{i,j}$, cell-centered quantities are denoted $q^{i+1/2,j+1/2}$, and edge-based quantities are denoted $q^{i+1/2,j}$ and $q^{i,j+1/2}$.  As shorthand for averages between pairs of adjacent cell-based quantities, we shall use the following intuitive notation:
\begin{equation}
q^{i+1/2,j} := \frac{1}{2}(q^{i+1/2,j-1/2}+q^{i+1/2,j+1/2}),
\end{equation}
and similarly for averages between pairs of adjacent edge-based quantities and pairs of adjacent vertex-based quantities.
In some instances, products of compatible quantities will be combined if there is no danger of ambiguity, e.g.
\begin{equation}
(q r)^{i+1/2,j} := q^{i+1/2,j}r^{i+1/2,j}
\end{equation}
In a slight abuse of notation, subscripts will be used for both matrix/vector indexing and temporal indexing.  While context should alleviate most ambiguities, we shall nonetheless reserve the subscript $k$ for temporal indexing; any other subscripted letter shall indicate componentwise indexing of matrices and vectors.

Finally, the discrete curls of vector fields $\mathbf{u}$ of the form $\mathbf{u}=(u,v)$ shall play an important role; these vertex-based quantities are defined as follows:
\begin{equation}
\omega^{i,j}(u,v) = \frac{ u^{i,j-1/2} + v^{i+1/2,j} - u^{i,j+1/2} - v^{i-1/2,j} }{\epsilon}.
\end{equation}

\begin{figure}[t]
\centering
\begin{tabular}{cc} \hspace{0in}
\vcent{\includegraphics[scale=0.3]{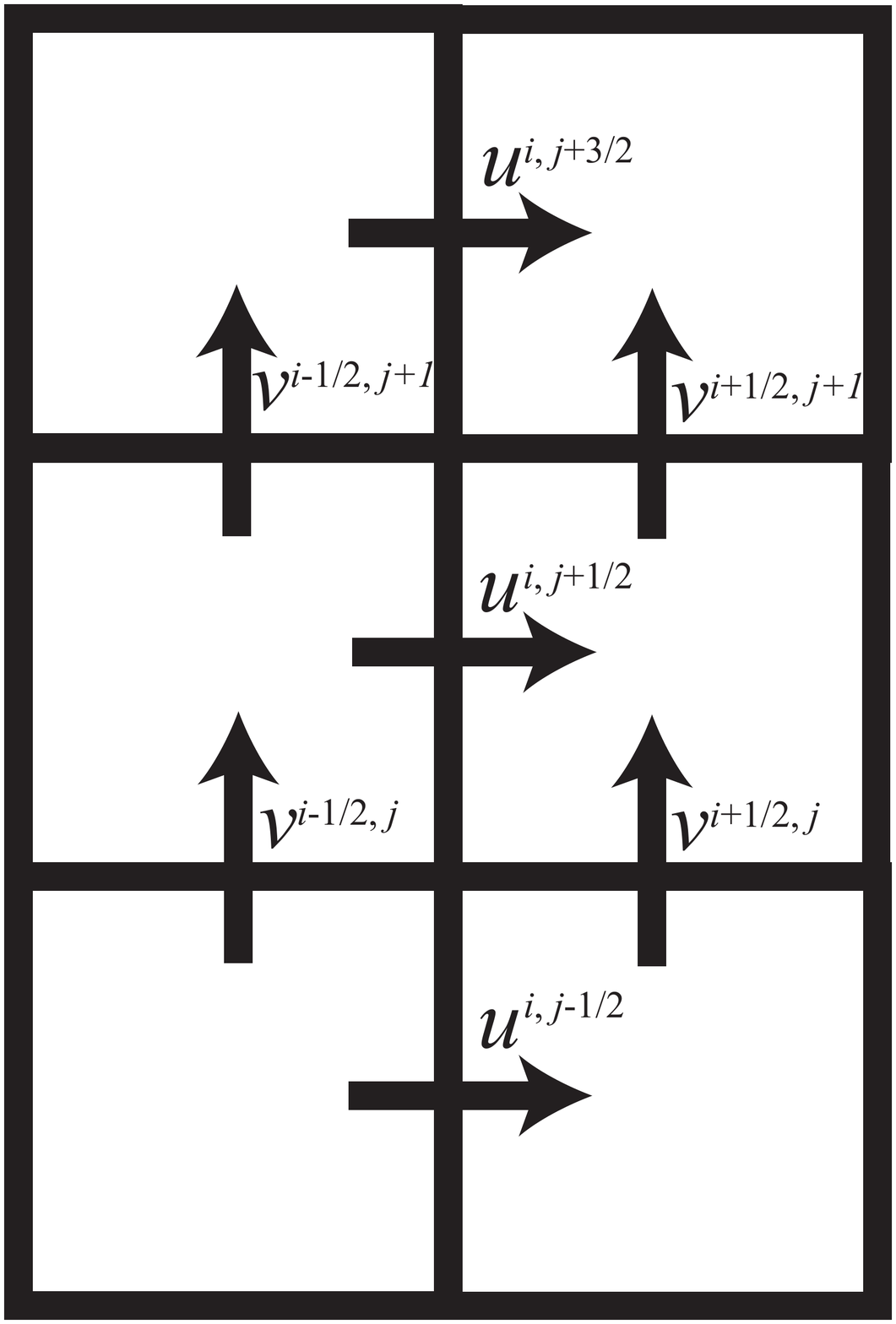}} & \hspace{0in}
\vcent{\includegraphics[scale=0.3]{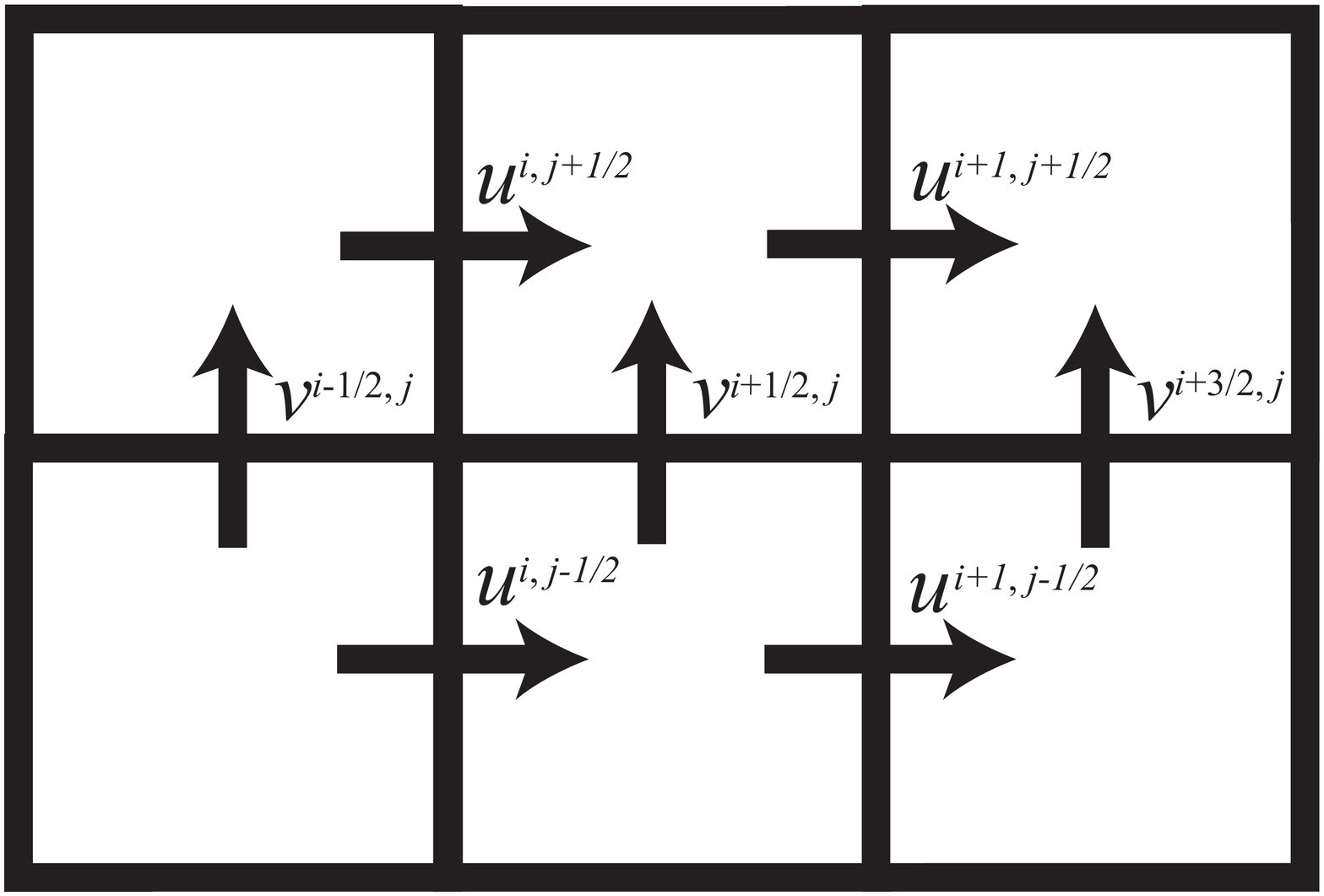}} \\
(a) &\hspace{0in} (b) \\
\end{tabular}
\caption{(a) Stencil for computations involving horizontally adjacent cells. (b) Stencil for computations involving vertically adjacent cells.}
\label{fig:sixcells}
\end{figure}

Fig.~\ref{fig:sixcells} provides a useful reference for the subsequent discussion.

\subsection{Frequently Encountered Quantities and their Cartesian Realizations}

Below we summarize the cartesian realizations of some frequently encountered quantities arising in discretizations of flows on the diffeomorphism group.  Only results are stated here; derivations of these results are presented in Appendix~\ref{appendix:cartesian}.

\subsubsection{The Lie Derivative of a Discrete One-Form} \label{section:Lie1form}

We will begin by considering the cartesian realization of quantities of the form $\pounds_A C^\flat$ with $A,C \in \mathcal{S} \subset \mathfrak{d}(\mathbb{M})$.  This quantity has, not surprisingly, appeared in every continuum theory discretization presented in this report. It is the basic object governing flows on the discrete diffeomorphism group, serving as the discrete counterpart to continuous quantities of the form $\pounds_\mathbf{u} \mathbf{w}^\flat$ with $\mathbf{u},\mathbf{w} \in \mathfrak{X}_{\mathrm{div}}(M)$.

Assume that the discrete vector fields $A,C \in \mathcal{S}$ approximate continuous vector fields $\mathbf{u}=(u,v),\mathbf{w}=(r,s) \in \mathfrak{X}_{\mathrm{div}}(M)$, respectively, so that the entries $A_{mn}$ and $C_{mn}$ for a pair of horizontally adjacent cells $\mathcal{C}_m$ and $\mathcal{C}_n$ centered at $(i-1/2,j+1/2)$ and $(i+1/2,j+1/2)$ are, in accordance with~(\ref{flux_relation}), given by
\begin{align}
A_{mn} &= -\frac{1}{2\epsilon} u^{i,j+1/2} \label{flux1} \\
C_{mn} &= -\frac{1}{2\epsilon} r^{i,j+1/2}, \label{flux2}
\end{align}
Likewise, the entries $A_{mn}$ and $C_{mn}$ for a pair of vertically adjacent cells $\mathcal{C}_m$ and $\mathcal{C}_n$ centered at $(i+1/2,j-1/2)$ and $(i+1/2,j+1/2)$ are given by
\begin{align}
A_{mn} &= -\frac{1}{2\epsilon} v^{i+1/2,j} \label{flux3} \\
C_{mn} &= -\frac{1}{2\epsilon} s^{i+1/2,j}. \label{flux4}
\end{align}

Fix a pair of vertically adjacent cells $\mathcal{C}_m$ and $\mathcal{C}_n$.  Then it can be shown (see Appendix~\ref{appendix:cartesian}) that, modulo a discrete differential,
\begin{equation}
(\pounds_A C^\flat)_{mn} =
-\frac{\epsilon}{2}\left(\omega^{i,j}(r,s)\left(\frac{u^{i,j-1/2}+u^{i,j+1/2}}{2}\right) + \omega^{i+1,j}(r,s)\left(\frac{u^{i+1,j-1/2}+u^{i+1,j+1/2}}{2}\right)  \right). \label{LACflat_y}
\end{equation}

Likewise, for a pair of horizontally adjacent cells $\mathcal{C}_m$ and $\mathcal{C}_n$ centered at $(i-1/2,j+1/2)$ and $(i+1/2,j+1/2)$, we have
\begin{equation}
(\pounds_A C^\flat)_{mn} =
\frac{\epsilon}{2}\left(\omega^{i,j}(r,s)\left(\frac{v^{i-1/2,j}+v^{i+1/2,j}}{2}\right) + \omega^{i,j+1}(r,s)\left(\frac{v^{i-1/2,j+1}+v^{i+1/2,j+1}}{2}\right)  \right). \label{LACflat_x}
\end{equation}

Notice that we have recovered, at the discrete level, the two-dimensional manifestation of the vector calculus identity
\begin{equation}
\pounds_\mathbf{u} \mathbf{w}^\flat = ((\nabla \times \mathbf{w}) \times \mathbf{u})^\flat,
\end{equation}
which holds for vector fields $\mathbf{u},\mathbf{w}$ defined on $\mathbb{R}^3$, modulo a full differential.

\subsubsection{The Lie Derivative of a Discrete Vector Field} \label{section:Lie_vector_field}

In this section, we consider the cartesian realization of quantities of the form $\pounds_A B$ with $A,B \in \mathcal{S} \subset \mathfrak{d}(\mathbb{M})$.  Quantities of this form arise in the discrete Euler-Poincar\'{e} equations~(\ref{dEP1_MHD}-\ref{dEP2_MHD}) for magnetohydrodynamics; they serve as the discrete counterparts to continuous quantities of the form $\pounds_\mathbf{u} \mathbf{v}$ with $\mathbf{u},\mathbf{v} \in \mathfrak{X}_{\mathrm{div}}(M)$.

Assume that the discrete vector fields $A,B \in \mathcal{S}$ approximate continuous vector fields $\mathbf{u}=(u,v),\mathbf{v}=(p,q) \in \mathfrak{X}_{\mathrm{div}}(M)$, respectively, so that the entries $A_{mn}$ and $B_{mn}$ for a pair of horizontally adjacent cells $\mathcal{C}_m$ and $\mathcal{C}_n$ centered at $(i-1/2,j+1/2)$ and $(i+1/2,j+1/2)$ are given by
\begin{align}
A_{mn} &= -\frac{1}{2\epsilon} u^{i,j+1/2} \\
B_{mn} &= -\frac{1}{2\epsilon} p^{i,j+1/2},
\end{align}
Likewise, the entries $A_{mn}$ and $B_{mn}$ for a pair of vertically adjacent cells $\mathcal{C}_m$ and $\mathcal{C}_n$ centered at $(i+1/2,j-1/2)$ and $(i+1/2,j+1/2)$ are given by
\begin{align}
A_{mn} &= -\frac{1}{2\epsilon} v^{i+1/2,j} \\
B_{mn} &= -\frac{1}{2\epsilon} q^{i+1/2,j}.
\end{align}

Fix a pair of vertically adjacent cells $\mathcal{C}_m$ and $\mathcal{C}_n$.  Then, in the abbreviated notation discussed in this section's introduction, it can be shown (see Appendix~\ref{appendix:cartesian}) that
\begin{equation}
(\pounds_A B)^\downarrow_{mn} = -\frac{1}{2\epsilon} \left(
 \frac{u^{i+1,j}q^{i+1,j} - u^{i,j}q^{i,j}}{\epsilon}
-\frac{p^{i+1,j}v^{i+1,j} - p^{i,j}v^{i,j}}{\epsilon} \right). \label{LAB_y}
\end{equation}

Similarly, for a pair of horizontally adjacent cells $\mathcal{C}_m$ and $\mathcal{C}_n$ centered at $(i-1/2,j+1/2)$ and $(i+1/2,j+1/2)$, we have
\begin{equation}
(\pounds_A B)^\downarrow_{mn} = -\frac{1}{2\epsilon} \left(
 -\frac{u^{i,j+1}q^{i,j+1} - u^{i,j}q^{i,j}}{\epsilon}
+\frac{p^{i,j+1}v^{i,j+1} - p^{i,j}v^{i,j}}{\epsilon} \right). \label{LAB_x}
\end{equation}

Notice that here we have recovered, at the discrete level, the two-dimensional manifestation of the vector calculus identity
\begin{equation}
\pounds_\mathbf{u} \mathbf{v} = \nabla \times (\mathbf{v} \times \mathbf{u}),
\end{equation}
which holds for divergence-free vector fields $\mathbf{u},\mathbf{v}$ defined on $\mathbb{R}^3$.

\subsubsection{The Antisymmetrization of an Outer Product} \label{section:antisymmetrization}

Below we consider the cartesian realization of quantities of the form $\mathrm{skew}(F E^T) \in \Omega_d^1(\mathbb{M})/d\Omega_d^0(\mathbb{M})$, where $E$ and $F$ are discrete zero-forms.  As the reader of Sections~\ref{section:nematic}-\ref{section:microstretch} is well aware, such quantities arise profusely in discretizations of scalar field advection by way of the diamond operation.  See for instance, the Euler-Poincar\'{e} equations~(\ref{dEP1_nematic}-\ref{dEP3_nematic}) for two-dimensional nematic liquid crystal flow and the Euler-Poincar\'{e} equations~(\ref{dEP1_microstretch}-\ref{dEP7_microstretch}) for two-dimensional microstretch fluid flow.

Assume that the discrete zero-forms $E,F \in \Omega_d^0(\mathbb{M})$ approximate continuous scalar fields $\alpha,\beta \in \mathcal{F}(M)$, respectively, so that the entries $E_n$ and $F_n$ for a cell $\mathcal{C}_n$ centered at $(i+1/2,j+1/2)$ are given by
\begin{align}
E_n &= \alpha^{i+1/2,j+1/2} \\
F_n &= \beta^{i+1/2,j+1/2}.
\end{align}
Recall the correspondence between $\mathrm{skew}(F E^T)$ and its continuous counterpart:
\renewcommand\arraystretch{1.5}
\begin{equation}
\begin{array}[h]{cc}
			\underline{\mathrm{Continuous}} & \underline{\mathrm{Discrete}} \\ \vspace{-0.15in}
			[-\beta d\alpha] & [\mathrm{skew}(FE^T)] \\
			\rotatebox{-90}{$\in$} & \rotatebox{-90}{$\in$} \\
			\Omega^1(M)/d\Omega^0(M) & \Omega_d^1(\mathbb{M})/d\Omega_d^0(\mathbb{M})
	\label{skew_correspondence}
\end{array}
\end{equation}
Here, we have explicitly included the brackets to emphasize the distinction between cosets and their representives.

Computing the $(m,n)$ entry of $\mathrm{skew}(F E^T)$ for vertically adjacent cells $\mathcal{C}_m$ and $\mathcal{C}_n$ centered at $(i+1/2,j-1/2)$ and $(i+1/2,j+1/2)$, respectively, one finds (see Appendix~\ref{appendix:cartesian}) that
\begin{equation}
\mathrm{skew}(F E^T)_{mn}
= -\frac{\epsilon}{2}\left(\alpha^{i+1/2,j}\left(\frac{\beta^{i+1/2,j+1/2}-\beta^{i+1/2,j-1/2}}{\epsilon}\right) - \beta^{i+1/2,j}\left(\frac{\alpha^{i+1/2,j+1/2}-\alpha^{i+1/2,j-1/2}}{\epsilon}\right)\right). \label{skewEF_y}
\end{equation}

Similarly, for a pair of horizontally adjacent cells $\mathcal{C}_m$ and $\mathcal{C}_n$ centered at $(i-1/2,j+1/2)$ and $(i+1/2,j+1/2)$, one obtains
\begin{equation}
\mathrm{skew}(F E^T)_{mn} = -\frac{\epsilon}{2}\left(\alpha^{i,j+1/2}\left(\frac{\beta^{i+1/2,j+1/2}-\beta^{i-1/2,j+1/2}}{\epsilon}\right) - \beta^{i,j+1/2}\left(\frac{\alpha^{i+1/2,j+1/2}-\alpha^{i-1/2,j+1/2}}{\epsilon}\right)\right). \label{skewEF_x}
\end{equation}

Evidently, the matrix $\mathrm{skew}(F E^T)$ encodes the continuous one-form $\frac{1}{2}(\alpha d\beta - \beta d\alpha)$.  This agrees with~(\ref{skew_correspondence}), up to the choice of representative for the coset $[-\beta d\alpha]$.

\subsection{Summary of the Cartesian Update Equations}

Having identified the cartesian realizations of the three most frequently encountered quantities appearing in the Section~\ref{section:applications}'s discretizations, we now record the cartesian update equations for ideal fluid flow, MHD, liquid crystal flow, and microstretch fluid flow in their most compact form, suitable for numerical implementation.

In anticipation of the ubiquity of terms of the form~(\ref{LACflat_y}-\ref{LACflat_x}) with $(u,v)=(r,s)$, define the following two operators, which take a cartesian (edge-based) discrete vector field $(u,v)$ to a cartesian (edge-based) discrete vector field $(\Psi_x,\Psi_y)$:
\begin{align}
\Psi_x^{i,j+1/2}(u,v) &= -\frac{1}{2}\left( \omega^{i,j}(u,v)\left(\frac{v^{i-1/2,j}+v^{i+1/2,j}}{2}\right) + \omega^{i,j+1}(u,v)\left(\frac{v^{i-1/2,j+1}+v^{i+1/2,j+1}}{2}\right) \right) \\
\Psi_y^{i+1/2,j}(u,v) &= \frac{1}{2}\left( \omega^{i,j}(u,v)\left(\frac{u^{i,j-1/2}+u^{i,j+1/2}}{2}\right) + \omega^{i+1,j}(u,v)\left(\frac{u^{i+1,j-1/2}+u^{i+1,j+1/2}}{2}\right) \right).
\end{align}

Likewise, for cartesian (edge-based) discrete vector fields $(u,v)$ and $(p,q)$, define
\begin{align}
\Phi_x^{i,j+1/2}(u,v,p,q) &= -\frac{u^{i,j+1}q^{i,j+1} - u^{i,j}q^{i,j}}{\epsilon}
+\frac{p^{i,j+1}v^{i,j+1} - p^{i,j}v^{i,j}}{\epsilon} \\
\Phi_y^{i+1/2,j}(u,v,p,q) &= \frac{u^{i+1,j}q^{i+1,j} - u^{i,j}q^{i,j}}{\epsilon}
-\frac{p^{i+1,j}v^{i+1,j} - p^{i,j}v^{i,j}}{\epsilon}.
\end{align}
Note that this definition employs the abbreviated notation for averages introduced in this section's introduction by writing, for instance, $u^{i,j}$ as shorthand for $\frac{1}{2}(u^{i,j-1/2}+u^{i,j+1/2})$.

In addition, for cartesian (cell-based) discrete zero-forms $\alpha$ and $\beta$, define the following cartesian (edge-based) discrete vector field $(\Lambda_x,\Lambda_y)$:
\begin{align}
\Lambda_x^{i,j+1/2}(\beta,\alpha) &= \frac{1}{2}\left(\alpha^{i,j+1/2}\left(\frac{\beta^{i+1/2,j+1/2}-\beta^{i-1/2,j+1/2}}{\epsilon}\right) - \beta^{i,j+1/2}\left(\frac{\alpha^{i+1/2,j+1/2}-\alpha^{i-1/2,j+1/2}}{\epsilon}\right)\right)  \\
\Lambda_y^{i+1/2,j}(\beta,\alpha) &= \frac{1}{2}\left(\alpha^{i+1/2,j}\left(\frac{\beta^{i+1/2,j+1/2}-\beta^{i+1/2,j-1/2}}{\epsilon}\right) - \beta^{i+1/2,j}\left(\frac{\alpha^{i+1/2,j+1/2}-\alpha^{i+1/2,j-1/2}}{\epsilon}\right)\right),
\end{align}
where once again we have employed abbreviated notation for pairwise averages between cells.

Lastly, let $\mathbf{\Delta}(\alpha)$ denote the cartesian (cell-based) discrete Laplacian of a discrete zero-form $\alpha$:
\begin{equation}
\mathbf{\Delta}^{i+1/2,j+1/2}(\alpha) = \frac{\alpha^{i-1/2,j+1/2}+\alpha^{i+1/2,j-1/2}+\alpha^{i+3/2,j+1/2}+\alpha^{i+1/2,j+3/2}-4\alpha^{i+1/2,j+1/2}}{\epsilon^2}.
\end{equation}

In terms of these operators, the update equations for ideal fluid flow, MHD, liquid crystal flow, and microstretch fluid flow on a two-dimensional cartesian grid with spacing $\epsilon$ and time-step $h$ are now reviewed.

\subsubsection{Ideal, Incompressible Fluid Flow}

Variables:
\begin{center}
		\begin{tabular}{c|c|c|c}
		  Variable & Meaning & Residence & Indexing \\ \hline
			$u$ & Velocity field, horizontal component & Vertical edges & $u^{i,j+1/2}$ \\
			$v$ & Velocity field, vertical component & Horizontal edges & $v^{i+1/2,j}$ \\
			$p$ & Pressure & Cell centers & $p^{i+1/2,j+1/2}$
		\end{tabular}
\end{center}

\medskip

\noindent Update equations:
\begin{align}
&\bullet \frac{u^{i,j+1/2}_k-u^{i,j+1/2}_{k-1}}{h} + \frac{\Psi_x^{i,j+1/2}(u_{k-1},v_{k-1})+\Psi_x^{i,j+1/2}(u_k,v_k)}{2}  = - \frac{p_k^{i+1/2,j+1/2} - p_k^{i-1/2,j+1/2}}{\epsilon} \label{uupdate_fluid} \\
&\bullet \frac{v^{i+1/2,j}_k-v^{i+1/2,j}_{k-1}}{h} + \frac{\Psi_y(u_{k-1}^{i+1/2,j},v_{k-1})+\Psi_y^{i+1/2,j}(u_k,v_k)}{2}  = - \frac{p_k^{i+1/2,j+1/2} - p_k^{i+1/2,j-1/2}}{\epsilon} \label{vupdate_fluid} \\
&\bullet \frac{u_k^{i+1,j+1/2}-u_k^{i,j+1/2}}{\epsilon} + \frac{v_k^{i+1/2,j+1}-v_k^{i+1/2,j}}{\epsilon} = 0
\end{align}

\subsubsection{Ideal, Incompressible Magnetohydrodynamics}

Variables:
\begin{center}
		\begin{tabular}{c|c|c|c}
		  Variable & Meaning & Residence & Indexing \\ \hline
			$u$ & Velocity field, horizontal component & Vertical edges & $u^{i,j+1/2}$ \\
			$v$ & Velocity field, vertical component & Horizontal edges & $v^{i+1/2,j}$ \\
			$p$ & Pressure & Cell centers & $p^{i+1/2,j+1/2}$ \\
			$B_x$ & Magnetic field, horizontal component & Vertical edges & $B_x^{i,j+1/2}$ \\
			$B_y$ & Magnetic field, vertical component & Horizontal edges & $B_y^{i+1/2,j}$
		\end{tabular}
\end{center}

\medskip

\noindent Update equations:
\begin{align}
&\bullet \frac{u^{i,j+1/2}_k-u^{i,j+1/2}_{k-1}}{h} + \frac{\Psi_x^{i,j+1/2}(u_{k-1},v_{k-1})+\Psi_x^{i,j+1/2}(u_k,v_k)}{2} \nonumber \\
&\hspace{0.5in}=
  \frac{\Psi_x^{i,j+1/2}(B_{x,k-1},B_{y,k-1})+\Psi_x^{i,j+1/2}(B_{x,k},B_{y,k})}{2}
- \frac{p_k^{i+1/2,j+1/2} - p_k^{i-1/2,j+1/2}}{\epsilon} \label{uupdate_MHD} \\
&\bullet \frac{v^{i+1/2,j}_k-v^{i+1/2,j}_{k-1}}{h} + \frac{\Psi_y^{i+1/2,j}(u_{k-1},v_{k-1})+\Psi_y^{i+1/2,j}(u_k,v_k)}{2} \nonumber \\
&\hspace{0.5in}=
  \frac{\Psi_y^{i+1/2,j}(B_{x,k-1},B_{y,k-1})+\Psi_y^{i+1/2,j}(B_{x,k},B_{y,k})}{2}
- \frac{p_k^{i+1/2,j+1/2} - p_k^{i+1/2,j-1/2}}{\epsilon} \label{vupdate_MHD} \\
&\bullet \frac{B^{i,j+1/2}_{x,k}-B^{i,j+1/2}_{x,k-1}}{h} + \Phi_x^{i,j+1/2}\left(u_{k-1},v_{k-1},\frac{B_{x,k-1}+B_{x,k}}{2},\frac{B_{y,k-1}+B_{y,k}}{2}\right) = 0 \label{Bxupdate_MHD} \\
&\bullet \frac{B^{i+1/2,j}_{y,k}-B^{i+1/2,j}_{y,k-1}}{h} + \Phi_y^{i+1/2,j}\left(u_{k-1},v_{k-1},\frac{B_{x,k-1}+B_{x,k}}{2},\frac{B_{y,k-1}+B_{y,k}}{2}\right) = 0 \label{Byupdate_MHD} \\
&\bullet \frac{u_k^{i+1,j+1/2}-u_k^{i,j+1/2}}{\epsilon} + \frac{v_k^{i+1/2,j+1}-v_k^{i+1/2,j}}{\epsilon} = 0 \label{uvconstraint_MHD}
\end{align}

\subsubsection{Nematic Liquid Crystals}

Variables:
\begin{center}
		\begin{tabular}{c|c|c|c}
		  Variable & Meaning & Residence & Indexing \\ \hline
			$u$ & Velocity field, horizontal component & Vertical edges & $u^{i,j+1/2}$ \\
			$v$ & Velocity field, vertical component & Horizontal edges & $v^{i+1/2,j}$ \\
			$p$ & Pressure & Cell centers & $p^{i+1/2,j+1/2}$ \\
			$\omega$ & Local angular velocity field & Cell centers & $\omega^{i+1/2,j+1/2}$ \\
			$\alpha$ & Director orientation & Cell centers & $\alpha^{i+1/2,j+1/2}$
		\end{tabular}
\end{center}

\medskip

\noindent Update equations:
\begin{align}
&\bullet \frac{u^{i,j+1/2}_k-u^{i,j+1/2}_{k-1}}{h} + \frac{\Psi_x^{i,j+1/2}(u_{k-1},v_{k-1})+\Psi_x^{i,j+1/2}(u_k,v_k)}{2} \nonumber \\
&\hspace{0.5in}=
  \Lambda_x^{i,j+1/2}(\mathbf{\Delta}\alpha_k,\alpha_k)
- \frac{p_k^{i+1/2,j+1/2} - p_k^{i-1/2,j+1/2}}{\epsilon} \label{uupdate_nematic} \\
&\bullet \frac{v^{i+1/2,j}_k-v^{i+1/2,j}_{k-1}}{h} + \frac{\Psi_y^{i+1/2,j}(u_{k-1},v_{k-1})+\Psi_y^{i+1/2,j}(u_k,v_k)}{2} \nonumber \\
&\hspace{0.5in}=
  \Lambda_y^{i+1/2,j}(\mathbf{\Delta}\alpha_k,\alpha_k)
- \frac{p_k^{i+1/2,j+1/2} - p_k^{i+1/2,j-1/2}}{\epsilon} \label{vupdate_nematic} \\
&\bullet \mathrm{Equations~(\ref{ddEPnematic2}-\ref{ddEPnematic3})}\mathrm{\footnotemark[1]} \\
&\bullet \frac{u_k^{i+1,j+1/2}-u_k^{i,j+1/2}}{\epsilon} + \frac{v_k^{i+1/2,j+1}-v_k^{i+1/2,j}}{\epsilon} = 0 \label{uvconstraint_nematic}
\end{align}
\footnotetext[1]{We do not recast these equations in cartesian form, as their structure does not admit such a representation.  (No rearrangement of the equations can eliminate the appearance of a matrix inverse.)  This is of no concern, as they are already in a form most suitable for efficient implementation; indeed, they are explicit updates.}

\subsubsection{Microstretch Fluids}

Variables:
\begin{center}
		\begin{tabular}{c|c|c|c}
		  Variable & Meaning & Residence & Indexing \\ \hline
			$u$ & Velocity field, horizontal component & Vertical edges & $u^{i,j+1/2}$ \\
			$v$ & Velocity field, vertical component & Horizontal edges & $v^{i+1/2,j}$ \\
			$p$ & Pressure & Cell centers & $p^{i+1/2,j+1/2}$ \\
			$\omega$ & Local angular velocity field & Cell centers & $\omega^{i+1/2,j+1/2}$ \\
			$R$ & Local stretch rate & Cell centers & $R^{i+1/2,j+1/2}$ \\
			$\alpha$ & Director orientation & Cell centers & $\alpha^{i+1/2,j+1/2}$ \\
			$\lambda$ & Logarithm of director length & Cell centers & $\lambda^{i+1/2,j+1/2}$ \\
			$j_r$ & Local rotational inertia & Cell centers & $j_r^{i+1/2,j+1/2}$ \\
			$j_s$ & Local stretch inertia & Cell centers & $j_s^{i+1/2,j+1/2}$ \\
			$\pi$ & $\pi = \frac{\delta\ell}{\delta \omega} = e^{j_r}\odot\omega$ & Cell centers & $\pi^{i+1/2,j+1/2}$ \\
			$Q$   & $Q = \frac{\delta\ell}{\delta R} = e^{j_s}\odot R$ & Cell centers & $Q^{i+1/2,j+1/2}$ \\
			$i_r$ & $i_r = \frac{\delta\ell}{\delta j_r} = \frac{1}{2}e^{j_r}\odot\omega\odot\omega$ & Cell centers & $i_r^{i+1/2,j+1/2}$ \\
			$i_s$ & $i_s = \frac{\delta\ell}{\delta j_s} = \frac{1}{2}e^{j_s}\odot R \odot R$ & Cell centers & $i_s^{i+1/2,j+1/2}$
		\end{tabular}
\end{center}

\medskip

\noindent Update equations:
\begin{align}
&\bullet \frac{u^{i,j+1/2}_k-u^{i,j+1/2}_{k-1}}{h} + \frac{\Psi_x^{i,j+1/2}(u_{k-1},v_{k-1})+\Psi_x^{i,j+1/2}(u_k,v_k)}{2} \nonumber \\
&\hspace{0.5in}=
  \Lambda_x^{i,j+1/2}(\pi_k,\omega_k)
+ \Lambda_x^{i,j+1/2}(Q_k,R_k)
+ \Lambda_x^{i,j+1/2}(i_{r,k},j_{r,k})
+ \Lambda_x^{i,j+1/2}(i_{s,k},j_{s,k}) \nonumber \\
&\hspace{0.6in}
+ \Lambda_x^{i,j+1/2}(\mathbf{\Delta}\alpha_k,\alpha_k)
- \frac{p_k^{i+1/2,j+1/2} - p_k^{i-1/2,j+1/2}}{\epsilon} \label{uupdate_microstretch} \\
&\bullet \frac{v^{i+1/2,j}_k-v^{i+1/2,j}_{k-1}}{h} + \frac{\Psi_y^{i+1/2,j}(u_{k-1},v_{k-1})+\Psi_y^{i+1/2,j}(u_k,v_k)}{2} \nonumber \\
&\hspace{0.5in}=
  \Lambda_y^{i+1/2,j}(\pi_k,\omega_k)
+ \Lambda_y^{i+1/2,j}(Q_k,R_k)
+ \Lambda_y^{i+1/2,j}(i_{r,k},j_{r,k})
+ \Lambda_y^{i+1/2,j}(i_{s,k},j_{s,k}) \nonumber \\
&\hspace{0.6in}
+ \Lambda_y^{i+1/2,j}(\mathbf{\Delta}\alpha_k,\alpha_k)
- \frac{p_k^{i+1/2,j+1/2} - p_k^{i+1/2,j-1/2}}{\epsilon} \label{vupdate_microstretch} \\
&\bullet \mathrm{Equations~(\ref{ddEP2_microstretch}-\ref{ddEP7_microstretch})}\mathrm{\footnotemark[2]}  \\
&\bullet \frac{u_k^{i+1,j+1/2}-u_k^{i,j+1/2}}{\epsilon} + \frac{v_k^{i+1/2,j+1}-v_k^{i+1/2,j}}{\epsilon} = 0 \label{uvconstraint_microstretch}
\end{align}
\footnotetext[2]{Although these equations do in fact admit cartesian representations, we do not modify them for they are already in a form most suitable for efficient implementation as explicit updates.}

\section{Numerical Results} \label{section:numerical}

To corroborate our theoretical results, we have implemented the variational update equations~(\ref{uupdate_MHD}-\ref{uvconstraint_MHD}),~(\ref{uupdate_nematic}-\ref{uvconstraint_nematic}),~(\ref{uupdate_microstretch}-\ref{uvconstraint_microstretch}) for MHD, nematic liquid crystal flow, and microstretch continua, respectively, on a 2-dimensional cartesian grid for a variety of test cases typically used in the literature. Owing to the prevalence of well-established numerical test cases for MHD relative to complex fluid flow, we focus primarily on comparing our novel MHD integrator with existing MHD integrators, and we give only a few examples of our complex fluid flow integrators for proof of concept.

\subsection{MHD Test Cases}

\subsubsection{Fluid Vortex in an Azimuthal Magnetic Field}

\begin{figure}[t]
\centering
\begin{tabular}{cc}
\includegraphics[width=0.5\textwidth]{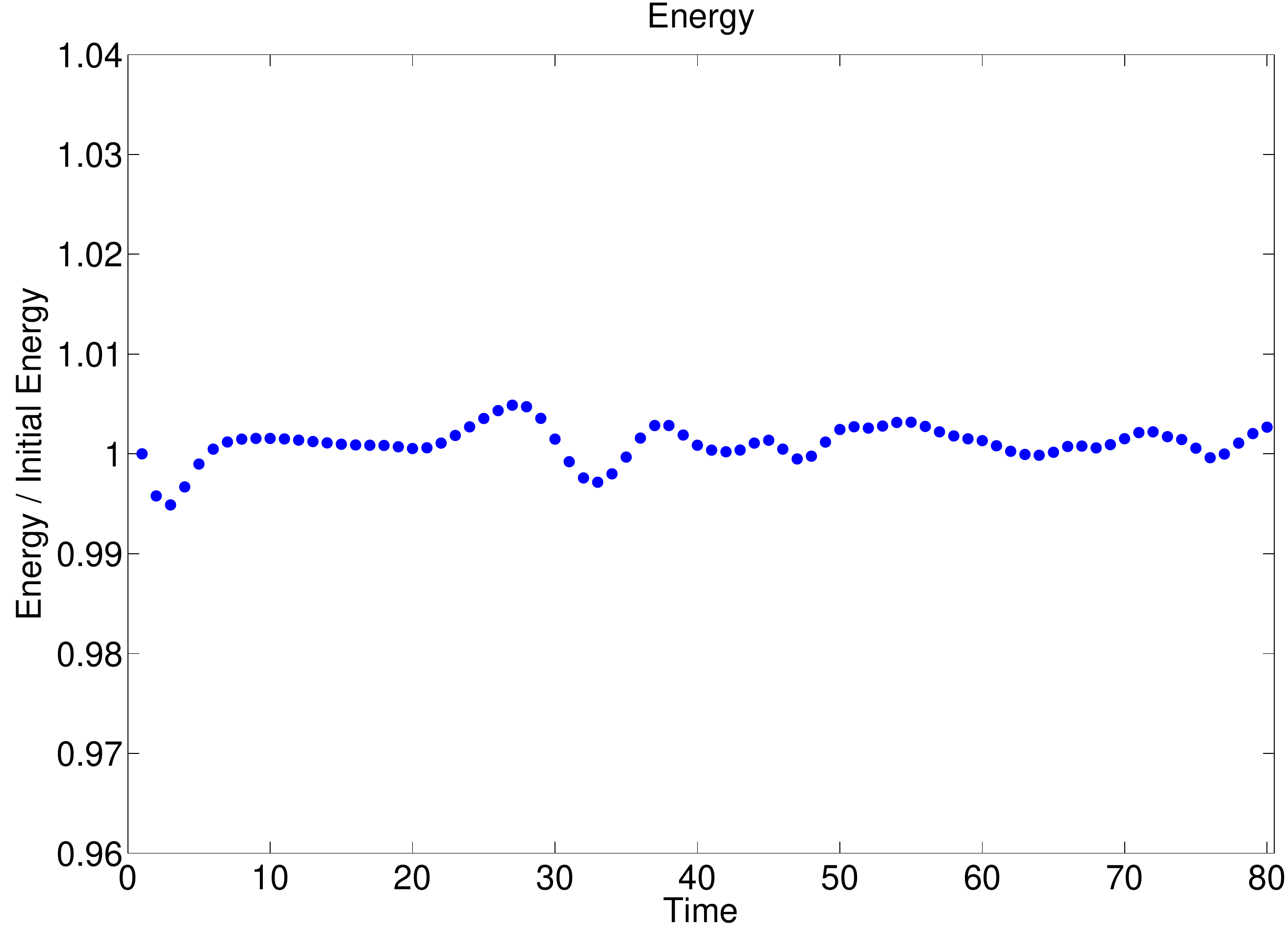} &
\includegraphics[width=0.5\textwidth]{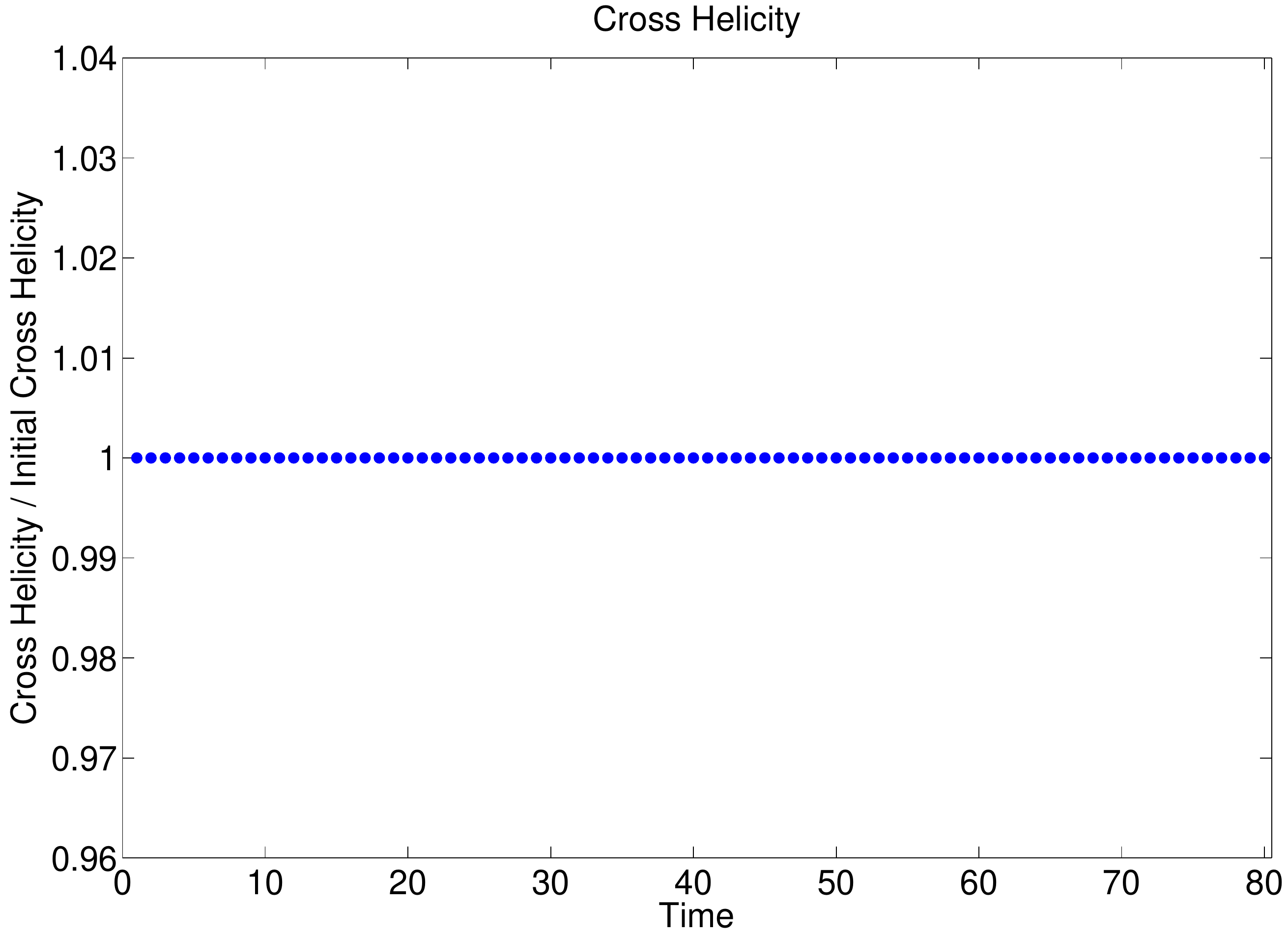} \\
(a) & (b) \\
\end{tabular}
\caption{(a) Energy vs. iteration number for an MHD simulation on a 2-dimensional cartesian grid using the variational integrator~(\ref{uupdate_MHD}-\ref{uvconstraint_MHD}). (b) Cross-helicity vs. iteration number for the same MHD simulation.  Notice that this integrator preserves the discrete cross-helicity~(\ref{discrete_cross_helicity}) exactly.}
\label{fig:MHDenergy}
\end{figure}

As our first test case, we consider the evolution of a localized fluid vortex coupled with an initially azimuthal magnetic field in a rectangular box with vanishing boundary conditions on the normal components of $\mathbf{v}$ and $\mathbf{B}$.  Our setup is adapted from a classic test case from computational fluid dynamics: the advection of a vortex in isentropic flow~\cite{Yee1999}.  We modify the setup by imposing no-transfer conditions across the boundaries of the domain, incompressibility of the initial velocity field, and uniformity of the fluid density.  The numerical details of the setup are given below:

\begin{list}{\labelitemi}{\leftmargin=1in}
\item{Domain: $[0,10] \times [0,12]$}
\item{Boundary Conditions: Tangential Velocity Field and Magnetic Field}
\item{Resolution: $20 \times 24$}
\item{Time Step: $h = 0.5$}
\item{Time Span: $0 \le t \le 80$}
\item{Initial Conditions:
\begin{align*}
u(x,y) &= u_0 + \frac{\beta}{2\pi}\exp\left(\frac{1-r^2}{2}\right)(y-y_0) \\
v(x,y) &= -\frac{\beta}{2\pi}\exp\left(\frac{1-r^2}{2}\right)(x-x_0) \\
B_x(x,y) &= -\sin\left(\frac{\pi x}{10}\right)\cos\left(\frac{\pi y}{12}\right) \\
B_y(x,y) &= \cos\left(\frac{\pi x}{10}\right)\sin\left(\frac{\pi y}{12}\right) \\
p(x,y) &= \frac{1}{\gamma}\left[1 - \frac{(\gamma-1)\beta^2}{8\gamma\pi}\exp(1-r^2) \right]^{\frac{\gamma}{\gamma-1}}
\end{align*}
with
\begin{equation*}
r = \sqrt{(x-x_0)^2+(y-y_0)^2}
\end{equation*}
}
\item{Parameters:
$ x_0 = 3, y_0 = 5.5,  u_0 = 0.5, \beta = 5, \gamma = 1.4.$
}
\end{list}

Since the initial velocity field defined above is inconsistent with the boundary conditions, we have used a numerical root-finding algorithm to obtain a nearby field that satisfies $\left. \mathbf{v} \cdot \hat{\mathbf{n}}\right|_{\partial M}  = 0$.

As Fig.~\ref{fig:MHDenergy} shows, the integrator exhibits excellent energy behavior and preserves the discrete analogue~(\ref{discrete_cross_helicity}) of the continuous cross-helicity $\langle \mathbf{v}^\flat, \mathbf{B} \rangle$ exactly.  Moreover, the divergence of the velocity field and the divergence of the magnetic field automatically vanish (modulo numerical roundoff errors) at each iteration.  An animation of this MHD simulation may be viewed at 
~\url{http://www.its.caltech.edu/~egawlik/MHD/MHDvortex.avi}.

\subsubsection{Magnetic Reconnection}

\begin{figure}[t]
\centering
\includegraphics[width=\textwidth]{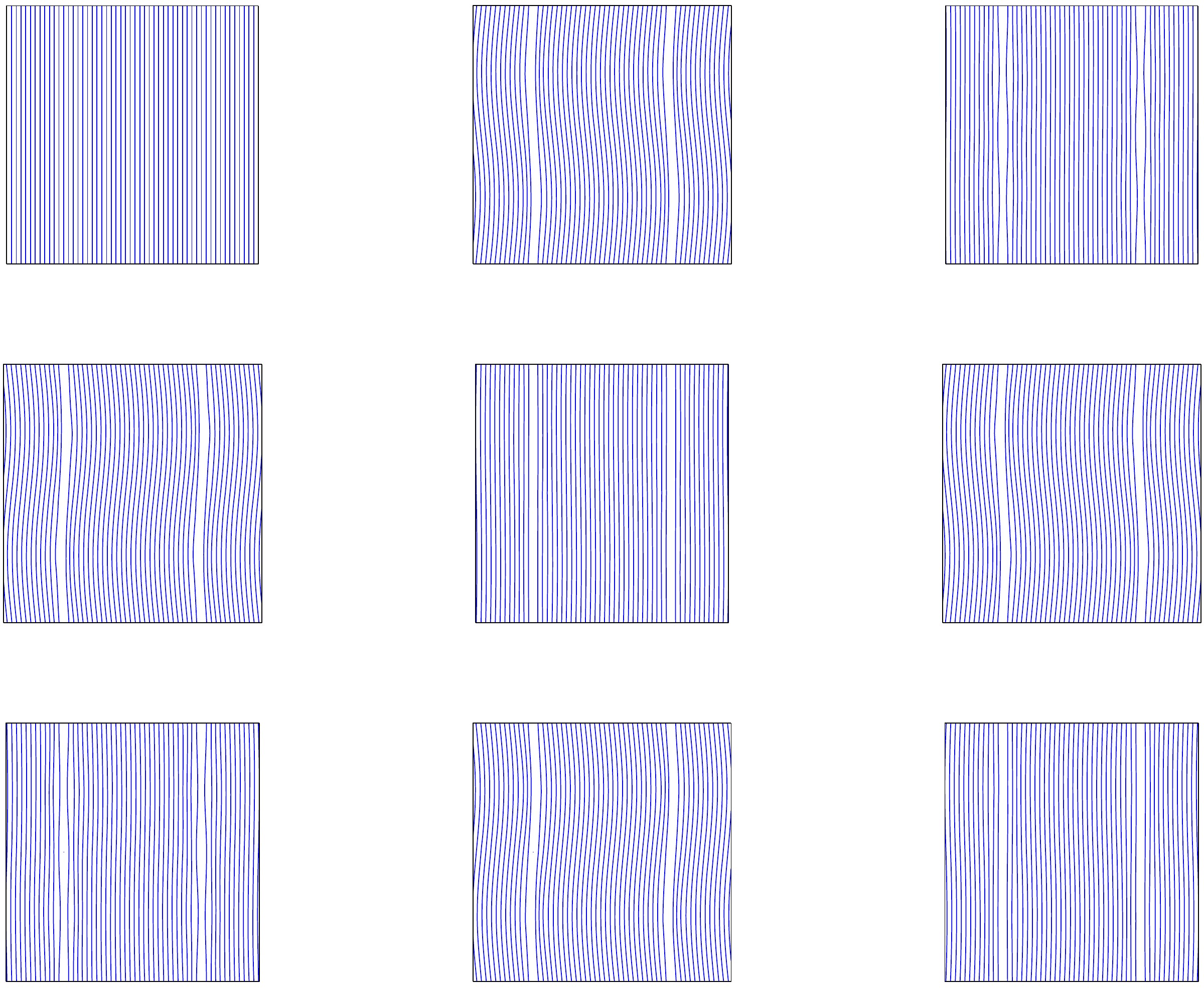}
\caption{Evolution of the magnetic field lines for an MHD current sheet simulation on a 2-dimensional cartesian grid using the variational integrator~(\ref{uupdate_MHD}-\ref{uvconstraint_MHD}).  The frames above are snapshots at times $t=0.0,0.5,1.0,1.5,2.0,2.5,3.0,3.5,4.0$, in normal reading order from left to right, top to bottom.}
\label{fig:MHDreconnection}
\end{figure}

\begin{figure}[t]
\centering
\includegraphics[width=0.7\textwidth]{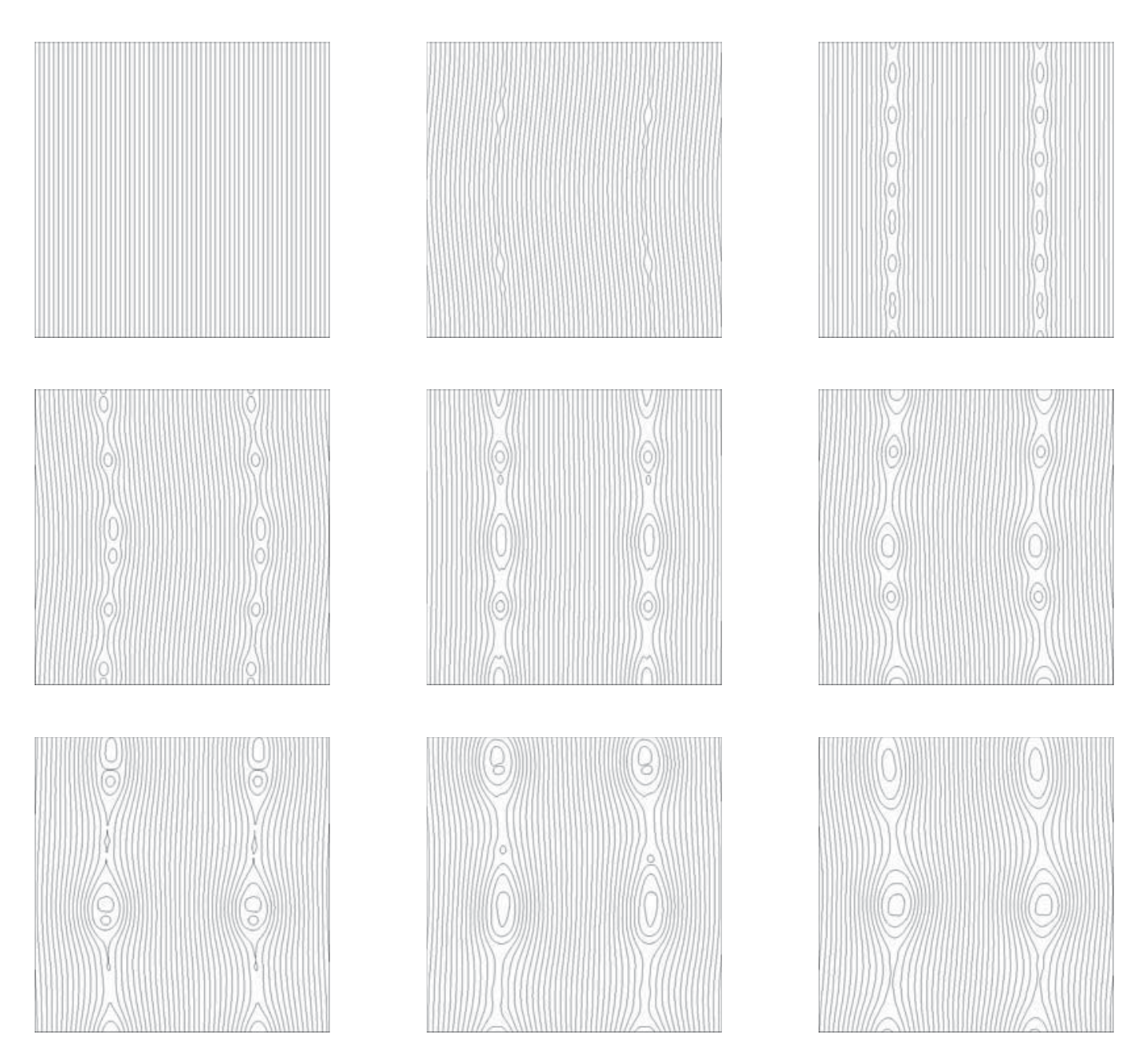}
\caption{(Borrowed from Gardiner and Stone~\cite{Gardiner2005}.) Evolution of the magnetic field lines for an MHD current sheet simulation on a 2-dimensional cartesian grid using a constrained transport Godunov scheme.  The frames above are snapshots at times $t=0.0,0.5,1.0,1.5,2.0,2.5,3.0,3.5,4.0$, in normal reading order from left to right, top to bottom.  Compared with our variational integrator (see Fig.~\ref{fig:MHDreconnection}), the Godunov scheme exhibits a noticeably larger amount of artificial magnetic reconnection due to numerical resistivity.}
\label{fig:MHDreconnection_reference}
\end{figure}

A good testament to the structure-preserving nature of a numerical MHD scheme is its ability to conserve the topology of the magnetic field lines over the course of a simulation.  Indeed, the pure advection of the magnetic field in ideal MHD gives rise to topological invariants that eliminate the possibility of $\emph{magnetic reconnection}$, a phenomenon whereby magnetic field lines break and recombine, altering the mutual connectivity of the field lines.   While such processes occur naturally in resistive MHD, where a magnetic resistivity term proportional to $\nabla^2 \mathbf{B}$ appears on the right-hand side of the magnetic field evolution equation~(\ref{MHD2}), the appearance of magnetic reconnection in an ideal MHD simulation is evidence of artificial numerical resistivity introduced by the numerical scheme.

Here we consider a setup proposed by Gardiner and Stone~\cite{Gardiner2005} in their studies of numerical methods for compressible MHD, whereby a pair of current sheets is juxtaposed with a perpendicular velocity field on a periodic domain.  The details of the setup are as follows:

\begin{list}{\labelitemi}{\leftmargin=1in}
\item{Domain: $[0,2] \times [0,2]$}
\item{Boundary Conditions: Periodic}
\item{Resolution: $30 \times 30$}
\item{Time Step: $h = 0.1$}
\item{Time Span: $0 \le t \le 8$}
\item{Initial Conditions:
\begin{align*}
u(x,y) &= u_0 \sin(\pi y) \\
v(x,y) &= 0 \\
B_x(x,y) &= 0 \\
B_y(x,y) &= \left\{\begin{array}{rl}
B_0 & \text{if } x < x_1 \\
-B_0 & \text{if } x_1 \le x \le x_2 \\
B_0 & \text{if } x_2 < x
\end{array} \right. \\
p(x,y) &= 0.1
\end{align*}
}
\item{Parameters:
$ x_1 = 0.5,  x_2 = 1.5, u_0 = 0.1, B_0 = 1, \theta = \tan^{-1}(0.5).$
}
\end{list}

Snapshots of the magnetic field lines over the course of the simulation are displayed in Fig.~\ref{fig:MHDreconnection}.  Notice that our integrator exhibits virtually no artificial reconnection, even on a relatively coarse grid.  The reader is invited to compare with Fig. 12 of Gardiner and Stone's study~\cite{Gardiner2005} (copied here in Fig.~\ref{fig:MHDreconnection_reference}), where the application of a constrained-transport Godunov scheme has led to visible reconnection in the form of magnetic field islands along the current sheets, despite their use of a high-resolution $256 \times 256$ grid.

\subsubsection{Field Loop Advection}

\begin{figure}[t]
\centering
\begin{tabular}{cc}
\vcent{\includegraphics[width=0.5\textwidth]{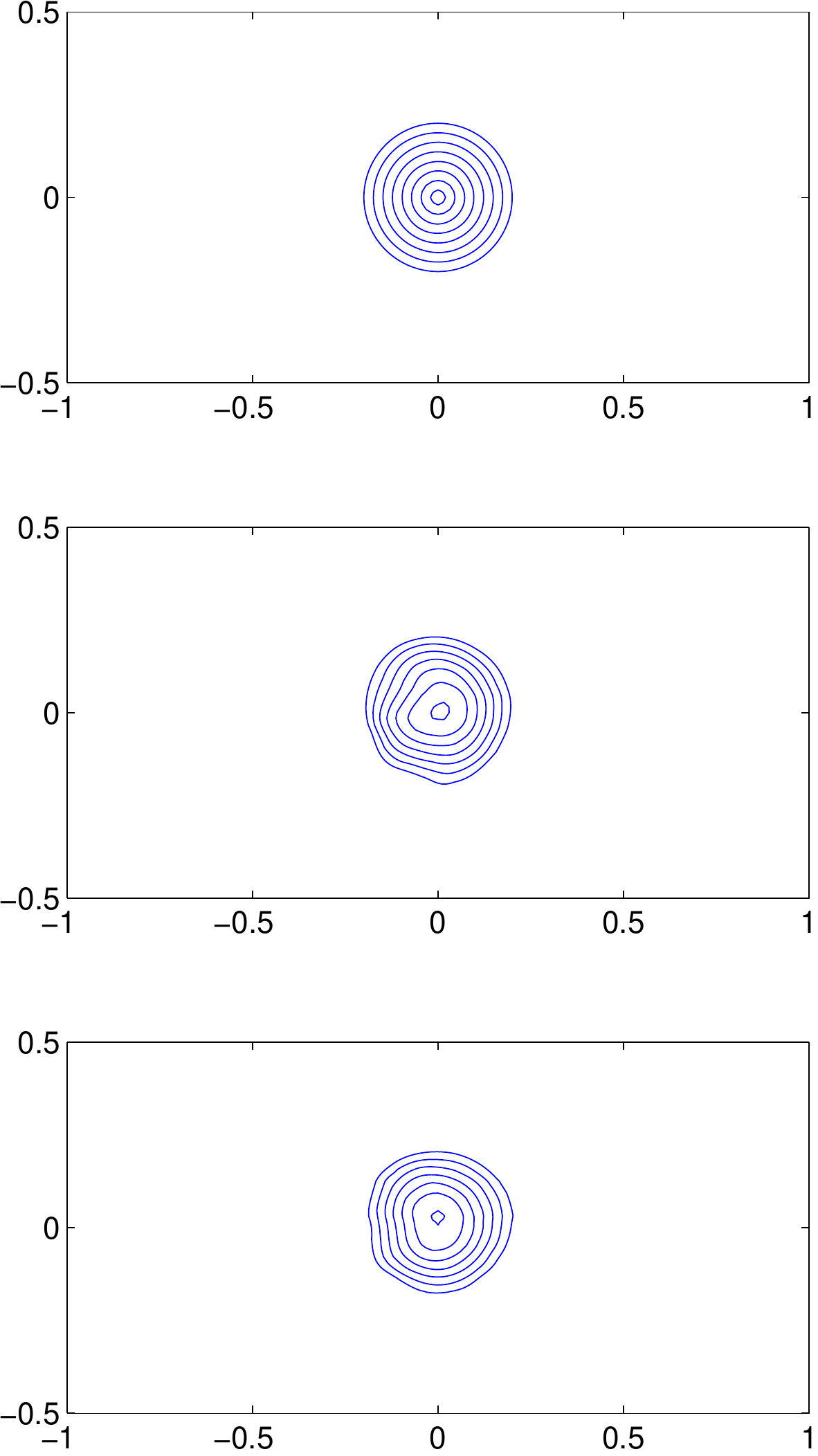}} &
\vcent{\includegraphics[width=0.5\textwidth]{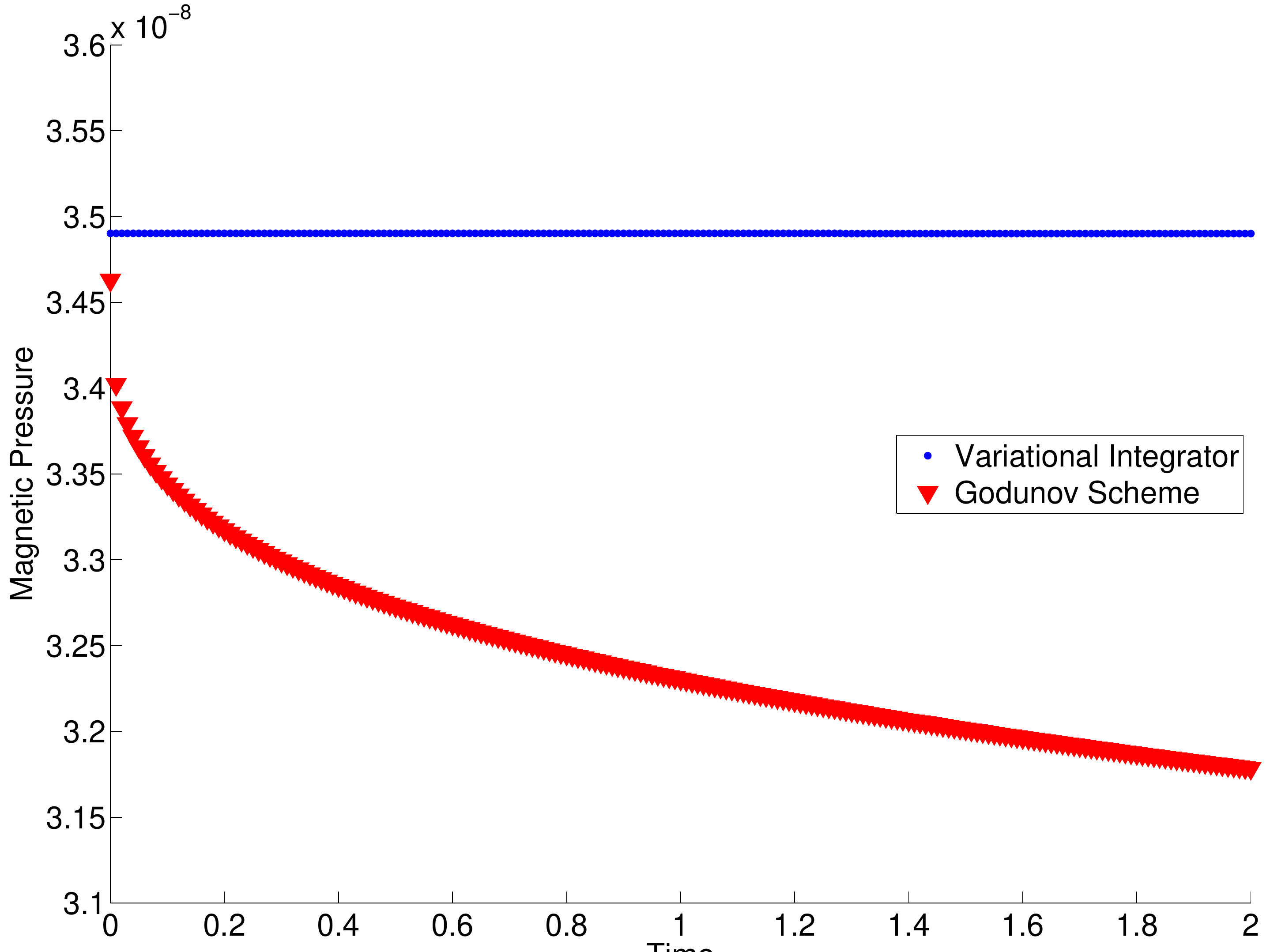}} \\
(a) & (b) \\
\end{tabular}
\caption{(a) Evolution of the magnetic field lines for an advected magnetic field loop using the variational integrator~(\ref{uupdate_MHD}-\ref{uvconstraint_MHD}).  The frames above, from top to bottom, are snapshots at times $t=0,1,2$. (b) Decay of the volume-averaged magnetic pressure $B_x^2+B_y^2$ during advection of the field loop.  Unlike the Godunov scheme proposed by Gardiner \& Stone~\cite{Gardiner2005}, our integrator exhibits no noticeable decay in net magnetic pressure during what amounts to a virtually passive advection of the magnetic field loop.}
\label{fig:MHDfieldloop}
\end{figure}

For sufficiently weak magnetic fields, the ideal MHD equations~(\ref{MHD1}-\ref{MHD4}) may be viewed as a nearly decoupled system whereby the magnetic field travels passively with the fluid and exerts negligible influence on the flow.  Here we consider the advection of a weak magnetic field loop in the presence of a uniform velocity field on a periodic domain.  The setup, proposed by Gardiner and Stone~\cite{Gardiner2005}, is designed in such a way that the magnetic field loop, centered at the origin at $t=0$, returns to the origin at fixed intervals of length $\Delta t = 1$.  The details of our simulation are as follows:

\begin{list}{\labelitemi}{\leftmargin=1in}
\item{Domain: $[-1,1] \times [-0.5,0.5]$}
\item{Boundary Conditions: Periodic}
\item{Resolution: $128 \times 64$}
\item{Time Step: $h = 0.01$}
\item{Time Span: $0 \le t \le 2$}
\item{Initial Conditions:
\begin{align*}
u(x,y) &= u_0 \cos(\theta) \\
v(x,y) &= u_0 \sin(\theta) \\
A(x,y) &= A_0 (R-\sqrt{x^2+y^2}) \\
p(x,y) &= 1
\end{align*}
}
\item{Parameters:
$ v_0 = \sqrt{5}, A_0 = 10^{-3}, R = 0.3, \theta = \tan^{-1}(0.5).$
}
\end{list}

In the specifications above, $A(x,y)$ corresponds to the magnetic vector potential, related to the magnetic field via $B_x=\frac{\partial A}{\partial y}$, $B_y=-\frac{\partial A}{\partial x}$.

In Fig.~\ref{fig:MHDfieldloop}, we display snapshots of the magnetic field lines at $t=0$, $t=1$, and $t=2$.  The field loop retains its structure and returns to the origin at the expected instants in time, although it suffers some distortion, presumably due to our magnetic update scheme's~(\ref{Bxupdate_MHD}-\ref{Byupdate_MHD}) low order of accuracy and absence of an upwind bias.  Note, however, that our scheme introduces no artificial decay in the volume-averaged magnetic pressure $B_x^2+B_y^2$ during the advection of the field loop, in stark contrast with the power-law decay in magnetic pressure observed by Gardiner and Stone~\cite{Gardiner2005} in their Godunov scheme simulations of the same system.

\subsubsection{MHD Rotor}

\begin{figure}[t]
\centering
\includegraphics[width=\textwidth]{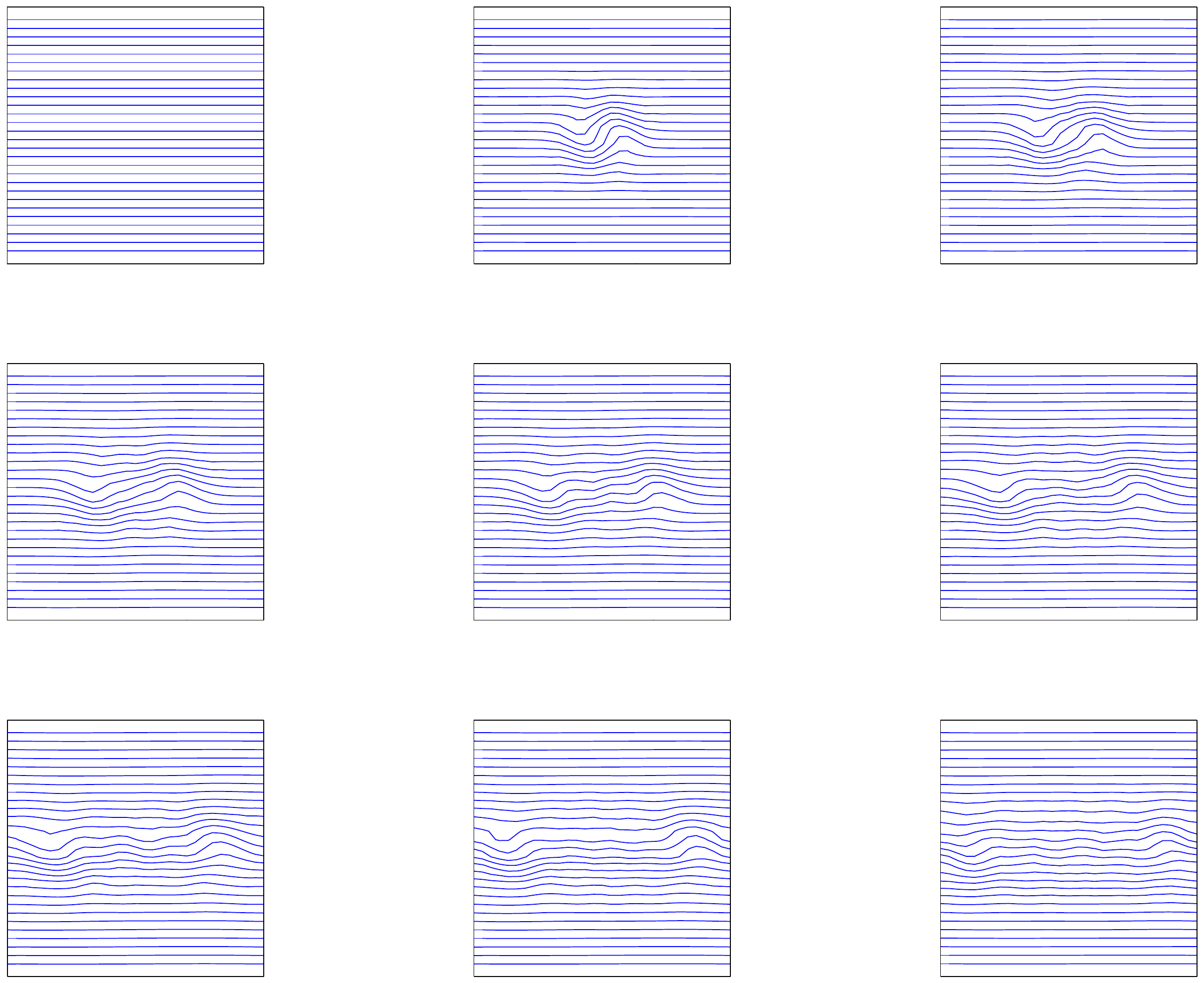}
\caption{Evolution of the magnetic field lines for a simulation of the incompressible MHD rotor on a 2-dimensional cartesian grid using the variational integrator~(\ref{uupdate_MHD}-\ref{uvconstraint_MHD}).  The frames above are snapshots at times $t=0.000,0.042,0.084,0.126,0.168,0.210,0.252,0.294,0.336$, in normal reading order from left to right, top to bottom.}
\label{fig:MHDrotor}
\end{figure}

As a model for star formation, the compressible MHD rotor provides an intriguing test bed for numerical simulation.  The classic MHD rotor setup consists of a dense rotating disk of fluid in an initially uniform magnetic field.  Rotation of the fluid induces winding of the magnetic field lines, generating magnetic tension that ultimately leads to a reduction in the disk's angular momentum due to the emission of torsional Alfv\'{e}n waves.  Here we adapt the setup for the MHD rotor problem studied by Balsara and Spicer~\cite{Balsara1999} as follows:

\begin{list}{\labelitemi}{\leftmargin=1in}
\item{Domain: $[0,1] \times [0,1]$}
\item{Boundary Conditions: Periodic}
\item{Resolution: $30 \times 30$}
\item{Time Step: $h = 0.003$}
\item{Time Span: $0 \le t \le 0.36$}
\item{Initial Conditions:
\begin{align*}
u(x,y) &= \left\{\begin{array}{rl}
-u_0 f(r)(y-0.5)/r_0 & \text{if } r < r_0 \\
-u_0 f(r)(y-0.5)/r & \text{if } r_0 \le r \le r_1 \\
0 & \text{if } r_1 < r
\end{array} \right. \\
v(x,y) &= \left\{\begin{array}{rl}
u_0 f(r)(x-0.5)/r_0 & \text{if } r < r_0 \\
u_0 f(r)(x-0.5)/r & \text{if } r_0 \le r \le r_1 \\
0 & \text{if } r_1 < r
\end{array} \right. \\
B_x(x,y) &= 5/\sqrt{4\pi} \\
B_y(x,y) &= 0 \\
p(x,y) &= 1
\end{align*}
with
\begin{align*}
r &= \sqrt{(x-0.5)^2+(y-0.5)^2} \\
f(r) &= \frac{r_1-r}{|r-r_0|}
\end{align*}
}
\item{Parameters:
$u_0 = 2, r_0 = 0.1, r_1 = 0.115, \theta = \tan^{-1}(0.5).$
}
\end{list}

An animation of the evolution of the velocity and magnetic fields for this MHD rotor simulation may be viewed at 
~\url{http://www.its.caltech.edu/~egawlik/MHD/MHDrotor.avi}.  Snapshots of the magnetic field lines, displayed in Fig.~\ref{fig:MHDrotor}, highlight the manner in which the magnetic field is ``frozen'' into the fluid.

\subsubsection{Orszag-Tang Vortex} \label{section:Orszag}

\begin{figure}[t]
\centering
\includegraphics[width=\textwidth]{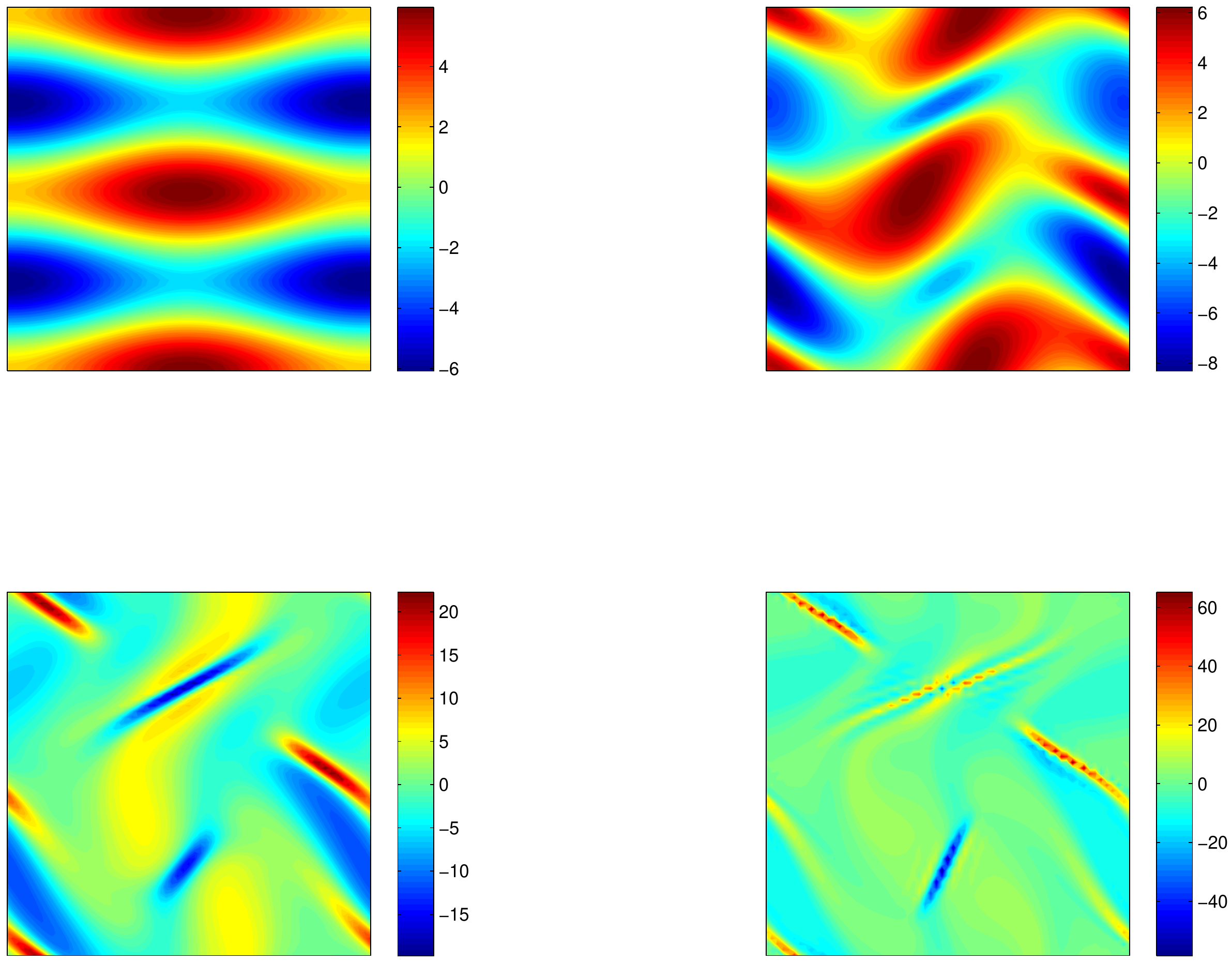}
\caption{Current density contours for a simulation of the incompressible Orszag-Tang vortex on a 2-dimensional cartesian grid using the variational integrator~(\ref{uupdate_MHD}-\ref{uvconstraint_MHD}).  The frames above are snapshots at times $t=0.00,0.25,0.50,0.75$, in normal reading order from left to right, top to bottom.}
\label{fig:MHDOrszagTang}
\end{figure}

\begin{figure}[t]
\centering
\includegraphics[width=0.7\textwidth]{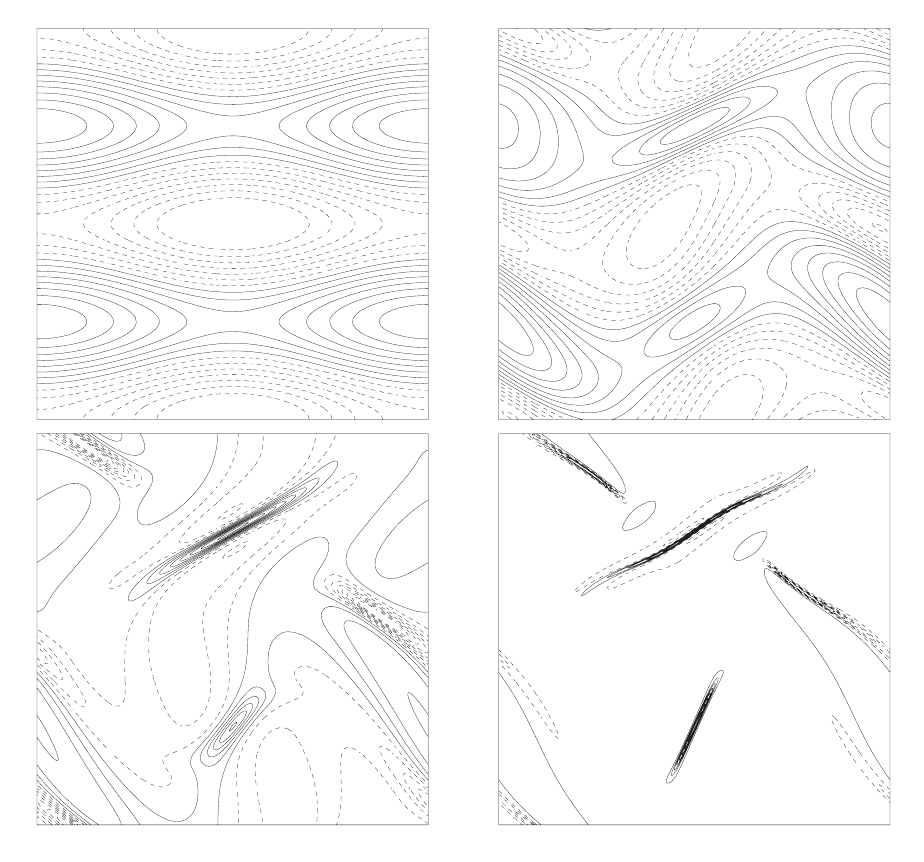}
\caption{(Borrowed from Cordoba and Marliani~\cite{Cordoba2000}.) Current density contours for a simulation of the incompressible Orszag-Tang vortex on a 2-dimensional cartesian grid using an adaptive mesh refinement scheme.  The frames above are snapshots at times $t=0.00,0.25,0.50,0.75$, in normal reading order from left to right, top to bottom.}
\label{fig:MHDOrszagTang_reference}
\end{figure}

In this example, we study the evolution of current sheets in an Orszag-Tang vortex, a classic test for two-dimensional MHD codes. This particular example illustrates the growth of MHD current sheets: curves along which the magnitude of the current density $\mathbf{j} = \nabla \times \mathbf{B}$ is large due to the presence of antiparallel magnetic field lines.  Here we consider the incompressible Orszag-Tang vortex studied by Cordoba and Marliani~\cite{Cordoba2000}:

\begin{list}{\labelitemi}{\leftmargin=1in}
\item{Domain: $[0,2\pi] \times [0,2\pi]$}
\item{Boundary Conditions: Periodic}
\item{Resolution: $64 \times 64$}
\item{Time Step: $h = 0.01$}
\item{Time Span: $0 \le t \le 0.75$}
\item{Initial Conditions:
\begin{align*}
\psi(x,y) &= 2\sin(y) - 2\cos(x) \\
A(x,y) &= \cos(2y) - 2\cos(x) \\
p(x,y) &= 1
\end{align*}
}
\end{list}

Once again, in the specifications above, $A(x,y)$ corresponds to the magnetic vector potential, related to the magnetic field via $B_x=\frac{\partial A}{\partial y}$, $B_y=-\frac{\partial A}{\partial x}$.  Similarly, $\psi(x,y)$ corresponds to the fluid stream function, related to the fluid velocity field via $u=\frac{\partial \psi}{\partial y}$, $v=-\frac{\partial \psi}{\partial x}$.

A comparison of our integrator's performance (Fig.~\ref{fig:MHDOrszagTang}) with a reference solution computed by Cordoba and Marliani~\cite{Cordoba2000} (Fig.~\ref{fig:MHDOrszagTang_reference}) illustrates that our integrator produces qualitatively accurate results, even at a low level of resolution.  For comparison, Cordoba and Marliani's plots~\cite{Cordoba2000} were generated with an adaptive mesh refinement scheme using an \emph{initial} resolution of $1024 \times 1024$; by the end of the simulation, the adaptive scheme employs an additional five levels of hierarchical refinement.  Our simulation employed a modest $64 \times 64$ grid, only lacking resolution toward the final stages of current sheet development.

\subsubsection{Convergence Rate Analysis}

\begin{figure}[t]
\centering
\includegraphics[width=0.5\textwidth]{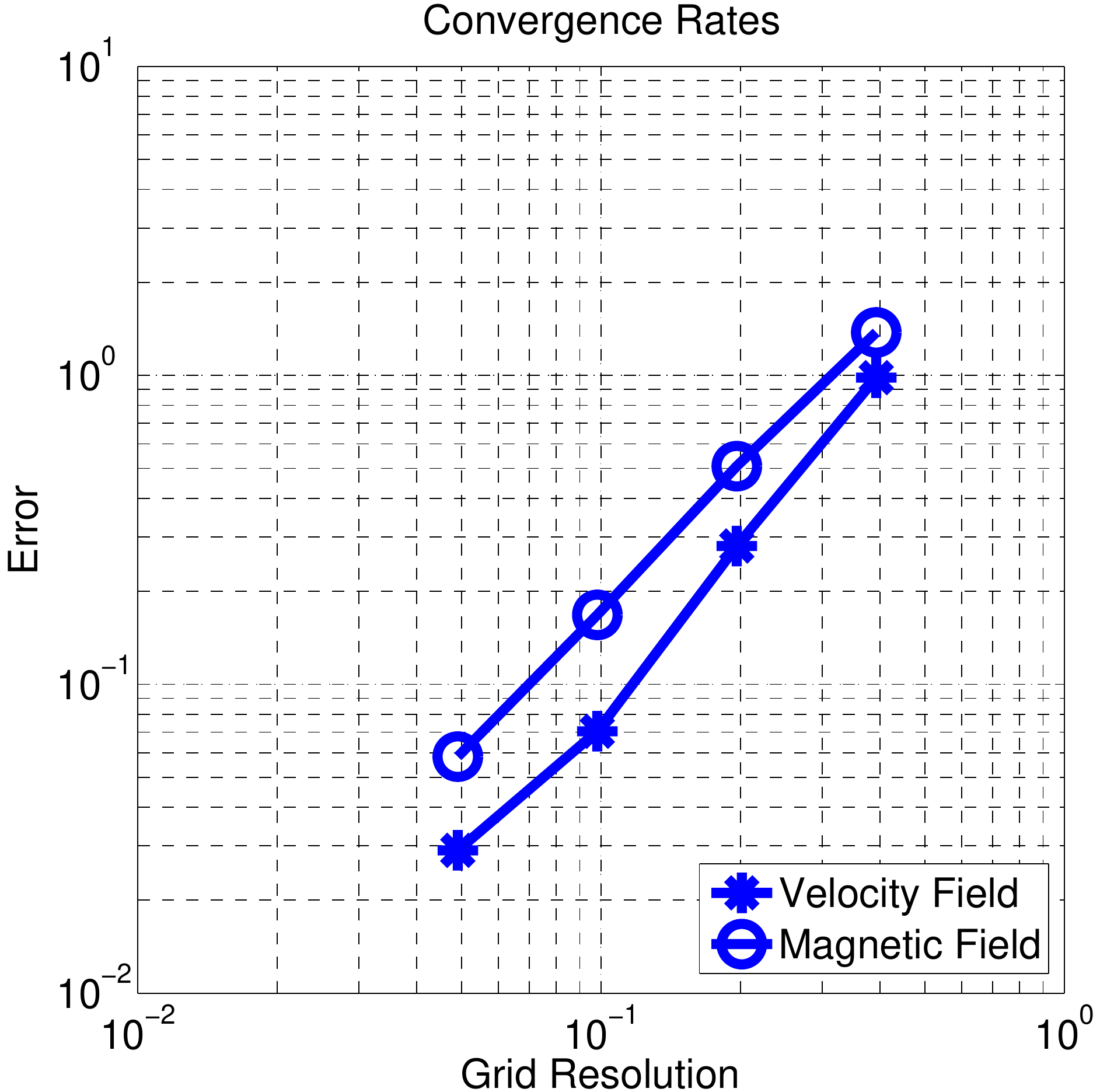}
\caption{Convergence rates for a simulation of the incompressible Orszag-Tang vortex on a 2-dimensional cartesian grid using the variational integrator~(\ref{uupdate_MHD}-\ref{uvconstraint_MHD}).  Data points represent the $L_2$-norm of the difference in vector fields at $t = 0.25$ between successive power-of-two refinements of the grid.  The reported grid resolution at each data point corresponds to the finer of the two resolutions under scrutiny.}
\label{fig:MHDconvergence}
\end{figure}

As a final MHD numerical test, we study the convergence rate of our MHD variational integrator~(\ref{uupdate_MHD}-\ref{uvconstraint_MHD}).  We use initial conditions identical to those in Section~\ref{section:Orszag}, but with varying spatial resolutions ($8 \times 8$, $16 \times 16$, $32 \times 32$, $64 \times 64$, and $128 \times 128$) and a fixed ratio of spatial to temporal resolutions ($\epsilon/h \approx 7.85$).  Fig.~\ref{fig:MHDconvergence} plots the $L_2$-norm of the difference in velocity fields at $t = 0.25$ between successive refinements of the grid, as well as the difference in magnetic fields between successive refinements of the grid.  The plot confirms the linear accuracy of the integrator.

\subsection{Complex Fluid Test Cases}

\subsubsection{A Nematic Liquid Crystal Test Case: Diffusion of a Gyrating Disk}

\begin{figure}[t]
\centering
\begin{tabular}{cc}
\includegraphics[width=0.5\textwidth]{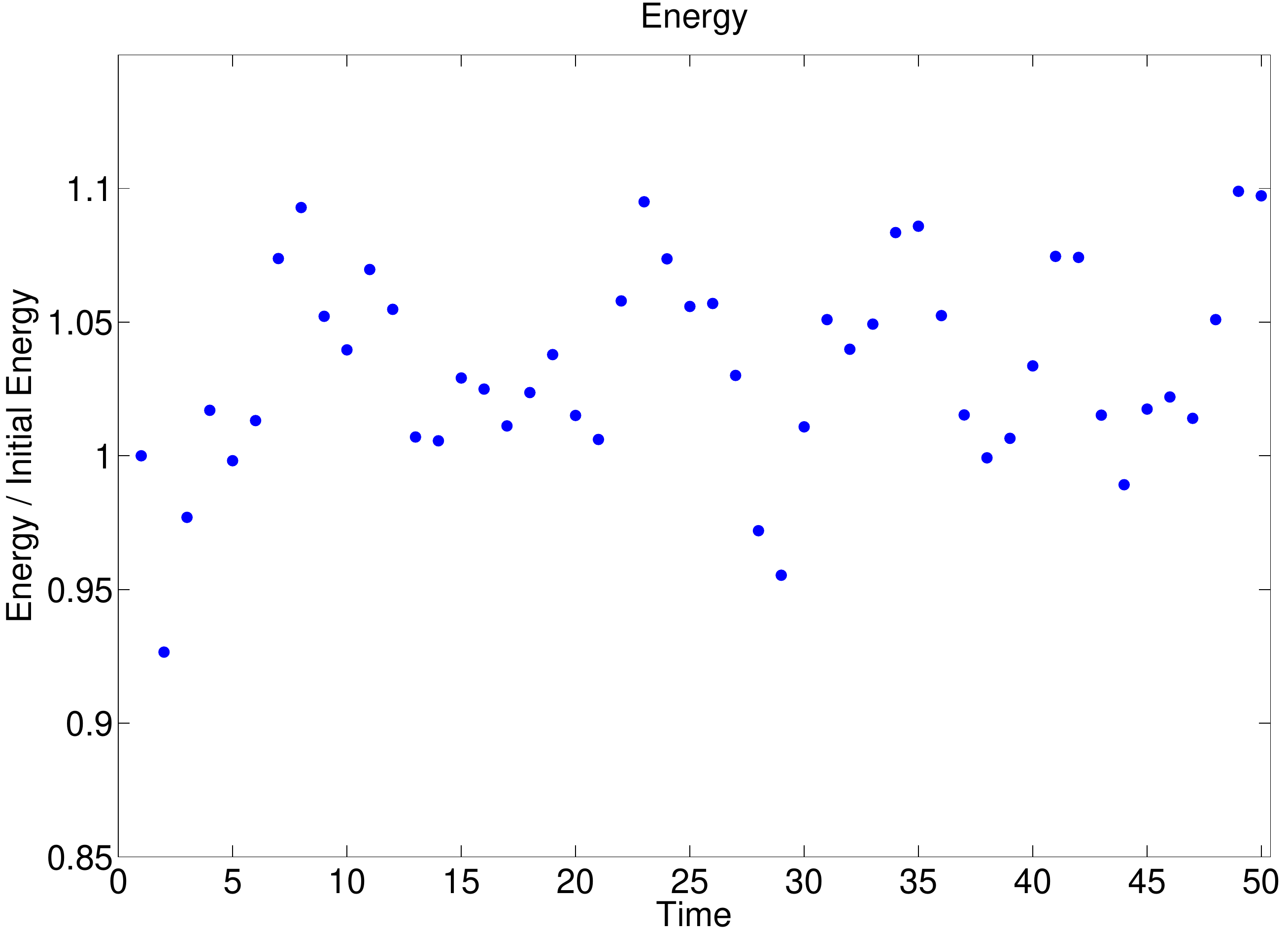} &
\includegraphics[width=0.5\textwidth]{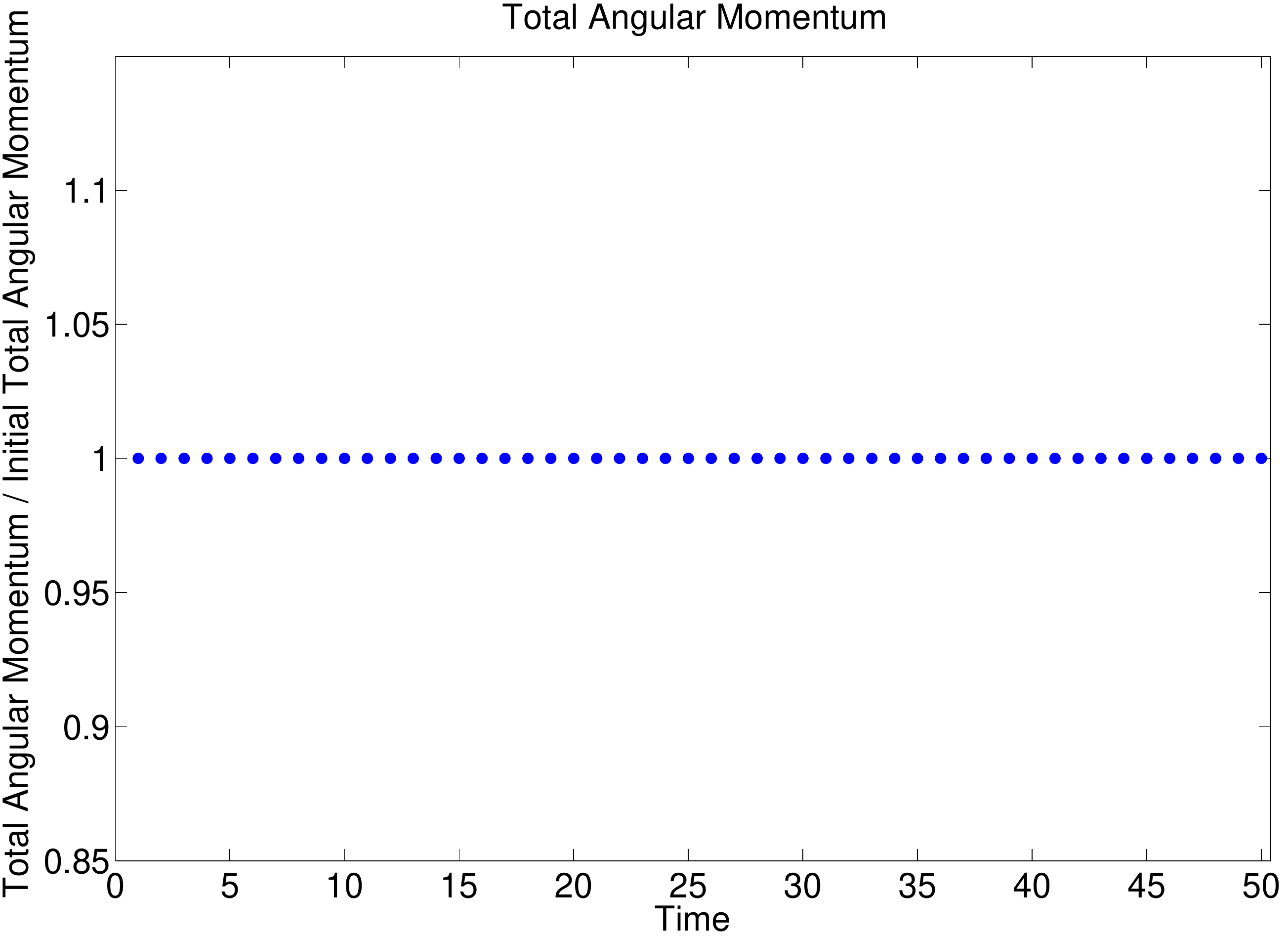} \\
(a) & (b) \\
\end{tabular}
\caption{(a) Energy vs. iteration number for a nematic liquid crystal simulation on a 2-dimensional cartesian grid using the variational integrator~(\ref{uupdate_nematic}-\ref{uvconstraint_nematic}). (b) Total angular momentum vs. iteration number for the same nematic liquid crystal fluid simulation.  Notice that this integrator preserves the total angular momentum exactly.}
\label{fig:nematic_energy}
\end{figure}

As a numerical test of our nematic liquid crystal integrator, we consider a disk of gyrating directors immersed in a steady state fluid flow:

\begin{list}{\labelitemi}{\leftmargin=1in}
\item{Domain: $[0,10] \times [0,10]$}
\item{Boundary Conditions: Tangential velocity field, vanishing director gradient}
\item{Resolution: $10 \times 10$}
\item{Time Step: $h = 0.4$}
\item{Time Span: $0 \le t \le 50$}
\item{Initial Conditions:
\begin{align*}
\psi(x,y) &= \sin\left(\frac{\pi x}{10}\right) \sin\left(\frac{\pi y}{10}\right) \\
\omega(x,y) &= \left\{\begin{array}{rl}
1 & \text{if } r < \frac{5}{2} \\
0 & \text{if } r \ge \frac{5}{2}
\end{array} \right. \\
\alpha(x,y) &= 0 \\
p(x,y) &= 1
\end{align*}
with
\begin{equation*}
r = \sqrt{(x-5)^2+(y-5)^2},
\end{equation*}
}
\end{list}
where $\psi$ is related to the velocity field $(u,v)$ via $u=\frac{\partial \psi}{\partial y}$, $v=-\frac{\partial \psi}{\partial x}$.

Plots of energy and angular momentum vs. time for this simulation are given in Fig.~\ref{fig:nematic_energy}.  The integrator exhibits good energy behavior and captures the qualitative dynamics of the system convincingly, as evidenced in the following animation:~\url{http://www.its.caltech.edu/~egawlik/Complex\%20Fluids/nematic.avi}.

\subsubsection{A Microstretch Fluid Test Case: Diffusion of a Gyrating Disk}

\begin{figure}[t]
\centering
\begin{tabular}{cc}
\includegraphics[width=0.5\textwidth]{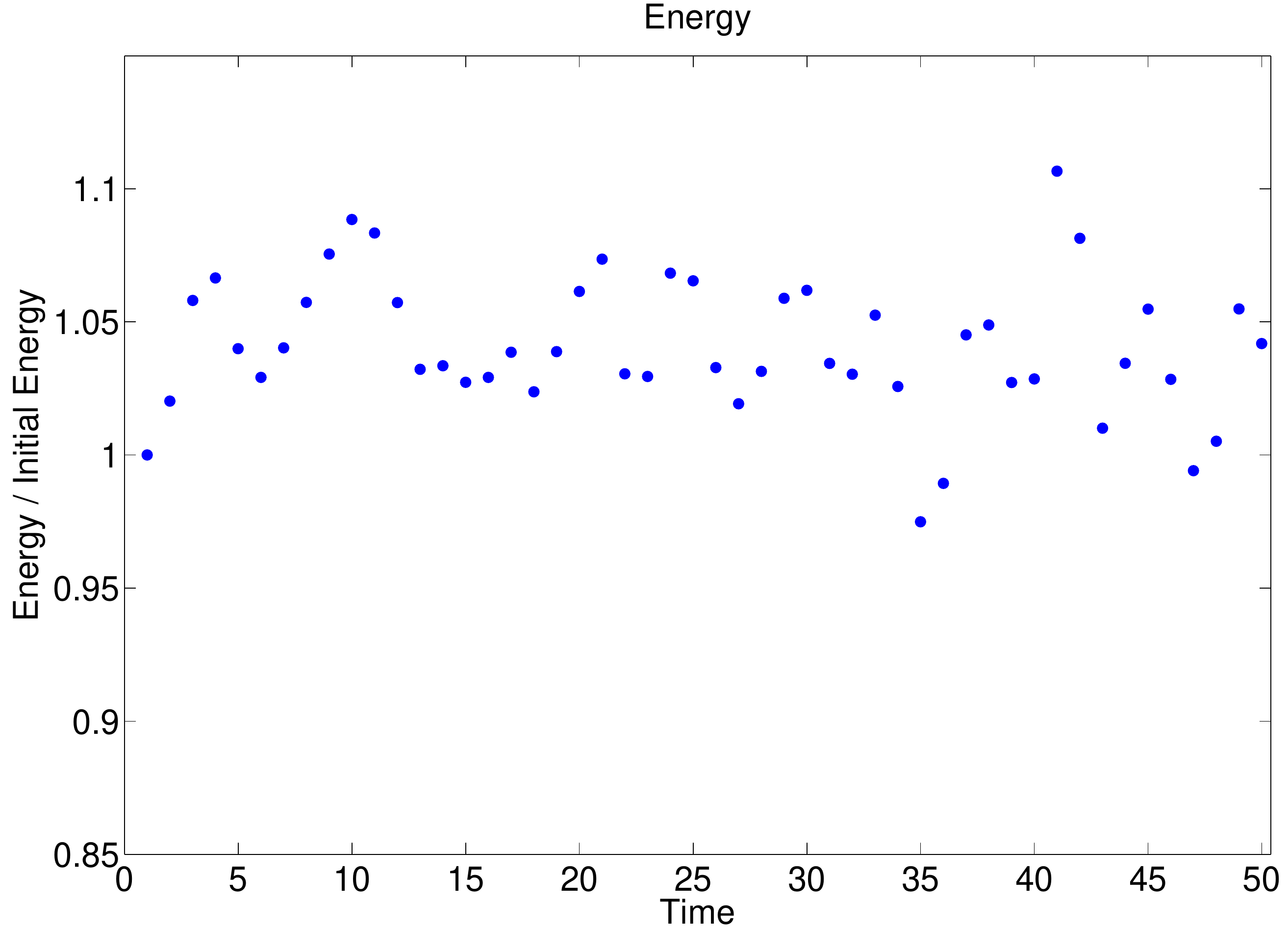} &
\includegraphics[width=0.5\textwidth]{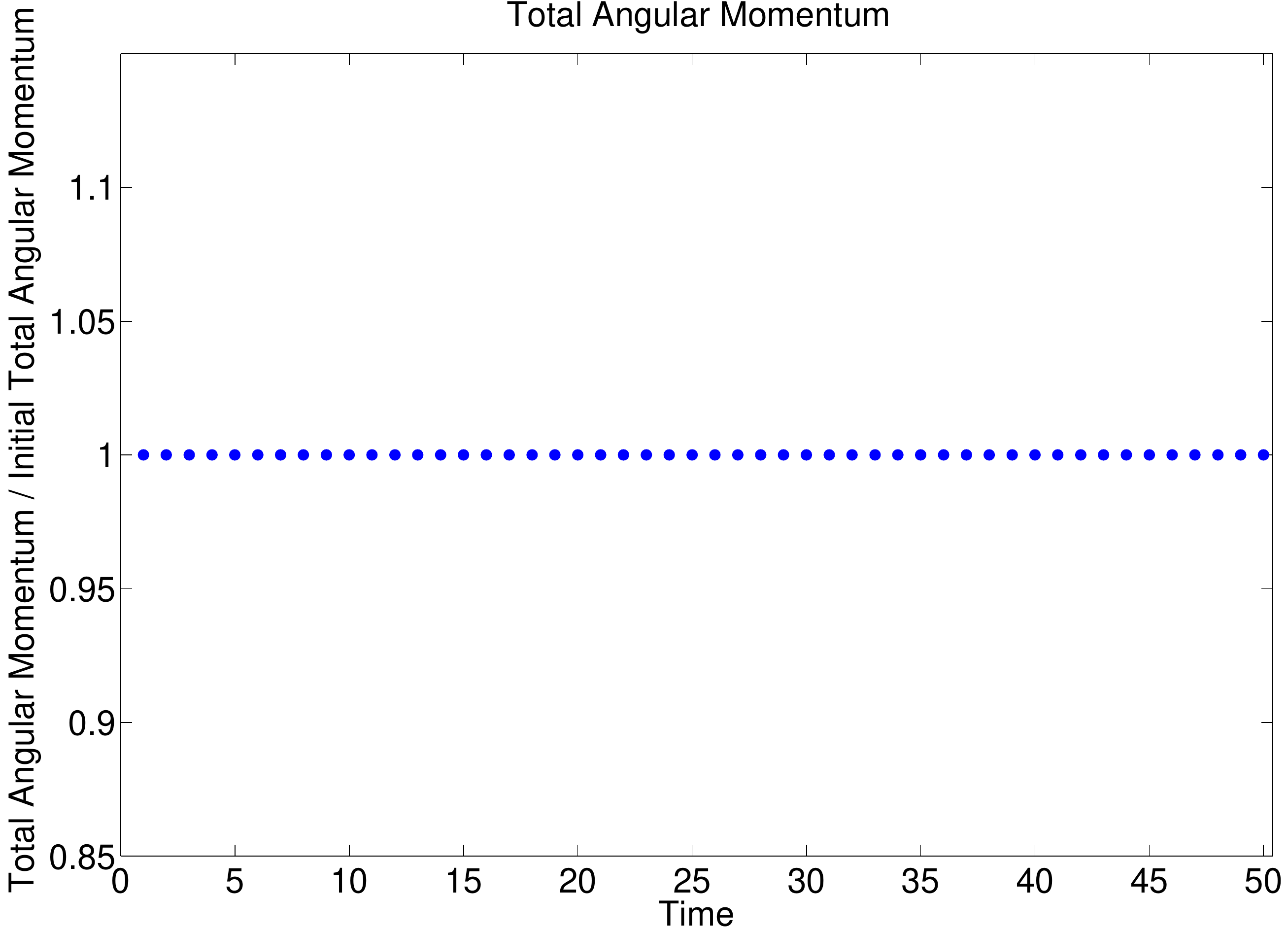} \\
(a) & (b) \\
\end{tabular}
\caption{(a) Energy vs. iteration number for a microstretch fluid simulation on a 2-dimensional cartesian grid using the variational integrator~(\ref{uupdate_microstretch}-\ref{uvconstraint_microstretch}). (b) Total angular momentum vs. iteration number for the same microstretch fluid simulation.  Notice that this integrator preserves the total angular momentum exactly.}
\label{fig:microstretch_energy}
\end{figure}

To test our microstretch continua integrator, we consider precisely the same setup as above; the evolution of the system will differ from the nematic liquid crystal case due to the extra flexibility in the length of the director.  The setup is generalized as follows:

\begin{list}{\labelitemi}{\leftmargin=1in}
\item{Domain: $[0,10] \times [0,10]$}
\item{Boundary Conditions: Tangential velocity field, vanishing director gradient}
\item{Resolution: $10 \times 10$}
\item{Time Step: $h = 0.4$}
\item{Time Span: $0 \le t \le 50$}
\item{Initial Conditions:
\begin{align*}
\psi(x,y) &= \sin\left(\frac{\pi x}{10}\right) \sin\left(\frac{\pi y}{10}\right) \\
\omega(x,y) &= \left\{\begin{array}{rl}
1 & \text{if } r < \frac{5}{2} \\
0 & \text{if } r \ge \frac{5}{2}
\end{array} \right. \\
R(x,y) &= 0 \\
\alpha(x,y) &= 0 \\
\lambda(x,y) &= 0 \\
j_r(x,y) &= 0 \\
j_s(x,y) &= 0 \\
p(x,y) &= 1
\end{align*}
with
\begin{equation*}
r = \sqrt{(x-5)^2+(y-5)^2},
\end{equation*}
}
\end{list}
where $\psi$ is related to the velocity field $(u,v)$ via $u=\frac{\partial \psi}{\partial y}$, $v=-\frac{\partial \psi}{\partial x}$.

Plots of energy and angular momentum vs. time for this simulation are given in Fig.~\ref{fig:microstretch_energy}.  Like the previous integrator, the microstretch fluid flow integrator exhibits good energy behavior and captures the qualitative dynamics of the system convincingly; see~\url{http://www.its.caltech.edu/~egawlik/Complex\%20Fluids/microstretch.avi} for an animation.  Notice the heightened intricacy in the dynamics of microstretch continua over nematic liquid crystals owing to the flexibility of the director field.

\section{Conclusions and Future Directions}

This study has merged techniques from variational mechanics and discrete exterior calculus to design structured integration algorithms for magnetohydrodynamics and complex fluid flow.  These integrators exhibit several desirable features that distinguish them from conventional integration schemes: they are symplectic, exhibit good long-term energy behavior, and conserve momenta arising from symmetries exactly.  Moreover, their formulation in the framework of DEC ensures that Stoke's theorem holds at the discrete level, leading to an automatic satisfaction of divergence-free constraints on the velocity and magnetic fields in the resulting numerical schemes.

Numerical tests of these structured integrators on cartesian grids confirmed our theoretical results and indicated that in addition to the properties listed above, the integrators faithfully respect topological features of the flow in MHD (e.g. the forbiddance of magnetic reconnection) and are predictive at remarkably coarse resolutions, despite being of low order.

In the process of designing structured integrators for MHD and complex fluid flow, we uncovered a general framework for the design of variational integrators for continuum systems described by Euler-Poincar\'{e} equations with advected parameters, a particular class of differential equations obtained from a variational principle.  This finding is tremendously encouraging given the plethora of physical systems arising in mechanics, fluid dynamics, and plasma physics that are governed by such equations; see references~\cite{Holm1998} and~\cite{Marsden1984} for examples.  To name one such example, nearly all of the governing equations for geophysical fluid dynamics---a field that plays an instrumental role in the modeling of global climate change---are of this type~\cite{Holm2005}.  We envision the exterior calculus based variational framework serving as a cornerstone for the design of structured integrators for these families of physical systems.

\textbf{Future Work.}
A number of numerical and theoretical challenges are left unaddressed in this study. Incorporation of compressibility, viscosity, resistivity, and inhomogeneity into the discrete variational formulation of fluid flows are obvious extensions. We are also developing a more general approach to handle Lie group valued functions in arbitrary dimension to resolve the issue mentioned in Section~\ref{section:nematic}. On the practical side, we have yet to develop high order versions of our variational update equations valid on arbitrary grids. Finally, we are interested in developing other structure-preserving integrators for, e.g., relativistic physical systems.

\section{Acknowledgments}
During the summer of 2009, Evan Gawlik was supported by the Caltech Summer Undergraduate Research Fellowship Program and the Aerospace Corporation.  We wish to thank Francois Gay-Balmaz for his many valuable comments on the geometric formulation of complex fluid flow, and for his help in editing this paper.

\appendix
\section{Appendix}

\subsection{Lie Groups and Lie Algebras} \label{appendix:lie}

In this section, we present those aspects of Lie group theory most relevant to our development of discretizations of continuum theories.  The notations appearing here are used consistently throughout the paper.  For a more complete treatment of Lie groups, Lie algebras, and group actions, the reader is referred to~\cite{Abraham1988}.

Let $G$ be a Lie group; that is, $G$ is a smooth manifold equipped with a group structure for which the group multiplication and inversion are smooth maps.  The \emph{Lie algebra} $\mathfrak{g}$ of $G$ is by definition the tangent space of $G$ at the identity $e \in G$:
\begin{equation}
\mathfrak{g} = T_e G = \left\{ \xi \mid \xi = \left.\frac{d}{dt}\right|_{t=0}g(t) \text{ for some curve } g : (-b,b)\subset\mathbb{R} \rightarrow G \text{ with } g(0)=e \right\}
\end{equation}
We denote the space dual to $\mathfrak{g}$ by $\mathfrak{g}^*$; this is the space of linear maps from $\mathfrak{g}$ into $\mathbb{R}$.  We use angular brackets to denote the natural pairing $\langle \cdot, \cdot \rangle : \mathfrak{g}^* \times \mathfrak{g} \rightarrow \mathbb{R}$.

We use the following notation for left and right translation of tangent vectors in $TG$:
\begin{align}
TL_g \eta &= \left.\frac{d}{ds}\right|_{s=0} g h(s) \\
TR_g \eta &= \left.\frac{d}{ds}\right|_{s=0} h(s) g,
\end{align}
where $g \in G$, $\eta \in T_{h(0)}G$, and $h : (-b,b)\subset\mathbb{R} \rightarrow G$ is a curve in $G$ with $h'(0)=\eta$.  If there is no danger of confusion, we simply denote these maps via concatenation: $g\eta = TL_g \eta$ and $\eta g = TR_g \eta$.

\subsubsection{Group Actions}
Let $V$ be a vector space.  A \emph{left action} or \emph{left representation} of $G$ on $V$ is a smooth map $\Phi : G \times V \rightarrow V$ for which
\begin{enumerate}
\item{$\Phi(e,a) = a$ for every $a \in V$.}
\item{$\Phi(g,\Phi(h,a))=\Phi(gh,a)$ for every $g,h \in G, a \in V$.}
\end{enumerate}
We denote such an action via concatenation, or if there is danger of confusion, we insert a binary operator:
\begin{equation}
ga = \Phi(g,a) = g \cdot a.
\end{equation}
The \emph{induced infinitesimal action} of $\mathfrak{g}$ on $V$ is given by differentiating the action of $G$ on $V$ at the identity:
\begin{equation}
\xi \cdot a = \left.\frac{d}{dt}\right|_{t=0} g(t) \cdot a,
\end{equation}
where $g : (-b,b)\subset\mathbb{R} \rightarrow G$ is a curve in $G$ with $g(0)=e, g'(0)=\xi$.

A corresponding notion is that of a \emph{right action} or \emph{right representation} of $G$ on $V$, which is a smooth map $\Psi : V \times G \rightarrow V$ for which
\begin{enumerate}
\item{$\Psi(a,e) = a$ for every $a \in V$.}
\item{$\Psi(\Psi(a,g),h)=\Psi(a,gh)$ for every $g,h \in G, a \in V$.}
\end{enumerate}
We denote such an action via concatenation or with a binary operator:
\begin{equation}
ag = \Psi(a,g) = a \cdot g.
\end{equation}
The infinitesimal right action of $\mathfrak{g}$ on $V$ is given by
\begin{equation}
a \cdot \xi = \left.\frac{d}{dt}\right|_{t=0} a \cdot g(t),
\end{equation}
where $g : (-b,b)\subset\mathbb{R} \rightarrow G$ is a curve in $G$ with $g(0)=e, g'(0)=\xi$.

All of these notions make sense if $V$ is replaced by a manifold $M$; in such a scenario, we say that there is an action of $G$ on $M$.  (The term \emph{representation}, however, is reserved for group actions on vector spaces only.)

\subsubsection{Adjoint and Coadjoint Actions}

For any Lie group $G$, there is a natural representation $\mathrm{Ad} : G \times \mathfrak{g} \rightarrow \mathfrak{g}$ of $G$ on its Lie algebra $\mathfrak{g}$ called the \emph{adjoint action}, given by conjugation:
\begin{equation}
\mathrm{Ad}_g \eta = TL_g \cdot TR_{g^{-1}} \cdot \eta = \left.\frac{d}{ds}\right|_{s=0} g h(s) g^{-1},
\end{equation}
where $g \in G$ and $h : (-b,b)\subset\mathbb{R} \rightarrow G$ is a curve in $G$ with $h(0)=e, h'(0)=\eta$.  Its dual is denoted $\mathrm{Ad}_g^*$ and is referred to as the \emph{coadjoint action} of $G$ on $\mathfrak{g}^*$ when $g$ is replaced by $g^{-1}$.  Explicitly,
\begin{equation}
\langle \mathrm{Ad}_g^* \mu, \eta \rangle = \langle  \mu, \mathrm{Ad}_g \eta \rangle
\end{equation}
and the coadjoint action for fixed $g \in G$ is the map $\mathrm{Ad}_{g^{-1}}^* : \mathfrak{g}^* \rightarrow \mathfrak{g}^*$.  The inversion of $g$ is a convention ensuring that this action is a left action on $\mathfrak{g}^*$ rather than a right action.

The \emph{bracket}, or \emph{commutator}, on $\mathfrak{g}$ is the differential of the adjoint action:
\begin{equation}
[\xi,\eta] = \left.\frac{d}{dt}\right|_{t=0} \mathrm{Ad}_{g(t)} \eta = \left.\frac{d}{dt}\frac{d}{ds}\right|_{s,t=0} g(t) h(s) g(t)^{-1},
\end{equation}
where $g$ and $h$ are curves in $G$ passing through the identity at $t=0$ and with $g'(0)=\xi,h'(0)=\eta$.

The \emph{infinitesimal adjoint action} is the map $\mathrm{ad}: \mathfrak{g} \times \mathfrak{g} \rightarrow \mathfrak{g}$ defined by
\begin{equation}
\mathrm{ad}_\xi \eta := [\xi,\eta].
\end{equation}
This notation permits the definition of the \emph{infinitesimal coadjoint action} $\mathrm{ad}^*$ of $\mathfrak{g}$ on $\mathfrak{g}^*$, given by the dual of the infinitesimal adjoint action:
\begin{equation}
\langle \mathrm{ad}_\xi^* \mu, \eta \rangle = \langle  \mu, \mathrm{ad}_\xi \eta \rangle.
\end{equation}

\subsubsection{The Exponential Map}

Given any Lie group $G$, there is a unique map $\exp: \mathfrak{g} \rightarrow G$ satisfying
\begin{equation}
\left\{\begin{array}{l}
\frac{d}{dt} \exp(t\xi) = TR_{\exp(t\xi)} t\xi = TL_{\exp(t\xi)} t\xi \\
\exp(0) = e
\end{array}\right.
\end{equation}
for all $\xi \in \mathfrak{g}, t \in \mathbb{R}$.  We call this map the \emph{exponential map}.  It is shown in standard texts on differential geometry that the exponential map is a local diffeomorphism that maps a neighborhood of $0 \in \mathfrak{g}$ to a neighborhood of $e \in G$.

For matrix groups, the exponential coincides with the usual matrix exponential:
\begin{equation}
\exp(tA) = \sum_{i=0}^\infty \frac{(tA)^i}{i!}.
\end{equation}

\subsection{Exponentiation on the Nematic Liquid Crystal Configuration Space} \label{appendix:nematic}

This section presents proofs of four lemmas stated in Section~\ref{section:nematic}, where a study of exponentiation on the spatially discretized nematic liquid crystal configuration space $G = \mathcal{D}(\mathbb{M}) \, \circledS \, \Omega_d^0(\mathbb{M},S^1)$ was conducted.  We restate the lemmas themselves for the reader's convenience.

\begin{lemma*} \emph{(cf. Lemma~\ref{lemma:nematic1})}
The exponential map $\exp : \mathfrak{g} \rightarrow G$ for the group $G = \mathcal{D}(\mathbb{M}) \, \circledS \, \Omega_d^0(\mathbb{M},S^1)$ is given by
\begin{equation*}
\exp(t(A,\omega)) = (e^{tA}, A^{-1}(I-e^{-tA})\omega), \tag{\ref{exp_nematic}}
\end{equation*}
where $t \in \mathbb{R}$, $A \in \mathfrak{d}(\mathbb{M})$, $\omega \in \Omega_d^0(\mathbb{M})$, $e^{tA}$ is the usual matrix exponential, and $A^{-1}(I-e^{-tA})$ is to be regarded as a power series
\begin{equation*}
A^{-1}(I-e^{-tA}) =  tI - \frac{t^2 A}{2} + \frac{t^3 A^2}{6} - \dots
\end{equation*}
(so that it is defined even for $A$ not invertible).
\end{lemma*}

\begin{proof}
Set $t=0$ to check that
\begin{equation*}
\exp(0,0) = (I,0).
\end{equation*}
Now differentiate~(\ref{exp_nematic}) with respect to $t$ to confirm that
\begin{align*}
\frac{d}{dt}\exp(t(A,\omega))
&= t(Ae^{tA},e^{-tA}\omega)  \\
&= TR_{\exp(t(A,\omega))} (tA,t\omega)
\end{align*}
\end{proof}

\begin{lemma*} \emph{(cf. Lemma~\ref{lemma:nematic2})}
The map $\tau : \mathfrak{g} \rightarrow G$ given by
\begin{equation}
\tau(A,\omega) = (\mathrm{cay}(A),\mathrm{cay}(-A/2)\omega) \tag{\ref{tau_nematic}}
\end{equation}
is a group difference map for the group $G = \mathcal{D}(\mathbb{M}) \, \circledS \, \Omega_d^0(\mathbb{M},S^1)$.  That is, $\tau$ is a local approximant to $\exp$ with $\tau(0,0)=(I,0)$, and $\tau$ satisfies
\begin{equation*}
\tau(A,\omega)^{-1} = \tau(-A,-\omega) \quad \forall (A,\omega) \in \mathfrak{g}. \tag{\ref{tau_inv_nematic}}
\end{equation*}
\end{lemma*}
\begin{proof}
Comparison of the series expansion of~(\ref{exp_nematic}) with that of~(\ref{tau_nematic}) confirms that $\tau$ is a local approximant to $\exp$.  The proof of~(\ref{tau_inv_nematic}) is by direct calculation.
\end{proof}

\begin{lemma*} \emph{(cf. Lemma~\ref{lemma:nematic3})}
The inverse right-trivialized tangent $d\tau^{-1} : \mathfrak{g} \times \mathfrak{g} \rightarrow \mathfrak{g}$ of the map~(\ref{tau_nematic}) is given by
\begin{equation*}
d\tau^{-1}_{(A,\omega)} (B,\psi) = \left( d\mathrm{cay}^{-1}_A B, \: \mathrm{cay}(-A/2)\psi+\frac{1}{2}d\mathrm{cay}_{A/2}(d\mathrm{cay}^{-1}_A B)\omega \right). \tag{\ref{dtau_inv_nematic}}
\end{equation*}
\end{lemma*}
\begin{proof}
By formulae~(\ref{dtauinv_formula}),~(\ref{exp_nematic}), and~(\ref{tau_nematic}),
\begin{align*}
d\tau^{-1}_{(A,\omega)} (B,\psi)
&= \left.\frac{d}{d\varepsilon}\right|_{\varepsilon=0} \tau^{-1}\left( \exp(\varepsilon(B,\psi))\tau(A,\omega) \right) \nonumber \\
&= \left.\frac{d}{d\varepsilon}\right|_{\varepsilon=0} \tau^{-1}\left( (e^{\varepsilon B},B^{-1}(I-e^{-\varepsilon B})\psi) (\mathrm{cay}(A), \: \mathrm{cay}(-A/2)\omega) \right) \nonumber \\
&= \left.\frac{d}{d\varepsilon}\right|_{\varepsilon=0} \tau^{-1} \left( e^{\varepsilon B}\mathrm{cay}(A), \: \mathrm{cay}(-A)B^{-1}(I-e^{-\varepsilon B})\psi + \mathrm{cay}(-A/2)\omega \right) \nonumber \\
&= \left.\frac{d}{d\varepsilon}\right|_{\varepsilon=0} \left( \mathrm{cay}^{-1}(e^{\varepsilon B}\mathrm{cay}(A)), \right. \\
& \hspace{1.7cm} \left. \mathrm{cay}\left(\frac{\mathrm{cay}^{-1}(e^{\varepsilon B}\mathrm{cay}(A))}{2}\right) \left[ \mathrm{cay}(-A)B^{-1}(I-e^{-\varepsilon B})\psi + \mathrm{cay}(-A/2)\omega\right] \right) \nonumber \\
&= \left( d\mathrm{cay}^{-1}_A B, \: \mathrm{cay}(A/2)[\mathrm{cay}(-A)\psi] + \frac{1}{2}d\mathrm{cay}_{A/2}(d\mathrm{cay_A^{-1}}B)\omega \right) \nonumber \\
&= \left( d\mathrm{cay}^{-1}_A B, \: \mathrm{cay}(-A/2)\psi + \frac{1}{2}d\mathrm{cay}_{A/2}(d\mathrm{cay_A^{-1}}B)\omega \right)
\end{align*}
\end{proof}

\begin{lemma*} \emph{(cf. Lemma~\ref{lemma:nematic4})}
For the map $\tau$ given by~(\ref{tau_nematic}), the dual of the operator $d\tau^{-1}$ is given by
\begin{equation*}
(d\tau^{-1}_{(A,\omega)})^* (C^\flat,\pi) = \left( (d\mathrm{cay}^{-1}_A)^*C^\flat + \frac{1}{2}(d\mathrm{cay}^{-1}_A)^*(d\mathrm{cay}_{A/2})^*\mathrm{skew}(\pi\omega^T), \: \mathrm{cay}(A/2)\pi \right).
\end{equation*}
Note that if $\pi$ is a scalar multiple of $\omega$, then this reduces to
\begin{equation*}
(d\tau^{-1}_{(A,\omega)})^* (C^\flat,\pi) = \left( (d\mathrm{cay}^{-1}_A)^*C^\flat, \: \mathrm{cay}(A/2)\pi \right). \tag{\ref{dtauinvstar_nematic_simplified}}
\end{equation*}
\end{lemma*}
\begin{proof}
One checks that the above formula satisfies
\begin{equation*}
\left\langle (d\tau^{-1}_{(A,\omega)})^* (C^\flat,\pi), (B,\psi) \right\rangle = \left\langle (C^\flat,\pi), d\tau^{-1}_{(A,\omega)}(B,\psi) \right\rangle \quad \forall (B,\psi) \in \mathfrak{g}.
\end{equation*}
\end{proof}

\subsection{Derivations of Cartesian Realizations} \label{appendix:cartesian}

This section presents derivations of the results reported in Section~\ref{section:cartesian}.  We employ the notation of Section~\ref{section:cartesian} and make use of Fig.~\ref{fig:sixcells} throughout the discussion.

\subsubsection{The Lie Derivative of a Discrete One-Form}

Let $A,C \in \mathcal{S}$ be discrete vector fields that approximate continuous vector fields $\mathbf{u}=(u,v),\mathbf{w}=(r,s) \in \mathfrak{X}_{\mathrm{div}}(M)$, respectively, as in Section~\ref{section:Lie1form}.

The entries of the matrix $C^\flat \in \Omega_d^1(\mathbb{M})/d\Omega_d^0(\mathbb{M})$ are then related to those of $C$ via~(\ref{cartesian_flat_operator}).  Note the disparity between the sparsity patterns of $C$ and $C^\flat$: $C$ has nonzero entries associated with cell-to-neighbor transfers only; $C^\flat$ has nonzero entries for cell-to-neighbor transfers \emph{and} for transfers between cells and their neighbors' neighbors.

Fix a pair of vertically adjacent cells $\mathcal{C}_m$ and $\mathcal{C}_n$.  For $l \in \{1,2,\dots,N\}$, let $N(l)$ denote the set of cells sharing an edge with the cell $\mathcal{C}_l$.  Taking into account the sparsity patterns of $A$ and $C^\flat$, we have
\begin{align}
(\pounds_A C^\flat)_{mn}
&= [C^\flat,A]_{mn} \nonumber \\
&= \displaystyle\sum_{l \in N(N(m))\cap N(n)} C^\flat_{ml}A_{ln} \:\:\: - \sum_{l \in N(m)\cap N(N(n))} A_{ml}C^\flat_{ln}.
\end{align}
Now, with the aid of Fig.~\ref{fig:sixcells}(b), substitute the values of the entries of $C^\flat$ and $A$ in accordance with~(\ref{flux1}-\ref{flux4}) and~(\ref{cartesian_flat_operator}):
\begin{align}
(\pounds_A C^\flat)_{mn} =
& -\frac{1}{2}\left( s^{i+1/2,j}+s^{i+1/2,j+1} \right)v^{i+1/2,j+1} \nonumber \\
& + \frac{1}{2}\left( s^{i+1/2,j-1}+s^{i+1/2,j} \right)v^{i+1/2,j-1} \nonumber \\
& -\frac{1}{2}\left( \frac{\epsilon\omega^{i+1,j}(r,s)}{2} + r^{i+1,j+1/2} + s^{i+1/2,j} \right) u^{i+1,j+1/2} \nonumber \\
& -\frac{1}{2}\left( \frac{\epsilon\omega^{i+1,j}(r,s)}{2} - r^{i+1,j-1/2} + s^{i+1/2,j} \right) u^{i+1,j-1/2} \nonumber \\
& -\frac{1}{2}\left( \frac{\epsilon\omega^{i,j}(r,s)}{2} + r^{i,j+1/2} - s^{i+1/2,j} \right) u^{i,j+1/2} \nonumber \\
& -\frac{1}{2}\left( \frac{\epsilon\omega^{i,j}(r,s)}{2} - r^{i,j-1/2} - s^{i+1/2,j} \right) u^{i,j-1/2}. \label{LACflat1}
\end{align}
Here, we have expressed the entries of $C^\flat$ in terms of the quantity $\omega(r,s)$ in anticipation of the role played by the discrete curl of the vector field $\mathbf{w}=(r,s)$:
\begin{equation}
\omega^{i,j}(r,s) = \frac{ r^{i,j-1/2} + s^{i+1/2,j} - r^{i,j+1/2} - s^{i-1/2,j} }{\epsilon} \\
\end{equation}

Rearranging, we obtain
\begin{align}
(\pounds_A C^\flat)_{mn} =
& -\frac{\epsilon}{2}\left(\omega^{i,j}(r,s)\left(\frac{u^{i,j-1/2}+u^{i,j+1/2}}{2}\right) + \omega^{i+1,j}(r,s)\left(\frac{u^{i+1,j-1/2}+u^{i+1,j+1/2}}{2}\right)  \right) \nonumber \\
& -\frac{1}{2}s^{i+1/2,j} \left(v^{i+1/2,j+1}-v^{i+1/2,j}+u^{i+1,j+1/2}-u^{i,j+1/2}\right) \nonumber \\
& -\frac{1}{2}s^{i+1/2,j} \left(v^{i+1/2,j}-v^{i+1/2,j-1}+u^{i+1,j-1/2}-u^{i,j-1/2}\right) \nonumber \\
& -\frac{1}{2}\left( (sv)^{i+1/2,j+1} + (ru)^{i+1,j+1/2} + (sv)^{i+1/2,j} + (ru)^{i,j+1/2} \right) \nonumber \\
& +\frac{1}{2}\left( (sv)^{i+1/2,j} + (ru)^{i+1,j-1/2} + (sv)^{i+1/2,j-1} + (ru)^{i,j-1/2} \right).
\end{align}
The second and third terms in the expression above are zero by the discrete divergence-freeness of $\mathbf{u}$.  The last two terms constitute a full discrete differential.  Thus, modulo a discrete differential,
\begin{equation*}
(\pounds_A C^\flat)_{mn} =
-\frac{\epsilon}{2}\left(\omega^{i,j}(r,s)\left(\frac{u^{i,j-1/2}+u^{i,j+1/2}}{2}\right) + \omega^{i+1,j}(r,s)\left(\frac{u^{i+1,j-1/2}+u^{i+1,j+1/2}}{2}\right)  \right) \tag{\ref{LACflat_y}}
\end{equation*}

Proceeding similarly for a pair of horizontally adjacent cells $\mathcal{C}_m$ and $\mathcal{C}_n$ centered at $(i-1/2,j+1/2)$ and $(i+1/2,j+1/2)$, one obtains
\begin{equation*}
(\pounds_A C^\flat)_{mn} =
\frac{\epsilon}{2}\left(\omega^{i,j}(r,s)\left(\frac{v^{i-1/2,j}+v^{i+1/2,j}}{2}\right) + \omega^{i,j+1}(r,s)\left(\frac{v^{i-1/2,j+1}+v^{i+1/2,j+1}}{2}\right)  \right). \tag{\ref{LACflat_x}}
\end{equation*}

\subsubsection{The Lie Derivative of a Discrete Vector Field}

Let $A,B \in \mathcal{S}$ be discrete vector fields that approximate continuous vector fields $\mathbf{u}=(u,v),\mathbf{v}=(p,q) \in \mathfrak{X}_{\mathrm{div}}(M)$, respectively, as in Section~\ref{section:Lie_vector_field}.  Below we derive the cartesian realization of $\pounds_A B = -[A,B] \in [\mathcal{S},\mathcal{S}]$.

Before doing so, recall from Section~\ref{section:nonholonomic} that weak equalities involving elements of $[\mathcal{S},\mathcal{S}]$ have solutions in $\mathcal{S}$ that are expressible in terms of the sparsity operator $^\downarrow$ defined in~(\ref{sparsity}).  Let us therefore compute the entries of the matrix $\pounds_A B = -[A,B] \in [\mathcal{S},\mathcal{S}]$ and apply the sparsity operator~(\ref{cartesian_sparsity}) to recover a matrix $(\pounds_A B)^\downarrow \in \mathcal{S}$ that is weakly equal to $\pounds_A B$.

Note that for any cells $\mathcal{C}_a$ and $\mathcal{C}_b$ separated by a distance of two,
\begin{equation}
(\pounds_A B)_{ab} = \sum_{c \in N(a) \cap N(b)} B_{ac}A_{cb} - A_{ac}B_{cb}.
\end{equation}
Thus, in terms of the vector fields $\mathbf{u}=(u,v)$ and $\mathbf{v}=(p,q)$ corresponding to $A$ and $B$, respectively, the $(m,n)$ entry of the matrix $(\pounds_A B)^\downarrow$ for any pair of vertically adjacent cells $\mathcal{C}_m$ and $\mathcal{C}_n$ centered at $(i+1/2,j-1/2)$ and $(i+1/2,j+1/2)$ is given by
\begin{align}
(\pounds_A B)^\downarrow_{mn} =
\frac{1}{4\epsilon^2}
&\left[ q^{i+1/2,j}(u^{i,j+1/2}+v^{i+1/2,j}-u^{i+1,j+1/2})-v^{i+1/2,j}(p^{i,j+1/2}+q^{i+1/2,j}-p^{i+1,j+1/2}) \right. \nonumber \\
&    + (q^{i+1/2,j}+p^{i+1,j-1/2}-p^{i,j-1/2})v^{i+1/2,j} - (v^{i+1/2,j}+u^{i+1,j-1/2}-u^{i,j-1/2})q^{i+1/2,j}    \nonumber \\
&   +\frac{1}{2} \{      q^{i+1/2,j} u^{i+1,j+1/2} - v^{i+1/2,j}p^{i+1,j+1/2} + p^{i+1,j-1/2}v^{i+3/2,j} - u^{i+1,j-1/2}q^{i+3/2,j}  \nonumber \\
&   \;\;\;\;\;         + p^{i,j-1/2}v^{i+1/2,j} - u^{i,j-1/2}q^{i+1/2,j} + q^{i-1/2,j}u^{i,j+1/2} - v^{i-1/2,j}p^{i,j+1/2}  \nonumber \\
&   \;\;\;\;\;         + p^{i+1,j+1/2}v^{i+3/2,j} - u^{i+1,j+1/2}q^{i+3/2,j} + q^{i+1/2,j} u^{i+1,j-1/2} - v^{i+1/2,j}p^{i+1,j-1/2} \nonumber \\
&   \;\;\;\;\; \left.  + p^{i,j+1/2}v^{i+1/2,j} - u^{i,j+1/2}q^{i+1/2,j} + q^{i-1/2,j}u^{i,j-1/2} - v^{i-1/2,j}p^{i,j-1/2} \} \right]. \label{sparsityv}
\end{align}

Note that in the first two lines of the equation above, we invoked the discrete divergence-freeness of $\mathbf{u}=(u,v)$ and $\mathbf{v}=(p,q)$ to rewrite the fluxes across the top center and bottom center edges in Fig.~\ref{fig:sixcells}(b) in terms of the fluxes across labeled edges in the figure.

Upon simplification,~(\ref{sparsityv}) reduces to
\begin{align}
(\pounds_A B)^\downarrow_{mn} = -\frac{1}{2\epsilon} \left\{ \frac{1}{\epsilon}\left[ \left(\frac{u^{i+1,j-1/2} + u^{i+1,j+1/2}}{2}\right) \left(\frac{q^{i+3/2,j} + q^{i+1/2,j}}{2}\right) \nonumber \right.\right.& \\
- \left. \left(\frac{u^{i,j-1/2} + u^{i,j+1/2}}{2}\right) \left(\frac{q^{i+1/2,j} + q^{i-1/2,j}}{2}\right) \right]&  \nonumber \\
\left. - \frac{1}{\epsilon}\left[ \left(\frac{p^{i+1,j-1/2} + p^{i+1,j+1/2}}{2}\right) \left(\frac{v^{i+3/2,j} + v^{i+1/2,j}}{2}\right) \right.\right. \nonumber \\
\left.\left. - \left(\frac{p^{i,j-1/2} + p^{i,j+1/2}}{2}\right) \left(\frac{v^{i+1/2,j} + v^{i-1/2,j}}{2}\right) \right] \right\}&.
\label{RLAB}
\end{align}

In the abbreviated notation discussed in this Section~\ref{section:cartesian}'s introduction,~(\ref{RLAB}) reads
\begin{equation*}
(\pounds_A B)^\downarrow_{mn} = -\frac{1}{2\epsilon} \left(
 \frac{u^{i+1,j}q^{i+1,j} - u^{i,j}q^{i,j}}{\epsilon}
-\frac{p^{i+1,j}v^{i+1,j} - p^{i,j}v^{i,j}}{\epsilon} \right). \tag{\ref{LAB_y}}
\end{equation*}

Proceeding similarly for a pair of horizontally adjacent cells $\mathcal{C}_m$ and $\mathcal{C}_n$ centered at $(i-1/2,j+1/2)$ and $(i+1/2,j+1/2)$, one obtains
\begin{equation*}
(\pounds_A B)^\downarrow_{mn} = -\frac{1}{2\epsilon} \left(
 -\frac{u^{i,j+1}q^{i,j+1} - u^{i,j}q^{i,j}}{\epsilon}
+\frac{p^{i,j+1}v^{i,j+1} - p^{i,j}v^{i,j}}{\epsilon} \right). \tag{\ref{LAB_x}}
\end{equation*}

\subsubsection{The Antisymmetrization of an Outer Product}

Let $E,F \in \Omega_d^0(\mathbb{M})$ be discrete zero-forms approximate continuous scalar fields $\alpha,\beta \in \mathcal{F}(M)$, respectively, as in Section~\ref{section:antisymmetrization}.

Let us compute the $(m,n)$ entry of $\mathrm{skew}(F E^T)$ for vertically adjacent cells $\mathcal{C}_m$ and $\mathcal{C}_n$ centered at $(i+1/2,j-1/2)$ and $(i+1/2,j+1/2)$, respectively:
\begin{align}
\mathrm{skew}(F E^T)_{mn}
&= \frac{1}{2}\left((FE^T)_{mn}-(EF^T)_{mn}\right) \nonumber \\
&= \frac{1}{2}\left( F_m E_n - E_m F_n\right) \nonumber \\
&= \frac{1}{2}(\beta^{i+1/2,j-1/2}\alpha^{i+1/2,j+1/2} - \alpha^{i+1/2,j-1/2}\beta^{i+1/2,j+1/2}) \nonumber \\
&= -\frac{\epsilon}{2}\left(\left(\frac{\alpha^{i+1/2,j-1/2}+\alpha^{i+1/2,j+1/2}}{2}\right)\left(\frac{\beta^{i+1/2,j+1/2}-\beta^{i+1/2,j-1/2}}{\epsilon}\right) \right. \nonumber \\
&\hspace{0.4in}\left. -\left(\frac{\beta^{i+1/2,j-1/2}+\beta^{i+1/2,j+1/2}}{2}\right)\left(\frac{\alpha^{i+1/2,j+1/2}-\alpha^{i+1/2,j-1/2}}{\epsilon}\right)\right) \nonumber \\
&= -\frac{\epsilon}{2}\left(\alpha^{i+1/2,j}\left(\frac{\beta^{i+1/2,j+1/2}-\beta^{i+1/2,j-1/2}}{\epsilon}\right) - \beta^{i+1/2,j}\left(\frac{\alpha^{i+1/2,j+1/2}-\alpha^{i+1/2,j-1/2}}{\epsilon}\right)\right). \tag{\ref{skewEF_y}}
\end{align}

Proceeding similarly for a pair of horizontally adjacent cells $\mathcal{C}_m$ and $\mathcal{C}_n$ centered at $(i-1/2,j+1/2)$ and $(i+1/2,j+1/2)$, one obtains
\begin{equation*}
\mathrm{skew}(F E^T)_{mn} = -\frac{\epsilon}{2}\left(\alpha^{i,j+1/2}\left(\frac{\beta^{i+1/2,j+1/2}-\beta^{i-1/2,j+1/2}}{\epsilon}\right) - \beta^{i,j+1/2}\left(\frac{\alpha^{i+1/2,j+1/2}-\alpha^{i-1/2,j+1/2}}{\epsilon}\right)\right). \tag{\ref{skewEF_x}}
\end{equation*}

\pdfbookmark[1]{References}{References}
\bibliographystyle{plain}
\bibliography{references}{}

\end{document}